{}

\documentclass{article}

\usepackage[utf8]{inputenc}
\usepackage{lmodern}
\usepackage[dvipsnames]{xcolor}
\usepackage{graphicx}
\usepackage{cite}

\usepackage{comment}
\usepackage{microtype}

\usepackage{tikz}
\usepackage{pgfplots}
\pgfplotsset{compat=1.18}
\usepackage{wrapfig}

\usepackage{geometry}
\usepackage{afterpage}
\usepackage{multirow}
\usepackage{tabularx} 
\usepackage{graphicx}

\usepackage{bm}
\usepackage{mathtools,slashed}
\usepackage{amsmath, amssymb}

\numberwithin{equation}{section}
\usepackage{mathrsfs}
\usepackage{physics} 
\usepackage{bigints}

\usepackage{url}
\usepackage[bottom,hang,flushmargin]{footmisc}
\usepackage{fancyhdr}
\usepackage{nicefrac}

\DeclareFontFamily{OT1}{mathc}{}
\DeclareFontShape{OT1}{mathc}{m}{it}{<-> mathc10}{}
\DeclareMathAlphabet{\mathabxcal}{OT1}{mathc}{m}{it}

\usepackage{subfiles}
\usepackage{enumitem}\setlist[enumerate]{font=\bfseries}

\usepackage{titlesec}
\usepackage[toc,page]{appendix}
\usepackage[breaklinks]{hyperref}
\usepackage{cleveref}

\titleformat*{\section}{\LARGE\bfseries}
\titleformat*{\subsection}{\Large\bfseries}
\titleformat*{\subsubsection}{\large\bfseries}

\usepackage{comment}

\newcommand{\SU}{\text{SU}}
\newcommand{\su}{\mathfrak{su}}

\newcommand{\Z}{\mathbb{Z}}
\newcommand{\R}{\mathbb{R}}
\newcommand{\Pexp}{\text{P}\overleftarrow{\text{exp}}}

\newcommand{\vp}{\varphi}
\newcommand{\p}{\partial}
\newcommand{\CP}{\mathbb{CP}^1}

\newcommand{\g}{\mathfrak{g}}
\newcommand{\Ac}{\mathcal{A}}
\newcommand{\Cas}{\mathabxcal{Cas}}
\newcommand{\ri}{{\rm i}}
\newcommand{\s}{\sigma}

\newcommand{\reg}{\text{reg}}
\usepackage{simpler-wick}

\DeclareFontEncoding{LS1}{}{}
\DeclareFontSubstitution{LS1}{stix}{m}{n}
\DeclareSymbolFont{stixsymbols}{LS1}{stixscr}{m}{n}
\SetSymbolFont{stixsymbols}{bold}{LS1}{stixscr}{b}{n}
\DeclareMathSymbol{\kay}{\mathalpha}{stixsymbols}{"6B}

\makeatletter
\let\@keywords\@empty
\let\@subject\@empty
\providecommand{\keywords}[1]{\gdef\@keywords{#1}}
\providecommand{\subject}[1]{\gdef\@subject{#1}}
\def\thetitle{\@title}
\def\theauthor{\@author}
\def\thesubject{\@subject}
\def\thedate{\@date}
\def\thekeywords{\@keywords}
\makeatother
\AtBeginDocument{
\hypersetup{pdftitle={\thetitle}}
\hypersetup{pdfauthor={\theauthor}}
\hypersetup{pdfsubject={\thesubject}}
\hypersetup{pdfkeywords={\thekeywords}}}

\usepackage{float}

\title{Integrable Structure of Wess-Zumino-Zitten Models}

\author{Sylvain Lacroix and Adrien Molines}

\newcommand{\tightoverset}[2]{%
  \mathop{#2}\limits^{\vbox to -.5ex{\kern-0.75ex\hbox{$#1$}\vss}}}

\newcommand{\Aq}{\mathsf{A}}
\newcommand{\Bq}{\mathsf{B}}
\newcommand{\Cq}{\mathsf{C}}
\newcommand{\Tq}{\mathsf{T}}
\newcommand{\Wq}{\mathsf{W}}
\newcommand{\Lq}{\mathsf{L}}
\newcommand{\Hq}{\mathsf{H}}

\newcommand{\Jq}{\mathsf{J}}
\newcommand{\Iq}{\mathsf{I}}
\newcommand{\Gq}{\mathsf{G}}
\newcommand{\Kq}{\mathsf{K}}
\newcommand{\Fq}{\mathsf{F}}
\newcommand{\kcl}{k_{\text{cl}}}
\newcommand{\KdV}{\text{KdV}}

\newcommand{\cl}{\text{cl}}
\newcommand{\Ec}{\mathcal{E}}
\newcommand{\Pc}{\mathcal{P}}

\newcommand{\BAE}{\text{BAE}}

\usetikzlibrary{decorations.pathreplacing,decorations.markings}

\begin{document}

\begin{titlepage}

\begin{center}

\vspace*{2cm}

\begingroup\Large\bfseries
On the Integrable Structure of the SU(2) Wess-Zumino-Novikov-Witten Model
\par\endgroup

\vspace{1.cm}

\begingroup
Sylvain Lacroix$^{a,}$\footnote{E-mail:~sylvain.lacroix@sorbonne-universite.fr} and
Adrien Molines$^{b,}$\footnote{E-mail:~amolines@phys.ethz.ch}
\endgroup

\vspace{1cm}

\begingroup
$^a$\it Sorbonne Université, CNRS, Laboratoire de Physique Théorique et Hautes \'Energies,\\ LPTHE, F-75005 Paris, France\\~\\

$^b$\it  Institut f\"ur Theoretische Physik, ETH Z\"urich,\\
Wolfgang-Pauli-Strasse 27, 8093 Z\"urich, Switzerland\\
\endgroup

\end{center}

\vspace{1cm}

\begin{abstract}
\noindent
This paper is devoted to the quantum integrable structure of Wess-Zumino-Novikov-Witten models, formed by an infinite number of commuting Integrals of Motion (IMs) in their current algebra. Focusing for simplicity on the $\SU(2)$ case, we obtain the first four commuting higher-spin local IMs, starting from a general SU(2)-invariant ansatz and imposing their commutativity. We further show evidence of their commutativity with quantum non-local IMs, which were already built in the literature as Kondo defects. We then investigate the diagonalization of these local operators on $\widehat{\su(2)}_k$ Verma modules: we explicitly find the first few eigenvectors and further discuss the affine Bethe ansatz and ODE/IQFT conjectures, which predict the full eigenstates and spectrum of the integrable structure. Our results show a perfect match between the direct diagonalization and these overarching conjectures. We conclude by discussing several outlooks, including multi-current generalisations, massive deformations and a general long-term program towards the first principle quantisation of 2-dimensional integrable sigma-models.
\end{abstract}
\end{titlepage}

\tableofcontents

\thispagestyle{empty}
\clearpage

\setcounter{page}{1}
\setcounter{footnote}{0} 

\section{Introduction}\label{sec:Intro}

\paragraph{First principle quantum integrability.} Integrable field theories are very valuable models in theoretical and mathematical physics, which find applications in condensed matter theory, high-energy physics, string theory and various other domains. They are characterised by the presence of an infinite number of symmetries, which strongly constrain their dynamics and allow for the exact computation of their physical observables (such as classical solutions, energy spectrum, correlation functions, ...). This integrability property is generally first established at the classical level. In this context, it more precisely consists in the existence of an infinite number of Integrals of Motion (IMs) which mutually Poisson-commute, thus encoding an infinite number of commuting symmetries. For relativistic (1+1)-dimensional field theories, these IMs include the Hamiltonian and momentum, generating space-time translations, but also more complicated quantities such as non-local charges and higher-spin local IMs, which encode hidden symmetries of the model. Given this classical setup, a natural route to study its quantisation goes through the following steps:\vspace{-3pt}
\begin{enumerate}\setlength\itemsep{0.5pt}
    \item quantise the Poisson structure to find the algebra of field operators and the Hilbert space of the quantum theory ;
    \item promote the classical IMs into commuting operators inside of this algebra ;
    \item diagonalise these commuting operators to find the eigenstates and spectrum of the model.
\end{enumerate}
The first step essentially defines the quantum theory, the second one establishes its quantum integrability and the third one solves it  (at least partially). We will refer to this program as that of \textit{first principle quantum integrability}. Despite its naturalness, this endeavour is generally quite difficult. In fact, even the first step of rigorously defining the algebra of field operators is a notoriously arduous task for a general interacting quantum field theory.

\paragraph{Integrability in CFT.} One potential strategy to make progress in this quantisation program was proposed by Zamolodchikov in the seminal papers~\cite{Zamolodchikov:1987jf,Zamolodchikov:1989hfa}. The main idea is to follow the renormalisation group flow of the model and to first investigate the integrable structure of its ultraviolet fixed-point, which defines a 2d Conformal Field Theory (CFT)~\cite{Belavin:1984vu}. The dynamics of such a theory decouples into two chiral sectors, formed by left-moving and right-moving fields respectively. 
Remarkably, the classical integrable structure typically exhibits a similar decoupling at the conformal point: namely, half of the Poisson-commuting IMs are built from left-moving fields only, while the other half is built from the right-moving fields. The quantisation of chiral sectors in a 2d CFT has been extensively studied in the literature, leading to the notion of chiral algebras. At the quantum level, the left-moving fields become operators on a Hilbert space, which we can multiply together to form new composite left-moving fields and which satisfy non-trivial commutation relations one with another. This is the structure captured by the left-moving chiral algebra $\Ac$, through the notion of Operator Product Expansions. The right-moving fields commute with the left-moving ones and form a similar algebra $\overline{\Ac}$ on their own. These chiral algebras $\Ac$ and $\overline{\Ac}$ depend on the 2d CFT under consideration and are the fundamental building blocks for its construction at the quantum level. Crucially for us, they provide a rigorous and well-controlled answer to the first step in our first principle quantisation program, namely the construction of the algebra of field operators.

We can thus proceed with the second step, \textit{i.e.} the search for an infinite number of commuting operators in this algebra, forming the quantum integrable structure of this conformal theory. If $\Aq$ is any left-moving field in the chiral algebra $\Ac$, it is clear that its integral $\int \Aq(x)\,\dd x$ over the spatial coordinate $x$ is a conserved local charge of the model. We thus get an infinite number of such local IMs, built from all the chiral fields in $\Ac$. This is reminiscent of integrability but we recall that the latter further requires the commutativity of these IMs. The integrals of two fields $\Aq$ and $\Bq$ in $\Ac$ will generally not commute one with another: however, we can obtain this commutativity for specific choices of $\Aq$ and $\Bq$, by requiring appropriate conditions on the commutator $[\Aq(x),\Bq(y)]$ (or equivalently on the OPE of $\Aq$ and $\Bq$). To obtain an integrable structure, we thus search for an infinite tower of local higher-spin IMs $\Iq^{(s)}$ (labelled by their spin $s$), built as integrals of left-moving densities chosen such that the $\Iq^{(s)}$  pairwise commute. Following the above discussion, this last condition is quite non-trivial and puts very strong constraints on these densities. In particular, it generally restricts the spins $s$ appearing in this tower to a specific model-dependent subset of integers. As one should expect, there also exists an infinite tower of local commuting IMs $\overline{\Iq}\null^{(s)}$ built from right-moving densities, with negative spins.

An important role in the CFT is played by its stress-energy tensor, which splits into a left-moving component $\Tq\in\Ac$ and a right-moving one $\overline{\Tq}\in\overline{\Ac}$. These fields exist for any 2d CFT and generate its conformal symmetries. They satisfy universal commutation relations and generate a subalgebra of $\Ac$ (resp. $\overline{\Ac}$) called the Virasoro algebra. For all CFT's integrable structures, the first local IMs $\Iq^{(1)}$ and $\overline{\Iq}\null^{(1)}$ have spin $\pm 1$ and are the light-cone Hamiltonians, whose densities are $\Tq$ and $\overline{\Tq}$. As the generators of space-time translations, these operators trivially commute with the integral of any other chiral field. The non-trivial part of the integrable structure is then the higher-spin IMs $\Iq^{(s)}$ and $\overline{\Iq}\null^{(s)}$, $s>1$, whose definition depends on the model and the chiral algebra that one works with. In the simplest cases, such as Liouville theory and minimal models, the Virasoro algebra in fact forms the whole chiral algebra: in other words, the stress-energy tensor $\Tq$ is the only independent local left-moving field. For these models, an integrable structure called Korteweg–De Vries (KdV) has been constructed in~\cite{Sasaki:1987mm,Eguchi:1989hs,Kupershmidt:1989bf,Bazhanov:1994ft}. The corresponding local IMs $\Iq^{(2p-1)}_\KdV$ have odd spins and their densities are built from normal-ordered products of $\Tq$ and its derivatives. 

For more general CFTs however, the chiral algebra is larger than the Virasoro algebra, \textit{i.e.} the model admits more local left-moving fields than the stress-energy tensor alone. We then say that the model has an extended conformal symmetry. Although the KdV local IMs $\Iq^{(2p-1)}_\KdV$ can be built in any of these models (inside their Virasoro subalgebra), this is generally not the most natural integrable structure to consider as it does not make use of the full extended conformal symmetry. In particular, these KdV charges would have degenerate spectrum and would typically not commute with non-local IMs or be preserved under massive deformations. We are thus led to search for different integrable structures depending on the choice of underlying chiral algebra. This has been done for instance for $\mathcal{W}$-algebras~\cite{Feigin:1993sb,Bazhanov:2001xm,Frenkel:2016gxg,Kudrna:2025bzg} (describing Toda CFTs). The corresponding local IMs take a different form than the KdV ones, involving more fields than $\Tq$ and following a different spin pattern.

\paragraph{Integrable structure of the WZNW model.} The goal of this article is to study the integrable structure of another very well-known CFT, the celebrated Wess-Zumino-Novikov-Witten (WZNW) model~\cite{Wess:1971yu,Witten:1983tw,Witten:1983ar,Novikov2007MULTIVALUEDFA}. This is a non-linear sigma-model, whose target space is a compact Lie group $G$. In this paper, we will restrict to the case of $G=\;$SU(2) for simplicity, but keep the discussion more general in the introduction, to explain the motivations in a broader context. The WZNW model is characterised by a specific chiral algebra $\Ac$ called a current algebra. It is generated by left-moving fields $\Jq^a(x+t)$, with $a=1,\dots,\dim G$ an index associated with a choice of orthonormal basis $\lbrace T^a \rbrace$ of the Lie algebra $\g=\text{Lie}(G)$. These fields have spin 1 and satisfy the commutation relations
\begin{equation}\label{eq:IntroCurrent}
    \bigl[\Jq^a(x),\Jq^b(y)\bigr]=\ri\pi \Big(2\ri f^{abc}\Jq^c(y)\delta(x-y)-k\,\delta^{ab}\,\delta' (x-y)\Big)\,,
\end{equation}
where $\delta(x-y)$ is the Dirac distribution, $f^{abc}$ are the structure constants of the basis $\lbrace T^a \rbrace$ and $k$ is the main parameter of the model called the level (generally taken to be a positive integer to ensure unitarity). The full chiral algebra $\Ac$ is formed by taking normal-ordered products of $\Jq^a$ and their derivatives (denoted with round brackets). For instance, the stress-energy tensor is defined, up to a proportionality factor, as the composite field $(\Jq^a\Jq^a)$. The current algebra $\Ac$ then contains a Virasoro subalgebra but also quite many more fields, which are built from $\Jq^a$ but cannot be written in terms of $\Tq$ alone. This is an example of extended conformal symmetry. As usual, the theory also admits right-moving currents $\overline{\Jq}\null^a(x-t)$, forming a similar algebra.

Following the general discussion above, we are interested in finding an integrable structure in the current algebra of the WZNW model.\footnote{Various integrability results for the WZNW model have already been studied, see for instance~\cite{Kobayashi:1989sg,Leblanc:1990fi,Bonora:1990hc,Durganandini:1991gma,Brazhnikov:1996fa,Haldane:1992sj,Bernard:1994wg,Bazhanov:2003ua,Lukyanov:2003rt,Lukyanov:2006cu}. We refer to the conclusion for a discussion of these works and their potential relations/differences with the present paper.} This is a quite natural question, due to the ubiquity of this model in many domains of theoretical physics. Moreover, it is also motivated and guided by various results on the so-called non-abelian Thirring or $\lambda$-model~\cite{Dashen:1973nhu,Balog:1993es,Sfetsos:2013wia}. This is a non-linear sigma-model obtained by perturbing the WZNW theory by the operator $\lambda\,\Jq^a\overline{\Jq}\null^a$, with $\lambda$ a deformation parameter. This recouples together the two chiralities of the theory, breaking the left- and right-moving dynamics of the currents. Accordingly, it also triggers an RG-flow for the parameter $\lambda$~\cite{Kutasov:1989dt,Itsios:2014lca} and thus breaks the conformal invariance.\footnote{Note that $\Jq\overline{\Jq}$-deformations by abelian currents are marginal and thus preserve the conformal symmetry. However, the present deformation is built from non-abelian currents, which is what triggers the RG-flow (at 2nd order in $\lambda$).} More precisely, it defines a relevant perturbation, with the WZNW model at $\lambda=0$ being the ultraviolet fixed-point of the RG-flow. Conjecturally, this theory becomes strongly coupled in the infrared and develops a dynamical mass-gap. This deformed model has been proven to be integrable at the classical level in~\cite{Balog:1993es,Sfetsos:2013wia,Itsios:2014vfa}. Its quantisation has been studied in~\cite{Ahn:1990gn,Hollowood:2015dpa}, where an integrable S-matrix has been proposed for the scattering of its massive excitations, in~\cite{Zamolodchikov:1991vg,Ravanini:1992fs,Hollowood:1993ac,Babichenko:2003rf,Evans:1994hi,Hegedus:2005bg} using the Thermodynamic Bethe Ansatz / integral equations and in~\cite{Bazhanov:1989yk,Appadu:2017fff} from lattice discretisations (see also~\cite{Konik:2000yg,nonne2011competing,may2020topology} for applications in condensed matter physics). Our motivation for this paper is to investigate the first principle quantum integrability of this theory starting from its UV fixed point, namely the WZNW CFT. In this context, the main perspective for future developments is then to study the deformation of this conformal integrable structure under the massive $\Jq^a\overline{\Jq}\null^a$-deformation. We refer to the conclusion for a more extensive discussion of this perspective.

The known results on the classical integrable structure of the $\lambda$-model will serve as a useful guide for our endeavour. Part of this integrable structure is obtained through the existence of a flat Lax connection satisfying a Maillet bracket~\cite{Maillet:1985fn,Maillet:1985ek}, as established in~\cite{Balog:1993es,Sfetsos:2013wia,Itsios:2014vfa}. This allows us to construct an infinite number of non-local Poisson-commuting IMs from the trace of the monodromy of this Lax connection, by considering its expansion in the auxiliary spectral parameter. In addition to these non-local IMs, the classical integrable structure of the $\lambda$-model is also composed by higher-spin local IMs, built from an alternative construction~\cite{Evans:1999mj,Evans:2000hx,Lacroix:2017isl}. For the purpose of this paper, we need to consider the behaviour of this integrable structure in the conformal limit $\lambda\to 0$, towards the WZNW model. There are in fact two ways of taking this limit in the Lax connection, by considering two different rescalings of the spectral parameter. Crucially, the first limit ends up being expressed purely in terms of the left-moving classical currents $J^a_\cl(x+t)$, while the second one is built from the right-moving currents $\overline{J}\null ^a_\cl(x-t)$. This is the chiral decoupling of the integrable structure advertised earlier in this introduction. More precisely, the ``left-moving'' limit produces the following generating function for the non-local IMs (working here with $L$-periodic fields):\vspace{-1pt}
\begin{equation}\label{eq:MonodromyWZW}
    \Tr\left[\Pexp\left( \ri\rho\int_0^L J_{\cl}^a(x)\,T^a\,\dd x \right)\right]\,,
\end{equation}
where $\rho$ is the spectral parameter and $\Pexp$ denotes the path-ordered exponential in the group $G$. In particular, this quantity is conserved and Poisson-commutes with itself for all values of the spectral parameter. Its quantisation into a commuting non-local combination of the quantum currents $\Jq^a(x)$ has been investigated in~\cite{Bachas:2004sy,Gaiotto:2020fdr,Gaiotto:2020dhf}. In this context, it is referred to as a chiral Kondo defect.

Our focus in the present paper is more on the higher-spin local IMs, which play an important role in integrable field theories~\cite{Zamolodchikov:1987jf,Zamolodchikov:1989hfa,Zamolodchikov:1978xm,Parke:1980ki} (see also the discussion on non-ultralocality in the conclusion). As above, one can take the $\lambda\to 0$ limit of the classical local IMs of the $\lambda$-model~\cite{Evans:1999mj,Evans:2000hx,Lacroix:2017isl}. As expected, one then observes that they split into two infinite towers $I^{(s)}_\cl$ and $\overline{I}\null^{(s)}_\cl$, expressed in terms of the left- and right-moving currents respectively. More precisely, the left-moving ones take the form
\begin{equation}\label{eq:IntroClassicalIM}
    I^{(s)}_\cl = \int_0^L \tau_{a_1 \dots a_{s+1}}\,J^{a_1}_\cl(x)\cdots J^{a_{s+1}}_\cl(x)\,\dd x\,,
\end{equation}
where $\tau_{a_1 \dots a_{s+1}}$ are well-chosen $(s+1)$-tensors on the Lie algebra $\g$~\cite{Evans:1999mj}. These tensors are characterised by strong constraints, which ensure the Poisson-commutativity of the local IMs one with another and with the non-local ones \eqref{eq:MonodromyWZW}, as well as their invariance under the global $G$-symmetry of the WZNW model (which acts on the current $J^a_\cl\,T^a$ by conjugation). It turns out that there exist solutions to these constraints only for specific degrees, thus fixing the spin pattern of the IMs $I^{(s)}_\cl$. Namely, the allowed spins are the affine exponents of $\g$~\cite{kac1989infinite} (an infinite set of positive integers related to the degrees of primitive invariant $\g$-tensors). 

One of the main goal of this paper is to study the quantisation of these local IMs in the quantum current algebra. The most direct approach to this question is to consider a very general ansatz for the densities of the quantum IMs $\Iq^{(s)}$ and to constrain it by imposing the commutativity of these operators. The above classical results serve as a useful guide to construct this ansatz: namely, it is natural to search for quantum IMs whose spins are the affine exponents of $\g$ and which are invariant under the global $G$-symmetry of the model. With these constraints, the density of $\Iq^{(s)}$ naturally includes the normal-ordered product $\tau_{a_1 \dots a_{s+1}}(\Jq^{a_1}\cdots \Jq^{a_{s+1}})$, which is the naive quantisation of the classical density in equation \eqref{eq:IntroClassicalIM}. However, it also contains many more terms of the form $\kappa_{a_1 \dots a_{p}}(\p^{m_1}\Jq^{a_1}\cdots \p^{m_p}\Jq^{a_{p}})$, built from a lower number of currents but also involving their derivatives. Indeed, we expect that the absence of such terms in the classical IM \eqref{eq:IntroClassicalIM} is an accident of the classical limit and that quantum corrections of this form are necessary to obtain commuting IMs at the quantum level. At a given spin, there are only a finite number of such possible corrections: a natural strategy to build $\Iq^{(s)}$ is then to start with a general ansatz containing all these terms with arbitrary coefficients and to constrain those by imposing the commutativity $\bigl[\Iq^{(s)},\Iq^{(s')}]=0$ spin by spin, using the fundamental commutation relation \eqref{eq:IntroCurrent} of the current algebra.

\paragraph{Main results for $\bm{G=}\,$SU(2).} Although the above quantisation program is conceptually straightforward, its concrete implementation is slightly technical. For simplicity, we thus restrict to the simplest case in this paper, by taking the Lie group $G$ to be SU(2). The affine exponents of this group are the odd positive integers, such that we expect local IMs $\Iq^{(2p-1)}$ of the same spins as the KdV ones. The search for these IMs is the main topic of Section \ref{sec:structure} of this paper. Let us briefly summarise our results. As one should expect, the commutativity of the IMs put very strong constraints on the parameters entering the ansatz for their densities. In fact, we find that these constraints only have two isolated solutions. The first one corresponds to densities which can be rewritten purely in terms of the stress-energy tensor $\Tq$ and thus recovers the KdV IMs in terms of this field. This is to be expected and serves as a consistency check of our procedure, since we know that the current algebra contains the KdV integrable structure inside of its Virasoro subalgebra.

Fortunately for us, our analysis produces a second independent solution, different from the KdV one. The corresponding local IMs $\Iq^{(2p-1)}$ cannot be expressed in terms of $\Tq$ alone and thus define a new integrable structure inherent to the WZNW model, which we expect to be the correct quantisation of the classical IMs \eqref{eq:IntroClassicalIM}. More precisely, we have found explicit expressions for the first IMs up to $\Iq^{(7)}$ and expect this tower to go on infinitely. As an illustration, we report on the first two here (up to inconsequential proportionality factors):
\begin{equation*}
    \Iq^{(1)} \propto \int_0^L \Tq\, \dd x = \frac{1}{k+2}\int_0^L (\Jq^a\Jq^a)\,\dd x \qquad\text{and} \qquad \Iq^{(3)}\propto\int_0^L \Bigl((\Tq^2) + \frac{3}{2(k+2)}(\p\Jq^a\p\Jq^a) \Bigr)\dd x\,.
\end{equation*}
In Section \ref{sec:Kondo}, we also give evidence that $\Iq^{(3)}$ commutes with the first non-local IM extracted from the Kondo defect, as one should hope for in view of the classical setup. Interestingly, the KdV charge $\Iq^{(3)}_\KdV$ does not have this property, making our new integrable structure quite more natural for the WZNW model and its $\Jq^a\overline{\Jq}\null^a$-deformation.

\paragraph{Diagonalization of the integrable structure.} So far, we have focused on the first two steps in our first principle quantum integrability program: the definition of the quantum algebra and the construction of the commuting integrable structure. The second half of this paper is devoted to the third step, namely the diagonalization of this integrable structure. To do so, we first have to translate the above construction of commuting IMs to concrete operators acting on the Hilbert space. For this, we put our theory on the cylinder and extract the Fourier modes $\Jq^a_n$ of the current $\Jq^a(x)$. These are the fundamental operators of the theory, forming an infinite dimensional Lie algebra called the affine $\widehat{\su(2)}_k$ algebra. In particular, the Hilbert space is then built from highest-weight representations of this affine algebra called Verma modules. In Section \ref{sec:diagonalization}, we investigate the diagonalization of the first two operators $\Iq^{(1)}$ and $\Iq^{(3)}$ on these modules. We start with a direct approach and find the ground and first few excited states' eigenvalues explicitly. We further compare these results with a more general but conjectural construction called the affine Bethe ansatz~\cite{Feigin:2007mr,Lacroix:2018fhf,Gaiotto:2020dhf,Schechtman:1991hgd}. The latter is expected to produce all the eigenstates in the Verma module and we check this conjecture for the first few states obtained by direct diagonalization.

Remarkably, there exists another conjecture which predicts the eigenvalues of the IMs on any Bethe eigenstate and for arbitrary spin. It fits in a quite more general framework based on so-called affine Gaudin models and a conjectural affinisation of the geometric Langlands correspondence~\cite{Feigin:2007mr,Lacroix:2018fhf}. It encodes the spectrum of $\Iq^{(2p-1)}$ in the WKB expansion of well-chosen Ordinary Differential Equations (ODE) and can thus be seen as an incarnation of the celebrated ODE/IQFT (or ODE/IM) correspondence, which relates the spectrum of Integrable QFTs with properties of ODEs and which was initially developed in~\cite{Dorey:1998pt,Bazhanov:1998wj} for the KdV integrable structure. For the WZNW model, this correspondence has been investigated in~\cite{Gaiotto:2020dhf}, where it was argued that the ODEs encode the spectrum of the non-local Kondo IMs. In the present paper, we propose an extension of this conjecture for the local IMs $\Iq^{(2p-1)}$ and perform several explicit checks by comparing with the direct diagonalization.

\paragraph{Plan of the paper.} Section \ref{sec:2D_CFT} is a review of 2d CFTs on the cylinder and in particular of the SU(2) WZNW model. This gives us the appropriate framework to investigate the commuting local IMs of this model in Section \ref{sec:structure}. Section \ref{sec:diagonalization} is then devoted to the diagonalization of these operators and the affine Bethe ansatz and ODE/IQFT conjectures. Moreover, Section \ref{sec:Kondo} contains a brief review of the non-local Kondo IMs and preliminary checks of their commutativity with the local ones. We finally conclude and discuss future perspectives in Section \ref{sec:Conc}.

\newpage
\section{2d CFT and WZNW model on the cylinder}
\label{sec:2D_CFT}

The main formalism used in this paper is that of 2-dimensional conformal field theories (CFTs) on the cylinder, and more precisely Wess-Zumino-Novikov-Witten models. We thus start with a brief refresher on these subjects, fixing the conventions used throughout the article. In many standard references, CFTs are studied on the sphere rather than the cylinder: for completeness, we briefly discuss the relation between these two formulations in Appendix \ref{app:Sphere}, focusing on the cylinder in the main text. The readers already familiar with this material are invited to skip to the next section.

\subsection[2d CFT on the cylinder]{2d CFT on the cylinder}
\label{subsec:2DCFT}

We consider a 2-dimensional cylinder, parameterised by a time coordinate $t\in\R$ and a $L$-periodic spatial coordinate $x\in \R/L\Z$. We will study relativistic CFTs and thus equip this cylinder with a flat Lorentzian metric $\dd t^2-\dd x^2$, with associated light-cone coordinates $x^{\pm}=t\pm x$. 

\paragraph{Chiral algebra and OPE.} A key feature of 2-dimensional conformal symmetry is the decoupling of the left-moving and right-moving chiralities. In particular, one of the most important characteristic of a 2d CFT is its chiral algebra $\mathcal{A}$, formed by all the left-moving local fields of the theory, \textit{i.e.} fields which depend only on $x^+=t+x$. In parallel, one can consider another chiral algebra $\overline{\Ac}$ formed by right-moving fields. Since these two chiral sectors decouple and have a very similar structure, we will mostly focus on the left-moving algebra $\mathcal{A}$. To lighten the notations, we will generally write only the spatial dependence of fields $\Aq(x)$ in $\mathcal{A}$, effectively working with fields on the circle: the time dependence can be trivially reinstated by substituting $x$ for $x+t$. The chiral algebra decomposes as $\Ac=\bigoplus_{n\in\Z_{\geq 0}} \Ac_n$, where $\Ac_n$ is the subspace of left-moving fields $\Aq$ of conformal dimension $h(\Aq)=n$.\footnote{For our purposes, it will be enough to restrict to local fields of bosonic CFTs, with only integer conformal dimensions.} This subspace generally has finite dimension, meaning that although there are infinitely many linearly independent chiral fields, there are only a finite number of them with a given conformal dimension. Note that if $\Aq\in\Ac$ is a left-moving field, so is its derivative $\p\Aq$, with conformal dimension $h(\p\Aq)=h(\Aq)+1$ (hence $\Ac$ is stable under derivation).

In QFT, products of fields usually diverge when the evaluation points collide, as formalised by the Operator Product Expansion (OPE). A key feature of the chiral algebra is that it is stable under the OPE. Namely, if $\Aq$ and $\Bq$ are chiral fields in $\Ac$, their OPE
\begin{equation}
    \Aq(x)\Bq(y)=\sum_{k\in \Z}\frac{\{\Aq\Bq\}^{(k)}(y)}{(x-y)^{k+1}}=\sum_{k = 0}^{h(\Aq)+h(\Bq)-1} \frac{\{\Aq\Bq\}^{(k)}(y)}{(x-y)^{k+1}}+\reg
    \label{ope_general}
\end{equation}
involves only other chiral fields $\{\Aq\Bq\}^{(k)}\in\Ac$, with conformal dimensions $h(\Aq)+h(\Bq)-k-1$. In the second equality, we have separated the OPE into a regular part `$\reg$' (finite when $x\to y$) and a divergent part. As made apparent by the range of summation of $k$, this divergent part only contains a finite number of terms and has a maximal degree of divergence, since $\{\Aq\Bq\}^{(k)}=0$ for all $k\geq h(\Aq)+h(\Bq)$. The OPE \eqref{ope_general} defines the main algebraic structure on $\Ac$ relevant to describe the operator formalism of our 2d CFT. Formally, it consists of an infinite number of bilinear products $(\Aq,\Bq)\mapsto\{\Aq\Bq\}^{(k)}$ on $\Ac$, indexed by integers $k\in\Z$ and obeying a number of constraints encoding the commutativity and associativity of the OPE.

A powerful use of the OPE is that its divergent part completely encodes the commutation relations of fields: 
\begin{equation}
    [\Aq(x),\Bq(y)]= 2\ri\pi\, \sum_{k=0}^{h(\Aq)+h(\Bq)-1}\frac{(-1)^k}{k!}\{\Aq\Bq\}^{(k)}(y)\,\delta^{(k)}(x-y)\,.
    \label{field_commutator}
\end{equation}
As expected from locality, this commutator is a finite linear combination of the Dirac distribution
\begin{equation}
    \delta(x-y)=\frac{1}{L}\sum_{n\in\mathbb{Z}}e^{2\ri\pi\, n(x-y)/L}
\end{equation}
on the circle and its derivatives $\delta^{(k)}(x-y)=\partial_x^k\delta(x-y)$. 

\paragraph{Fourier modes and their commutators.} We define the Fourier modes $(\Aq_n)_{n\in\Z}$ of a generic chiral field $\Aq\in\Ac$ through the expansion
\begin{equation}
    \Aq(x) = \left(\frac{2\ri\pi}{L}\right)^{h(\Aq)} \sum_{n\in\Z} \Aq_n\, e^{-2\ri\pi\,nx/L}\,, \qquad \Aq_n=\left(\frac{L}{2\ri\pi}\right)^{h(\Aq)}\,\int_0^{L}  e^{2\ri\pi\,nx/L}\,\Aq(x)\,\frac{\dd x}{L}\,.\label{mode_expansion_general}
\end{equation}
These modes are (dimensionless) operators acting on the Hilbert space of the theory, whose interpretation will be discussed later in this subsection. We note that $(\p\Aq)_n = -n\,\Aq_n$.

The divergent part of the OPE \eqref{ope_general} controls the commutation relation of the modes of $\Aq$ and $\Bq$. Indeed, by a few manipulations, the field commutator \eqref{field_commutator} translates to
\begin{equation}
     {[\Aq_m,\Bq_n]=\sum_{k=0}^{h(\Aq)+h(\Bq)-1}\frac{m^k}{k! } \{\Aq\Bq\}^{(k)}_{m+n}}\,.    \label{commutation_mode_ope}
\end{equation}

\paragraph{Stress-energy tensor.} An important field in the chiral algebra $\Ac$ is the (left-moving component of the) stress-energy tensor $\Tq(x)$. It is a distinguished field of conformal dimension $h(\Tq)=2$, present in every CFT, which generates conformal transformations on the chiral algebra. Its OPE is completely fixed by conformal invariance:
\begin{equation}
    \label{OPE_T}
     {\Tq(x)\Tq(y)=\frac{\partial \Tq(y)}{x-y}+\frac{2\Tq(y)}{(x-y)^2}+\frac{c/2}{(x-y)^4}+\reg}\,,
\end{equation}
up to the central charge $c\in\R$ which is model dependent. According to equation \eqref{field_commutator}, its commutator is then given by \vspace{-2pt}
\begin{equation}
    [\Tq(x),\Tq(y)]=-2\ri\pi\,\Big((\Tq(x)+\Tq(y))\delta'(x-y)+\frac{c}{12}\delta^{(3)}(x-y)\Big)\label{commutator_T}\,,
\end{equation}
One can also expand
\begin{equation}
    \Tq(x)= \left(\frac{2\ri\pi}{L}\right)^2\sum_{n\in\mathbb{Z}}\Tq_n\, e^{-2\ri\pi\,nx/L}\label{mode_T_cylinder}
\end{equation}
in modes, whose commutation relations then follow from equation \eqref{commutation_mode_ope} and the OPE \eqref{OPE_T}:
\begin{equation}
      [\Tq_m,\Tq_n]=(m-n)\,\Tq_{m+n}+\frac{c}{12}m^3\,\delta_{m+n,0}\,.\label{Virasoro-Cylinder}
\end{equation}
Although this resembles the Virasoro algebra, it differs from it by the expression of the central extension. One can obtain the standard Virasoro algebra by considering a shifted Fourier expansion
\begin{equation}
     \Tq(x) = \left(\frac{2\ri\pi}{L}\right)^2\left( \sum_{n\in\mathbb{Z}}\Lq_n\, e^{-2\ri\pi\,nx/L} - \frac{c}{24}\right)\,, \qquad\text{ with } \qquad  \Lq_n := \Tq_n + \frac{c}{24}\delta_{n,0}\,, \label{modes_T_spher_vs_cylinder}
\end{equation}
such that
\begin{equation}
    [\Lq_m,\Lq_n]=(m-n)\,\Lq_{m+n}+\frac{c}{12}m(m^2-1)\delta_{m+n,0}\,.\label{Virasoro-Sphere}
\end{equation}
As explained in Appendix \ref{app:Sphere}, the operators $\Lq_n$ in fact correspond to the usual modes of the stress-energy tensor on the sphere, rather than on the cylinder.

\paragraph{Hamiltonian and Hilbert space.} The Hamiltonian of the theory is defined as
\begin{equation}\label{eq:HP}
    \Hq = -\frac{1}{2\pi} \int_0^L \bigl( \Tq(x)+\overline{\Tq}(x)\bigr) \dd x =  \frac{2\pi}{L}\left(\Lq_0+\overline{\Lq}_0-\frac{c}{12}\right)\,,
\end{equation}
in terms of the left-moving and right-moving components $\Tq$ and $\overline{\Tq}$ of the stress-energy tensor. We note that the term in bracket is dimensionless, while the circumference $L$ of the cylinder provides the characteristic energy scale $1/L$ of the theory. The term $-\frac{\pi c}{6L}$ is a ``finite-size'' contribution to the energy, due to the theory being put on a cylinder, and is called the Casimir energy.

The Hilbert space of the model is built from representations of the chiral algebras $\Ac$ and $\overline{\Ac}$, on which the Fourier modes of the fields act, obeying the commutation relations \eqref{commutation_mode_ope}. The Hamiltonian $\Hq$ is diagonalisable on this Hilbert space, with an energy spectrum bounded below, as expected in a physical unitary QFT. Heuristically, the negative modes $(\Aq_{-n})_{n>0}$ are interpreted as creation operators on this Hilbert space, producing quanta of energy $\frac{2\pi}{L} n$, while the positive modes $(\Aq_n)_{n>0}$ behave as annihilation operators, destroying these excitations. An important property of these positive modes is then the following: for any state $|\psi\rangle$ in the Hilbert space, there exists a positive integer $M_{|\psi\rangle}$ such that
\begin{equation}\label{eq:Annihil}
    \Aq_n |\psi\rangle = 0 \quad \text{ if } n>M_{|\psi\rangle}\,,
\end{equation}
for any field $\Aq\in\Ac$. We refer to Section \ref{sec:diagonalization} for the description of an explicit example of Hilbert space, namely that of the SU(2) WZNW model.

\paragraph{Normal ordering.} Suppose we are given two chiral fields $\Aq,\Bq\in\Ac$. As illustrated by the OPE \eqref{ope_general}, the product $\Aq(x)\Bq(y)$ diverges when the points $x,y$ collide. Thus, one cannot define a ``composite field'' $\Aq\Bq(x)$ as the point-wise product of $\Aq(x)$ and $\Bq(x)$. In order to make sense of such an object, we will need the notion of normal ordered product. In the present context, it is simply defined as
\begin{equation}
     (\Aq\Bq)(x):=\{\Aq\Bq\}^{(-1)}(x)\,,
\end{equation}
\textit{i.e.} as the first regular term in the OPE \eqref{ope_general}. In particular, it can be understood as the limit $y\to x$ of $\Aq(x)\Bq(y)$ after removing the divergent terms, hence curing the aforementioned divergences. We note that this normal ordered product has conformal dimension $h(\Aq\Bq)=h(\Aq)+h(\Bq)$.

So far, it is not transparent how this construction relates to a notion of ordering of non-commutative operators. To make this link, one needs to recall that the true operators acting on the Hilbert space of the theory are extracted as the Fourier modes of the fields in $\Ac$. A natural question is then to determine how the Fourier expansion of the normal ordered product $(\Aq\Bq)$ is expressed in terms of the expansions of $\Aq$ and $\Bq$. The naive product $\Aq(x)\Bq(x)$ can be formally expanded as a Fourier series with coefficients $\sum_{m\in\Z} \Aq_{n-m}\,\Bq_{m}$. This involves an infinite sum and should thus be treated with suspicion. As mentioned above, the negative modes $(\Bq_{m})_{m<0}$ essentially behave as creation operators on the Hilbert space of the theory: the sum under consideration then contains an infinite number of creation operators (of increasing energy) acting on the right, making it an ill-defined object. Therefore, we are faced again with the presence of short-distance/large-energy divergences in the product $\Aq(x)\Bq(x)$. This is solved by the introduction of the normal ordered product $(\Aq\Bq)(x)$. A general formula for the Fourier expansion of this product was derived in~\cite{Prochazka:2023zdb} (see also~\cite{Dymarsky:2019iny} for another equivalent formulation). It reads
\begin{equation}
     (\Aq\Bq)(x)=\Bq(x)\Aq_+(x)+\Aq_-(x)\Bq(x)+\sum_{k=0}^{+\infty} \left( \frac{2\ri\pi}{L} \right)^{k+1} c_k\,\{ \Aq\Bq\}^{(k)}(x)\,,
    \label{NO_cyl}
\end{equation}
where
\begin{equation*}
    \Aq_+(x)=\left(\frac{2\ri\pi}{L}\right)^{\!h(\Aq)}\sum_{n=0}^{+\infty} \Aq_n\,e^{-2\ri\pi\,nx/L}\quad\text{ and }\quad \Aq_{-}(x)=\left(\frac{2\ri\pi}{L}\right)^{\!h(\Aq)}\sum_{n=-1}^{-\infty} \Aq_{n}\,e^{-2\ri\pi\,nx/L}
\end{equation*}
are the ``positive'' and ``negative'' parts of $\Aq$, while the numbers $c_k$ are the coefficients of the power series expansion
\begin{equation}
    \sum_{k=0}^\infty c_k\, u^k = \frac{1}{1-e^u}+\frac{1}{u}\,.
\end{equation}
In terms of Fourier modes, the formula \eqref{NO_cyl} for the normal ordered product becomes
\begin{equation}\label{NO_cyl_modes}
    (\Aq\Bq)_n=\sum_{m=0}^{+\infty}\Bq_{n-m}\,\Aq_m+\sum_{m=-1}^{-\infty}\Aq_m\,\Bq_{n-m}+\sum_{k=0}^{h(\Aq)+h(\Bq)-1} c_k\,\{ \Aq\Bq\}^{(k)}_n\,.
\end{equation}
Let us explain the precise content of the equations \eqref{NO_cyl} and \eqref{NO_cyl_modes}. The first two terms essentially correspond to the naive product $\Aq(x)\Bq(x)$ but with the positive modes of $\Aq$ put on the right and the negative modes put on the left. This is reminiscent of the standard notion of normal ordering in QFT, since these positive/negative modes are interpreted as annihilation/creation operators. In particular, this reordering ensures that these infinite sums eventually truncate on any state due to the property \eqref{eq:Annihil}, hence producing well-defined operators on the Hilbert space. The third term of equations \eqref{NO_cyl}--\eqref{NO_cyl_modes} is a specificity of CFTs on the cylinder and indicates the presence of ``finite-size'' corrections arising in this reordering procedure. Such a term would not be present when working with Laurent expansions on the sphere, but will play an important role in this paper. In practice, it is completely controlled by the singular part of the OPE $\Aq(x)\Bq(y)$ and only contains a finite number of contributions (for $k< h(\Aq)+h(\Bq)$). Finally, we observe that the coefficient $c_k$ is equal to $-B_{k+1}/(k+1)!$, where $B_n$ is the $n$-th Bernoulli number. For the reader's convenience, we gather below the explicit value of the first few numbers $c_k$:
\begin{equation}
    c_0 = \frac{1}{2}\,,\qquad c_1 = -\frac{1}{12}\,,\qquad c_2 = 0\,,\qquad c_3 = \frac{1}{720}\,,\qquad c_4=0\,,\qquad c_5=-\frac{1}{30240}\,.
\end{equation}

We end with a few general properties of the normal ordered product. We first note that it is neither commutative nor associative, in the sense that $(\Aq\Bq)\neq(\Bq\Aq)$ and $(\Aq(\Bq\mathsf{C}))\neq ((\Aq\Bq)\mathsf{C})$ in general. By convention, when the order of priority is not indicated, we will consider that normal orderings are taken from right to left, \textit{i.e.} $(\Aq^{(1)}\cdots\Aq^{(n)}):=(\Aq^{(1)}(\cdots(\Aq^{(n-1)}\Aq^{(n)})))$. Finally, the normal ordered product satisfies the Leibniz rule with respect to the spatial derivation: $\p(\Aq\Bq) = ( \p\Aq\,\Bq) + ( \Aq\,\p\Bq)$.

\paragraph{Generalised Wick theorem.} Let $\Aq,\Bq,\Cq\in\Ac$ be three generic chiral fields. A natural question is whether we can reconstruct the OPE of $\Aq$ with the normal-ordered product $(\Bq\Cq)$ from the OPEs of $\Aq$, $\Bq$ and $\Cq$. This is answered by the so-called generalised Wick theorem, which states
\begin{equation}
    \Aq(x)\,(\Bq\Cq)(y)= \mathop{\mathrm{Res}}_{z = y} \frac{1}{z-y} \left( \wick{\c1\Aq(x)\c1\Bq(z)}\Cq(y)+ \Bq(z)\wick{\c1\Aq(x)\c1\Cq(y)}\right)+\reg\,,\label{GWT}
\end{equation}
where the contraction \resizebox{14pt}{6pt}{$\bm{\sqcap}$} of two fields denotes the divergent part of their OPE. Let us explain how this formula works in practice. Our starting point is the contractions $\wick{\c1\Aq(x)\c1\Bq(z)}$ and $\wick{\c1\Aq(x)\c1\Cq(y)}$, which are finite linear combinations of fields evaluated at $z$ and $y$ respectively. We further take the OPEs of these fields with $\Cq(y)$ on the right and $\Bq(z)$ on the left respectively  (including regular terms). This produces a linear combination of fields evaluated at $y$, multiplied by rational functions of $(x,y,z)$. We finally eliminate the auxiliary variable $z$ by taking the residue at $z=y$. This results in a Laurent series in $(x-y)$, whose coefficients are fields at $y$, as expected from an OPE. Implementing this procedure carefully, we find that the term of order $(x-y)^{-(k+1)}$ ($k\geq0$) in the OPE \eqref{GWT} is
\begin{equation}\label{eq:ABC}
    \bigl\{ \Aq(\Bq\Cq) \bigr\}^{(k)} = \bigl(\{\Aq\Bq\}^{(k)}\Cq\bigr) + \bigl(\Bq\{\Aq\Cq\}^{(k)}\bigr) + \sum_{l=0}^{k-1} {k \choose l} \bigl\{ \{ \Aq\Bq \}^{(l)}\Cq\bigr\}^{(k-l-1)}\,.
\end{equation}
This formula is a ``generalised Leibniz rule'' for the distribution of the $k$-th product $\{\cdot\,\cdot \}^{(k)}$ with respect to the normal-ordered product $(\cdot\,\cdot)$, with the third term representing the corrections compared to a standard Leibniz rule.

\paragraph{Re-orderings.} For concrete manipulations, it is often useful to know the effect of re-ordering fields inside normal-ordered products. Once again, this is purely controlled by the (singular part of the) OPE of the two fields. Namely, for any fields $\Aq^{(1)},\dots,\Aq^{(p)}\in\Ac$, we have
\begin{align}\label{eq:Reordering}
    \bigl(\Aq^{(1)}\cdots \Aq^{(i)} \Aq^{(i+1)} \cdots\Aq^{(p)}\bigr) &= \bigl(\Aq^{(1)}\cdots \Aq^{(i+1)} \Aq^{(i)} \cdots\Aq^{(p)} \bigr) \\
    & \hspace{25pt} + \sum_{k\geq 1} \frac{(-1)^k}{k!} \bigl( \Aq^{(1)}\cdots\p^k\lbrace \Aq^{(i)}\Aq^{(i+1)}\rbrace^{(k-1)}\cdots\Aq^{(p)} \bigr)\,. \nonumber
\end{align}

\paragraph{Generators.} We say that a collection of fields $\lbrace \Gq^{(i)} \rbrace_{i\in I}$ forms generators of the chiral algebra if any element of $\Ac$ can be written as a linear combination of normal ordered products of the fields $\Gq^{(i)}$ and their derivatives. Such generators provide a very efficient way of doing computations with chiral fields. We note that the algebraic structure on $\Ac$ is completely fixed by the divergent parts of the OPEs of these generators, \textit{i.e.} by the data of $\lbrace \Gq^{(i)}\Gq^{(j)}\rbrace^{(k)}$ with $i,j\in I$ and $k\in\lbrace 0,\dots,h(\Gq^{(i)})+h(\Gq^{(j)})-1\rbrace$. Indeed, one can compute the OPE of any two fields in $\Ac$ starting from that of the generators, by repeated use of the generalised Wick theorem \eqref{GWT}, at least in principle. In practice, these computations can become quite involved as the complexity of the fields grows. Luckily, these rules are efficiently implemented in the Mathematica$^{\rm TM}$ package {\ttfamily OPEdefs} by Thielemans~\cite{mathematica}, which we used at various points in the preparation of this paper.

\subsection{SU(2)-WZNW Model on the cylinder}
\label{subsec:WZW_cyl}

We now specialise to a specific 2-dimensional CFT, the $\SU(2)$ Wess-Zumino-Novikov-Witten (WZNW) model~\cite{Wess:1971yu,Witten:1983tw,Witten:1983ar,Novikov2007MULTIVALUEDFA}.

\paragraph{Lie-algebraic conventions.} Let us fix our conventions for the Lie algebra $\su(2)$. We will work with generators $T^a=\frac{1}{2} \s^a$, $a\in\lbrace1,2,3\rbrace$, where $\s^a$ are the Pauli matrices. These generators are orthonormal with respect to the invariant scalar product $2\Tr(T^aT^b)=\delta^{ab}$. Since the basis is orthonormal, we will always work with fully raised Lie algebra indices and will sum over repeated ones. With these conventions, the $\su(2)$ structure constants coincide with the Levi-Civita symbol $\varepsilon^{abc}$ (completely antisymmetric, with $\varepsilon^{123}=1$), \textit{i.e.}
\begin{equation}
 [T^a,T^b] = \ri\,\varepsilon^{abc}\, T^c\,.
\end{equation}

\paragraph{The classical theory.} The classical $\SU(2)$-WZNW model is defined in terms of a 2d field $g(x^+,x^-)$ valued in the group $\SU(2)$, depending on the light-cone cylinder coordinates $x^\pm=t\pm x$. Its dynamics is governed by the action
\begin{equation}
    S[g]=\frac{\kcl}{16\pi}\int \dd^2x\, \Tr(\partial_+ g^{-1}\partial_- g) + \frac{\kcl}{24\pi}\int_B \dd^3 x\, \varepsilon^{\alpha\beta\gamma}\,\Tr(g^{-1}\partial_\alpha g\,g^{-1}\partial_\beta g\, g^{-1}\partial_\gamma g)\,,
    \label{S_WZW}
\end{equation}
where the classical level $\kcl\in\R_{> 0}$ is a constant parameter. In the second contribution, called the Wess-Zumino term, $B$ is a 3-dimensional manifold whose boundary $\p B$ is the cylinder and we have extended $g$ from the cylinder to $B$ (keeping the same symbol by a slight abuse of notation).\footnote{\label{fn:WZW}This extension is not unique: however, two different extensions lead to the same Wess-Zumino term up to a shift by $2\pi n\,\kcl$ with $n\in\Z$ an integer. At the classical level, this shift does not affect the equations of motion associated with the action $S[g]$ and the choice of extension of $g$ is thus inconsequential. In contrast, the consistency of the quantum theory will require a quantisation of the level (see below).} The action \eqref{S_WZW} is invariant under a pair of chiral symmetries $\SU(2)_L\times \SU(2)_R$ acting on the field as
\begin{equation}
    g(x^+,x^-)\longmapsto U^{-1}_L(x^+)g(x^+,x^-)U_R(x^-)\,,\label{group_invariance}
\end{equation}
where $U_L(x^+)$ and $U_R(x^-)$ are $\SU(2)$-valued matrices, depending arbitrarily on one of the light-cone coordinates. Consequently, the equations of motion of \eqref{S_WZW} can be rephrased as chirality equations
\begin{equation}
    \partial_- J_{\text{cl}} = 0 \qquad \text{ and } \qquad \partial_{+}\bar{J}_{\text{cl}} =0
    \label{eoms_WZW}
\end{equation}
for the currents
\begin{equation}
    J_{\text{cl}}:=-\frac{k_{cl}}{4} \partial_{+}g g^{-1}, \qquad \text{ and } \qquad \bar{J}_{\text{cl}}:=\frac{k_{cl}}{4}g^{-1}\partial_{-}g\,.
\end{equation}
In other words, $J_{\text{cl}}(x^+)$ is a left-moving field while $\bar{J}_{\text{cl}}(x^-)$ is right-moving. We decompose these currents along the basis of $\su(2)$ as $J_{\text{cl}}(x^+)=J_{\text{cl}}^a(x^+)\,T^a$ and $\bar J_{\text{cl}}(x^-)=\bar J_{\text{cl}}^a(x^-)\,T^a$ (where we recall that a summation over $a$ is implied).

The WZNW model is conformally invariant (so far at the classical level). Its stress-energy tensor splits into a left-moving component
\begin{equation}\label{classical_T}
    T_{\text{cl}}(x^+) = \frac{1}{\kcl} J_{\text{cl}}^a(x^+) J_{\text{cl}}^a(x^+)
\end{equation}
and a right-moving component built out of $\bar{J}_{\text{cl}}(x^-)$. Performing the canonical analysis of the theory, one finds that the left-moving current satisfies a closed Poisson algebra:
\begin{equation}
    \bigl\lbrace J^a_{\text{cl}}(x), J^b_{\text{cl}}(y) \bigr\rbrace = \pi \Big(2\ri\varepsilon^{abc}J_{\text{cl}}^c(y)\delta(x-y)-\kcl\,\delta^{ab}\,\delta' (x-y)\Big)\,.\label{poisson_kac_moody}
\end{equation}
The right-moving current $\bar{J}_{\text{cl}}$ has a similar Poisson bracket, which we will not describe here since we focus on the left-moving sector. Finally, the stress-energy tensor \eqref{classical_T} satisfies the Poisson algebra
\begin{equation}\label{eq:PBT}
    \bigl\lbrace T_{\text{cl}}(x), T_{\text{cl}}(y) \bigr\rbrace = -2\pi\bigl( T_{\text{cl}}(x) + T_{\text{cl}}(y) \bigr) \delta'(x-y)\,.
\end{equation}

\paragraph{Quantisation of the WZNW model.} So far, we have treated the WZNW model \eqref{S_WZW} as a classical field theory. There are several complementary approaches to its quantisation. For most 2-dimensional $\s$-models, the need to renormalise UV divergences in the quantum theory breaks the conformal invariance. In this respect, the WZNW model is quite special: it is a fixed-point of the renormalisation group and thus stays conformal at the quantum level. In the path integral formulation, this quantum model is described by integrals over the field configurations of $g$, with weights $e^{\ri S[g]/\hbar}$. As mentioned in footnote \ref{fn:WZW}, the action $S[g]$ is defined only up to shifts by $2\pi n\,\kcl$ with $n\in\Z$, due to the 3-dimensional origin of the Wess-Zumino term. The consistency of the path integral then requires that the weight $e^{\ri S[g]/\hbar}$ is invariant under this ambiguity and thus that the quantum level
\begin{equation}\label{eq:QuantumLevel}
    k = \frac{\kcl}{\hbar} \; \in \; \Z_{\geq 1}
\end{equation}
is an integer (which we also take to be positive to ensure the positivity of the kinetic term). An alternative and powerful approach to the quantisation of the WZNW model, which is the one that we will follow in this paper, is to use the operator formalism of 2d CFTs reviewed in Subsection \ref{subsec:2DCFT}.

\paragraph{Chiral algebra and quantised Kac-Moody currents.} In the operator approach, we start by quantising the local chiral fields of the theory. In the present case, recall that we have a left-moving current $J_{\text{cl}}(x^+)=J_{\text{cl}}^a(x^+)\,T^a$ and a right-moving one $\bar{J}_{\text{cl}}(x^-)=\bar J_{\text{cl}}^a(x^-)\,T^a$, valued in $\su(2)$. Focusing on the left-moving sector, the quantised components $\lbrace \Jq^a \rbrace_{a=1,2,3}$ of this current become the generators of the chiral algebra $\Ac$ of the quantum WZNW model. Thus, the fields in $\Ac$ are built as normal ordered products of $\Jq^a$ and their derivatives (see the end of the section for more details).

The main information required to define the algebraic structure on $\Ac$ is the OPE of the generators $\Jq^a$. From equation \eqref{field_commutator}, this is equivalent to the data of the commutator of these fields, which itself is guided by the classical theory. Indeed, this commutator should be a quantisation of the Poisson algebra \eqref{poisson_kac_moody} obeyed by the classical fields $J^a_{\text{cl}}$, in the sense that it should satisfy $[\cdot,\cdot]=\ri\hbar\,\lbrace\cdot,\cdot\rbrace + O(\hbar^2)$ in the classical limit $\hbar\to 0$. We then impose the following commutation relation for the quantised current:
\begin{equation}
    \bigl[\Jq^a(x),\Jq^b(y)\bigr]=\ri\pi \Big(2\ri\varepsilon^{abc}\Jq^c(y)\delta(x-y)-k\,\delta^{ab}\,\delta' (x-y)\Big)\,.\label{kac_moody_commutator}
\end{equation}
As required, this reproduces the classical Poisson bracket \eqref{poisson_kac_moody} when $\hbar\to 0$, recalling the relation \eqref{eq:QuantumLevel} between the quantum and classical levels and letting
\begin{equation}
    \Jq^a = \frac{J^a_{\text{cl}}}{\hbar}\,.
\end{equation}
This equation is simply a choice of normalisation of the current at the quantum level, which allows to eliminate the $\hbar$-dependence in the commutator \eqref{kac_moody_commutator} (the classical limit is then understood as taking $\hbar\to 0$ while keeping $J^a_{\text{cl}}=\hbar\,\Jq^a$ and $\kcl=\hbar\,k$ fixed). By equation \eqref{field_commutator}, the commutator \eqref{kac_moody_commutator} is equivalent to the OPE
\begin{equation}
    \Jq^a(x)\Jq^b(y)=\frac{\ri\, \varepsilon^{abc} \Jq^c(y)}{(x-y)} + \frac{k/2\,\delta^{ab}}{(x-y)^2}+\reg \label{kac_moody_OPE}.
\end{equation}

The Fourier modes $\Jq^a_n$ of the current are defined by
\begin{equation}
        \Jq^a(x) = \frac{2\ri\pi}{L} \sum_{n\in\Z} \Jq^a_n\, e^{-2\ri\pi\,nx/L}\,.
\end{equation}
By equation \eqref{commutation_mode_ope}, they satisfy the commutation relation
\begin{equation}
    \bigl[\Jq^a_m,J^b_n\bigr]=\ri\,\varepsilon^{abc}\Jq^c_{m+n}+\frac{k}{2}\,m\,\delta_{m+n,0}\,\delta^{ab}\,,\label{kac_moody_mode_commutator}
\end{equation}
defining the affine Kac-Moody algebra $\widehat{\su(2)}_k$ at level $k$. Accordingly, $\Jq(x)$ is often called a Kac-Moody current. We note that the algebra $\widehat{\su(2)}_k$ admits unitary representations only when $k$ is an integer (see section \ref{sec:diagonalization} for details on these representations and the Hilbert space of the theory): we thus recover the quantisation condition \eqref{eq:QuantumLevel} of the level $k$ from algebraic considerations.

\paragraph{Sugawara construction.} The stress-energy tensor of the WZNW model is built out of the current $\Jq(x)$ through the so-called Sugawara construction:
\begin{equation}
    \Tq(x) = \frac{(\Jq^a\Jq^a)(x)}{k+2}\,.
\end{equation}
This is to be compared with the classical stress-energy tensor \eqref{classical_T}. Recalling that the classical limit $\hbar\to 0$ is taken while keeping $J^a_{\text{cl}}=\hbar\,\Jq^a$ and $\kcl=\hbar\,k$ fixed, we get
\begin{equation}
    \Tq(x) = \frac{1}{\hbar} \frac{(J^a_{\text{cl}}J^a_{\text{cl}})(x)}{\kcl+2\hbar} = \frac{1}{\hbar} \bigl( T_{\text{cl}}(x) + O(\hbar) \bigr)\,.
\end{equation}
In particular, we see that the shift of the level $k$ by $2$ can be interpreted as a quantum correction, which does not contribute in the classical limit.

The OPE of the stress-energy tensor with the Kac-Moody current can be easily computed from that \eqref{kac_moody_OPE} of $\Jq$, using the generalised Wick theorem \eqref{GWT}. It reads
\begin{equation}\label{eq:OpeTJ}
    \Tq(x)\Jq^a(y) = \frac{\p\Jq^a(y)}{x-y} + \frac{\Jq^a(y)}{(x-y)^2} + \reg\,.
\end{equation}
As expected, this confirms that $\Tq$ generates conformal transformations on $\Jq^a$. More precisely, the latter behaves as a primary field of conformal dimension $h(\Jq^a)=1$.
Computing the OPE of $\Tq$ with itself yields the standard Virasoro OPE \eqref{OPE_T}, with central charge
\begin{equation}\label{eq:cWZW}
    c=\frac{3k}{k+2}\,.
\end{equation}
In particular, $\Tq(x)$ obeys the commutation relation \eqref{commutator_T}, including a term proportional to $c\,\delta^{(3)}(x-y)$. This is in contrast with the classical Poisson bracket \eqref{eq:PBT}, which does not contain such a term. Indeed, re-establishing the $\hbar$-dependence, one can check that this term disappears in the classical limit. This shows that the classical chiral algebra of the WZNW model is quite simpler than its quantum version. As we shall see in Section \ref{sec:structure}, this will also be reflected in the integrable structure of the model. 

Formally, the Sugawara construction provides an explicit embedding of the Virasoro chiral algebra inside of the current algebra of the WZNW model, so far in terms of fields. We can also express this embedding in terms of Fourier modes, using the formula \eqref{NO_cyl_modes} for the normal ordered product on the cylinder. We then find that $\Tq(x)$ admits the Fourier expansion \eqref{mode_T_cylinder}, with modes
\begin{equation}
    \Tq_n = \frac{1}{k+2}\left(\sum_{m\geq0 }\Jq^a_{n-m}\Jq^a_m+\sum_{m<0 }\Jq^a_m \Jq^a_{n-m} \right) - \frac{c}{24}\,.\label{mode_T_of_J}
\end{equation}
The first two contributions in equation \eqref{NO_cyl_modes} produce the terms involving $\Jq_m$, while the third contribution, which represents the finite size corrections in normal ordered products, yields the term proportional to the central charge. Comparing with the shifted Fourier expansion \eqref{modes_T_spher_vs_cylinder}, we thus recover the Casimir energy $-c/24$ on the cylinder (now from the construction of $\Tq(x)$ as a composite field) and identify
\begin{equation}\label{eq:SugModes}
    \Lq_n = \frac{1}{k+2}\left(\sum_{m\geq0 }\Jq^a_{n-m}\Jq^a_m+\sum_{m<0 }\Jq^a_m \Jq^a_{n-m} \right)
\end{equation}
as the Virasoro generators (see Appendix \ref{app:Sphere} for a more detailed computation and discussion). By construction, the modes $\Lq_n$ satisfy the Virasoro commutation relations \eqref{Virasoro-Sphere}, showing how the Virasoro algebra is built out from the generators of the affine algebra $\widehat{\su(2)}_k$.

\section{Local commuting IMs of SU(2) WZNW models}
\label{sec:structure}

\subsection{Integrable structures of 2d CFTs}

\paragraph{Generalities.} At the classical level, integrable field theories are characterised by the presence of an infinite number of Integrals of Motion (IMs) in involution with respect to the Hamiltonian structure of the model, \textit{i.e.} conserved quantities $I^{(p)}_{\text{cl}}$ satisfying $\lbrace I^{(p)}_{\text{cl}}, I^{(q)}_{\text{cl}} \rbrace = 0$. A natural manifestation of quantum integrability is thus the existence of an infinite number of commuting conserved operators $\Iq^{(p)}$, with $[\Iq^{(p)},\Iq^{(q)}]=0$, forming what we call the integrable structure of the model. In this context, our main goals are then to construct these commuting operators and simultaneously diagonalise them on the Hilbert space of the theory, \textit{i.e.} determine their spectrum and eigenvectors. These are generally complicated tasks, since even properly defining the algebra of quantum operators and the Hilbert space of a generic quantum field theory is arduous. In this respect, 2d CFTs are quite special since their operator formulation is well under control, as reviewed in the previous section. This offers a natural algebraic setup for the construction and diagonalization of quantum commuting IMs in CFTs, as initiated in~\cite{Zamolodchikov:1989hfa,Sasaki:1987mm,Eguchi:1989hs,Kupershmidt:1989bf,Feigin:1993sb,Bazhanov:1994ft}. The main goal of this paper is to discuss these questions for the specific case of the SU(2) WZNW model. We start by reviewing some general considerations.

The integrable structure usually decomposes into a set of higher-spin local IMs and a set of non-local IMs, all pairwise commuting. The former are the focus of this section, while the latter will be discussed later in Section \ref{sec:Kondo}. As reviewed in Subsection \ref{subsec:2DCFT}, the chiral algebra $\Ac$ of a 2d CFT is formed by its local left-moving fields. By construction, the spatial integral $\int_0^L\Aq(x)\,\dd x$ of any such field $\Aq(x)$, proportional to its zero mode $\Aq_0$, is a local conserved operator. However, these zero modes do not commute in general. Indeed, according to equation \eqref{commutation_mode_ope}, we have
\begin{equation}\label{eq:com_zero_modes}
    \bigl[ \Aq_0, \Bq_0 \bigr] = \lbrace \Aq\Bq \rbrace^{(0)}_0\,,
\end{equation}
where we recall that $\lbrace \Aq\Bq\rbrace^{(0)}$ is the coefficient of the simple pole in the OPE of $\Aq$ and $\Bq$. Our main task to construct local commuting IMs is then to find an infinite tower of fields $\Wq^{(s+1)} \in \Ac_{s+1}$ whose OPE coefficients
\begin{equation}\label{general_key_condition}
    \lbrace \Wq^{(s+1)}\,\Wq^{(s'+1)} \rbrace^{(0)} = \p(\dots)
\end{equation}
are total derivatives. Indeed, the zero mode of such a total derivative vanishes, such that equation \eqref{eq:com_zero_modes} ensures the commutativity of the corresponding local charges
\begin{equation}\label{eq:IM}
    \Iq^{(s)} = \Wq^{(s+1)}_0 =  \frac{L^s}{(2\ri\pi)^{s+1}}\int_0^L \Wq^{(s+1)}(x)\,\dd x\,,
\end{equation}
\textit{i.e.} $[\Iq^{(s)},\Iq^{(s')}]=0$. Here, we have chosen to label the densities $\Wq^{(s+1)}$ by their conformal dimensions $h(\Wq^{(s+1)})=s+1$, which also coincide with their Lorentz spins. The local IM $\Iq^{(s)}$ obtained after integration is then of spin $s$.\footnote{\label{fn:Spin}An operator $\mathrm{O}$ has spin $s$ if it transforms as $\mathrm{O}\mapsto e^{s\theta}\mathrm{O}$ under a Lorentz boost $\begin{pmatrix}
    t \\ x
\end{pmatrix} \mapsto \begin{pmatrix}
    \cosh(\theta) & \!\! -\sinh(\theta) \\ -\sinh(\theta) & \!\! \cosh(\theta)
\end{pmatrix} \begin{pmatrix}
    t \\ x
\end{pmatrix}$ of rapidity $\theta$. A left/right-moving field $\Aq(x \pm t)$ has spin $\pm h(\Aq)$, while its integral $\int_0^L \Aq(x)\dd x$ has spin $\pm (h(\Aq)-1)$.} Note that the condition \eqref{general_key_condition} ensuring the commutativity of the $\Iq^{(s)}$'s is quite constraining and generally does not allow for all the spins to appear. Typically, one obtains an infinite tower $\lbrace \Iq^{(s)} \rbrace_{s\in E}$ of local commuting IMs labelled by their spins $s$,\footnote{Note that the most general theories can have several commuting IMs of the same spin. This will not be the case for the models considered in this paper, so we can consistently label the IMs by their spin.} which take values in a specific subset $E\subset\Z_{\geq 1}$ of integers. This subset is model-dependent but always starts with spin $s=1$, corresponding to the IM
\begin{equation}\label{eq:I1}
    \Iq^{(1)} = -\frac{L}{4\pi^2} \int_0^L \Tq(x)\,\dd x = \Lq_0 - \frac{c}{24}\,,
\end{equation}
whose density is the stress-energy tensor. Indeed, the OPE coefficient $\lbrace \Tq \Aq\rbrace^{(0)}=\p\Aq$ is a total derivative for any field $\Aq\in\Ac$, so that the condition \eqref{general_key_condition} is trivially satisfied.

A few additional remarks are in order. We have focused here on the commuting local IMs of the left-moving sector. Of course, in an integrable relativistic CFT, there exists an analogue infinite tower of commuting charges $\lbrace\bar\Iq^{(s)}\rbrace_{s\in \bar E}$ built from right-moving densities and of negative Lorentz spins $-s$. Moreover, we note that the prefactor in equation \eqref{eq:IM} is such that $\Iq^{(s)}$ is a dimensionless operator. It is expressed purely in terms of fields' modes and the parameters of the CFT, without any dependence on the cylinder size $L$ -- see for instance $\Iq^{(1)}$ in equation \eqref{eq:I1}. Physically relevant quantities are then often obtained by reintroducing $L$-dependent factors: for example, the Hamiltonian of the model is $\frac{2\pi}{L}(\Iq^{(1)}+\bar\Iq^{(1)})$. Finally, it is clear that the IMs $\Iq^{(s)}$ are invariant under a shift of the densities $\Wq^{(s+1)}$ by spatial derivatives $\p(\dots)$. We are thus looking for solutions of the condition \eqref{general_key_condition} up to such derivatives.

\paragraph{Example: the quantum KdV integrable structure.} To illustrate the general ideas explained above, let us discuss the simplest example of integrable structure in a 2d CFT, called the quantum Korteweg-de Vries (KdV) integrable structure~\cite{Sasaki:1987mm,Eguchi:1989hs,Kupershmidt:1989bf,Bazhanov:1994ft}. It is built inside the simplest chiral algebra, namely the Virasoro one, which is generated by the stress-energy tensor only (physically, it is then relevant for the Liouville theory and minimal models). The local IMs $\Iq^{(2p-1)}_{\KdV}$ forming this integrable structure have odd spins $E=\lbrace 2p-1, p\in\Z_{\geq 1}\rbrace$. Their densities $\Wq^{(2p)}_{\KdV}$ are linear combinations of normal-ordered products of $\Tq$ and its derivatives, with coefficients depending on the central charge $c$. For instance, the first 3 densities read\vspace{-1pt}
\begin{equation}
    \Wq^{(2)}_{\KdV}=\Tq,\hspace{.9cm }\Wq^{(4)}_{\KdV}=(\Tq\Tq),\hspace{.9cm} \Wq^{(6)}_{\KdV}=(\Tq\Tq\Tq)+\frac{c+2}{12}\, (\partial \Tq\partial \Tq)\,.\vspace{-2pt}
    \label{KdV_densities}
\end{equation}
More generally, the density of spin $2p$ is of the form $\Wq^{(2p)}=(\Tq^p) + \dots$, where the dots represent corrections involving a smaller number of $\Tq$'s but also derivatives. These densities are uniquely fixed, up to total derivatives, by requiring the commutativity of the corresponding integrals $\Iq^{(2p-1)}_{\KdV}$, \textit{i.e.} by the condition \eqref{general_key_condition}. For instance, the factor $(c+2)/12$ appearing in $\Wq^{(6)}_{\KdV}$ is completely fixed by imposing that $\lbrace \Wq^{(4)}_{\KdV}\,\Wq^{(6)}_{\KdV}\rbrace^{(0)}$ is a total derivative.\vspace{-1pt}

\paragraph{Goal of this section.} The example of the KdV IMs mirrors the objective of this section. While it provides the integrable structure of Liouville theory and minimal models, whose chiral algebra is Virasoro, we are interested in this paper in the SU(2) WZNW model, whose local chiral fields form a current algebra $\widehat{\su(2)}_k$. We thus want to build commuting IMs from the Kac-Moody current of the theory. To guide us, we will start by reviewing some known results at the classical level.\footnote{More precisely, we will describe the conformal limit of the classical integrable structure of the so-called $\lambda$-model~\cite{Balog:1993es,Sfetsos:2013wia}, which is the $\Jq^a\overline{\Jq}\null^a$-deformation of the WZNW CFT. We note that the WZNW model also admits other integrable structures, corresponding to different massive deformations~\cite{Kobayashi:1989sg,Leblanc:1990fi,Bonora:1990hc,Durganandini:1991gma,Brazhnikov:1996fa} (see the conclusion for a brief discussion of those).}

\subsection{The classical integrable structure}
\label{subsec:EHMM}

The classical SU(2) WZNW model was described in equation \eqref{S_WZW} and below. The classical integrable structure which we now review is split into two halves: one built from the left-moving Kac-Moody current $J_{\cl}(x^+)$ and the other one built from the right-moving current $\bar{J}_{\cl}(x^-)$. We will focus here on the left-moving part, the other being treated in a similar fashion. As is standard, it is composed by both non-local and local IMs. The former are extracted from the path-ordered exponential of the Kac-Moody current, by taking its trace in a chosen representation $V$ of SU(2):\vspace{-1pt}
\begin{equation}\label{eq:Kcl}
    K^V_{\cl}(\lambda) = \Tr_V\left[\Pexp\left( \ri \lambda\int_0^L J_{\cl}(x)\,\dd x \right)\right]\,.\vspace{-1pt}
\end{equation}
This quantity Poisson-commutes with itself for all values $\lambda,\mu$ of the spectral parameter and all choices $V,W$ of representations, \textit{i.e.} $\big\lbrace K^V_{\cl}(\lambda),K^W_{\cl}(\mu)\big\rbrace=0$. Varying the representation and expanding in powers of the spectral parameter, we thus extract an infinite number of Poisson-commuting non-local IMs from this object. Their quantisation as so-called Kondo defects has been addressed in the works~\cite{Bachas:2004sy,Gaiotto:2020fdr,Gaiotto:2020dhf} and will be discussed in Section \ref{sec:Kondo}.

In this section, we focus on the local IMs of the theory. For the classical SU(2) WZNW model, these turn out to take a very simple form~\cite{Evans:1999mj,Evans:2000hx,Lacroix:2017isl}. Namely, they are defined as the integral of powers of the energy-momentum tensor:
\begin{equation}\label{eq:ClassicalIM}
    I^{(2p-1)}_{\text{cl}} = \frac{L^{2p-1}}{(2\ri\pi)^{2p}} \int_0^L W^{(2p)}_{\cl}(x)\,\dd x\,, \qquad \text{ with } \qquad W_{\cl}^{(2p)}(x) = T_{\cl}(x)^p\,,
\end{equation}
where we recall that $T_{\cl}(x)=\frac{1}{\kcl}J^a_\cl(x)\,J^a_\cl(x)$ in terms of the Kac-Moody current. One can check that these charges Poisson-commute between themselves and with the non-local IMs \eqref{eq:Kcl}:
\begin{equation}
    \bigl\lbrace I^{(2p-1)}_{\text{cl}}, I^{(2q-1)}_{\text{cl}} \bigr\rbrace=0 \qquad \text{ and } \qquad \bigl\lbrace I^{(2p-1)}_{\text{cl}}, K^V_\cl(\lambda) \bigr\rbrace = 0\,.
\end{equation}

As noted above, the local IMs $I^{(2p-1)}_{\text{cl}}$ are particularly simple, being written purely in terms of the stress-energy tensor $T_{\cl}(x)$. As such, they can be seen as forming a classical KdV integrable structure inside the Virasoro subalgebra of the theory. This is due to the simplicity of the classical SU(2) WZNW model and a few remarks are in order at this point to put back this apparent triviality in perspective. Firstly, the IMs \eqref{eq:ClassicalIM} are the relevant ones for the group SU(2): for a higher-rank group $G$, the local IMs would involve other (higher-degree) combinations of the Kac-Moody current than the stress-energy tensor and would follow a different spin pattern (more precisely the allowed spins would be the exponents~\cite{kac1989infinite} of the corresponding affine algebra $\hat{\g}$). Secondly, the classical integrable structure considered here is part of a quite more general construction~\cite{Feigin:2007mr,Vicedo:2017cge,Lacroix:2017isl} based on so-called affine Gaudin models (see~\cite{Lacroix:2023gig} for a review). These are integrable field theories built in terms of an arbitrary number of Kac-Moody currents and whose local IMs involve more complicated combinations of these currents (generally depending on arbitrary continuous parameters). Finally, and most importantly, we claim that the simplicity of the IMs \eqref{eq:ClassicalIM} is an accidental consequence of having taken the classical limit. Indeed, in the rest of this paper, we shall argue that the correct quantisation of this integrable structure will involve terms which cannot be written in terms of the stress-energy tensor only: this will define a new integrable structure, inherent to the WZNW model and different from the KdV one (see the paragraph \ref{par:Discussion} for more details).

Despite its simplicity, the classical integrable structure described above will still serve as a useful guide for our quantum investigations. In particular, we note the following two properties of the local IMs \eqref{eq:ClassicalIM} which we will aim to preserve under quantisation:
\begin{enumerate}
    \item Their densities $W^{(2p)}_\cl(x)$ are local left-moving fields of even spins $2p$ ($p\in\Z_{\geq 1}$).
    \item The $W^{(2p)}_\cl(x)$'s are invariant under the global SU(2)$_L$--symmetry $J_\cl(x) \mapsto U_L^{-1} J_\cl(x)U_L$.
\end{enumerate}

\subsection{The quantum integrable structure}
\label{subsec:quant-Hierarchy}

\paragraph{The Casimir subalgebra.} The goal of this section is to find the quantisation of the local IMs \eqref{eq:ClassicalIM} of the SU(2) WZNW model. To guide us in this endeavour, we will use the invariance of the the classical densities under the conjugation $J_\cl(x) \mapsto U_L^{-1}\,J_\cl(x)\,U_L$ (see above). This is a global symmetry of the WZNW model and it is therefore quite natural to search for an integrable structure which is itself invariant under this transformation. In particular, we wish to preserve this property under quantisation. We are thus led to consider the space of all local fields built from the quantum current $\Jq(x)$ which are invariant under conjugations. This forms a well-known subalgebra of the current algebra $\Ac$, called the Casimir subalgebra $\Cas$~\cite{Thierry-Mieg:1987tib,Bais:1987dc} and which we describe below.

At the quantum level, the SU(2)$_L$--symmetry is generated by the zero modes $\Jq^a_0$ of the Kac-Moody current. A left-moving field $\Cq(x)$ thus belongs to the Casimir subalgebra if and only if it commutes with these zero modes:
\begin{equation}\label{eq:Cas}
    \Cq(x) \in \Cas \hspace{8pt}\Longleftrightarrow\hspace{8pt} \bigl[ \Jq^a_0, \Cq(x) \bigr] = 0  \hspace{8pt}\Longleftrightarrow\hspace{8pt} \bigl\lbrace \Jq^a\Cq\rbrace^{(0)} = 0\,,
\end{equation}
where we recall that $\bigl\lbrace \Jq^a\Cq\rbrace^{(0)}$ is the first order pole in the OPE of $\Jq^a$ with $\Cq$. The space $\Cas$ is a subalgebra of the chiral algebra $\Ac$ in the sense that it is stable under linear combinations, derivatives, OPEs and normal-ordered products. It is in fact quite simple to find all the solutions to the ``Casimir'' condition \eqref{eq:Cas}. Namely, they take the form
\begin{equation}\label{eq:CasField}
    \tau^{a_1\cdots a_p}\,\bigl( \p^{m_1}\Jq^{a_1}(x) \cdots \p^{m_p}\Jq^{a_p}(x) \bigr)\,,
\end{equation}
where $m_1,\dots,m_p\in\Z_{\geq 0}$ are non-negative integers and $\tau^{a_1\cdots a_p}$ is an invariant $p$-tensor\footnote{More precisely, $\tau^{a_1\cdots a_p}$ should be invariant under the adjoint action of SU(2), \textit{i.e.} satisfying the condition 
\begin{equation}
    \label{casimir_condition}
    \varepsilon^{bca_1}\,\tau^{ca_2\cdots a_p}+\varepsilon^{bca_2}\,\tau^{a_1ca_3\cdots a_p}+\cdots+\varepsilon^{bca_p}\,\tau^{a_1\cdots a_{p-1}c}=0\, .
\end{equation}} on the Lie-algebra $\su(2)$  (not necessarily symmetric or skew-symmetric). Note that equation \eqref{eq:CasField} includes a normal-ordering and a sum over the repeated indices $a_k$. The above field is of spin $p+m_1+\dots+m_p$ and satisfies the condition \eqref{eq:Cas} due to the invariance \eqref{casimir_condition} of the tensor $\tau^{a_1\cdots a_p}$.

Although the construction \eqref{eq:CasField} gives all fields in the Casimir subalgebra $\Cas$, we stress that they are not all independent. Indeed,  taking into account the symmetry/skew-symmetry properties of the invariant tensors and re-ordering the currents inside the normal-ordered product (using equation \eqref{eq:Reordering}), we find various relations between these fields. Let us illustrate this idea for the first few degrees. The Lie algebra $\su(2)$ does not admit any invariant 1-tensor and possesses a unique (up to multiplication by a scalar) invariant 2-tensor, the symmetric bilinear form $\delta^{ab}$. The first element of the Casimir algebra built from the recipe \eqref{eq:CasField} is therefore the spin-2 field $(\Jq^a\Jq^a)$, which is simply proportional to the stress-energy tensor $\Tq$. The invariant bilinear form also allows us to construct a spin-3 Casimir field $(\Jq^a\p\Jq^a)$, which is proportional to the derivative $\p\Tq$ of the stress-energy tensor. Furthermore, the algebra $\su(2)$ possesses an invariant (completely skew-symmetric) 3-tensor, formed by the structure constants $\varepsilon^{abc}$. The corresponding spin-3 Casimir field $\varepsilon^{abc}(\Jq^a\Jq^b\Jq^c)$ turns out to be also proportional to $\p\Tq$ and thus does not provide a new independent element of $\Cas$. Indeed, the tensor $\varepsilon^{abc}$ being skew-symmetric, this field can be rewritten as $\frac{1}{2}\varepsilon^{abc}\bigl(\Jq^a(\Jq^b\Jq^c-\Jq^c\Jq^b)\bigr)=\frac{1}{2}\varepsilon^{abc}\varepsilon^{dbc}(\Jq^a \p\Jq^d)=(\Jq^a \p\Jq^a)$, where the first equality follows from the re-ordering identity \eqref{eq:Reordering}. Therefore, we get only one spin-2 field $\Tq$ and one spin-3 field $\p\Tq$.

We will denote by $\Cas_n$ the space of spin-$n$ fields in the Casimir subalgebra. The search for a basis of $\Cas_n$ is easily systematised for low values of $n$ using the Mathematica package {\ttfamily OPEdefs}~\cite{mathematica} and employing the same techniques as above. Moreover, there exists a quite useful analytical tool to help us keep track of the expected number of independent Casimir fields at each spin, the vacuum character of $\Cas$:
\begin{equation}
    \chi(q) := \Tr_{\Cas}(q^{\Lq_0}) =\sum_{n=0}^\infty d_n q^n\,,
\end{equation}
where $d_n=\dim(\Cas_n)$. It turns out that $\chi(q)$ admits the following simple expression~\cite{Feigin:2001yq}\footnote{More precisely, \cite[Equation (6.11)]{Feigin:2001yq} gives the vacuum character of the coset algebra $\widehat{\su(2)}_{k}\oplus \widehat{\su(2)}_{l}/\widehat{\su(2)}_{k+l}$ (before quotienting by potential null vectors). This coset algebra coincides with the Casimir one $\Cas$ in the limit $l\to\infty$ and its character is independent of $(k,l)$. It thus also provides the vacuum character of $\Cas$.}:
\begin{equation}
    \chi(q)=\frac{1}{\prod_{n\geq 1}(1-q^n)^3}\left(1+3\sum_{n\geq 1}(-1)^nq^{n(n+1)/2}+\sum_{n\geq 2}(-1)^nq^{n(n+1)/2-1}\right)\,,
        \label{character}
\end{equation}
whose first few orders in $q$ are
\begin{equation}
    \chi(q)=1+q^2+q^3+3q^4+3q^5+8q^6+9q^7+19q^8+\dots\quad .
    \label{character_su2}
\end{equation}

For our purposes, we will only need to know a basis for Casimir fields of even spins $\Cas_{2p}=\text{span}(\mathcal{B}_{2p})$. We give the first three below:
\begin{equation}\label{eq:BasesC246}
    \mathcal{B}_2=\Tq\,, \qquad\quad \mathcal{B}_4=\bigl( (\Tq\Tq)\,, \hspace{6pt} (\p\Jq^a\p\Jq^a)\,, \hspace{6pt} \p^2\Tq \bigr)\,,
\end{equation}
\begin{equation*}
    \mathcal{B}_6 =\bigl( (\Tq\Tq\Tq)\hspace{1pt},\hspace{3pt} (\p\Tq^2)\hspace{1pt},\hspace{3pt} (\partial^4 \Jq^{a}\Jq^a)\hspace{1pt},\hspace{3pt}
     (\partial^2\Jq^a\Jq^a\Jq^b\Jq^b)\hspace{1pt},\hspace{3pt} (\partial \Jq^a\partial \Jq^a\Jq^b\Jq^b)\hspace{1pt}, \hspace{3pt} \p^2(\Tq\Tq)\hspace{1pt},  \hspace{3pt} \p^2(\p\Jq^a\p\Jq^a)\hspace{1pt}, \hspace{3pt} \p^4\Tq \bigr)\,.
\end{equation*}
These are of size 1, 3 and 8, as expected from the character \eqref{character_su2}. In building these bases, we have chosen to include a maximal number of total derivatives (which we put in last) and to write as many fields as possible in terms of the stress-energy tensor. The other elements, such as $(\p\Jq^a\p\Jq^a)$, cannot be written in terms of $\Tq$ only: they belong to the Casimir algebra $\Cas$ but not to its Virasoro subalgebra. More generally, we introduce $d'_{2p}=d_{2p}-d_{2p-1}$ and fix a choice of basis
\begin{equation}\label{eq:C2p}
    \mathcal{B}_{2p} = \left(\Cq^{(2p,1)}=(\Tq^p),\Cq^{(2p,2)},\dots,\Cq^{(2p,d'_{2p})}, \p(\mathcal{B}_{2p-1})\right)\,,
\end{equation}
where the last $d_{2p-1}$ elements are derivatives of a basis $\mathcal{B}_{2p-1}$ of $\Cas_{2p-1}$, while the first $d_{2p}'$ elements cannot be written as derivatives, starting with the field $\Cq^{(2p,1)}=(\Tq^p)$.

\paragraph{Ansatz.} Now that we have fixed bases of the Casimir algebra spin by spin, we can make an ansatz for the densities $\Wq^{(2p)}$ of our local IMs. One important thing to keep track of is that we only care about the IMs themselves, \textit{i.e.} the integrals/$0$-modes of the densities. Therefore, we do not need to include total derivatives in the densities. Moreover, we note that we also have the freedom of rescaling the density by a constant factor. Beyond this, we will take a quite systematic approach and will keep the ansatz as general as possible. In the notations of equation \eqref{eq:C2p}, we then let
\begin{equation}
    \Wq^{(2p)}(x) = \bigl(\Tq(x)^p\bigr) + \sum_{i=2}^{d'_{2p}} \alpha_{2p,i}\,\Cq^{(2p,i)}(x)\,,
\end{equation}
where we have included only non-derivative fields and fixed the rescaling freedom by setting the coefficient of $(\Tq^p)$ to 1.\footnote{The classical limit \eqref{eq:ClassicalIM} of the densities guarantees that the term $(\Tq^p)$ appears with a non-vanishing coefficient.} This leaves $d'_{2p}-1$ parameters $\alpha_{2p,i}$, which are free for the time being and which we wish to fix by requiring the commutativity of the corresponding IMs. Before we come to this, let us quickly analyse the first few densities. As expected, the first one is completely fixed to be the stress-energy tensor itself: $\Wq^{(2)}=\Tq$. The next one is the spin-4 density and depends on 1 free parameter $\alpha_{4,2}$, which we reparameterise as $\frac{2}{3(k+2)}\alpha$ for future convenience:
\begin{equation}\label{eq:W4}
    \Wq^{(4)}(x) = \bigl(\Tq(x)^2\bigr) + \frac{2\alpha}{3(k+2)}\, \bigl( \p\Jq^a(x)\p\Jq^a(x) \bigr)\,.
\end{equation}
Further on, we get 4 parameters at spin-6, 9 at spin-8 and 19 at spin-10, showing the growing complexity of the ansatz.

\paragraph{Commutativity.} As explained above, we want to fix the parameters $\alpha_{2p,i}$ by requiring the commutativity of the local IMs
\begin{equation}
    \Iq^{(2p-1)} = \frac{L^{2p-1}}{(2\ri\pi)^{2p}} \int_0^L \Wq^{(2p)}(x)\,\dd x\,.
\end{equation}
Following equation \eqref{general_key_condition}, this is done by imposing the condition
\begin{equation}\label{eq:KeyCond2}
    \lbrace \Wq^{(2p)}\, \Wq^{(2q)} \rbrace^{(0)} = \p(\dots) \,,
\end{equation}
for all $p,q\in\Z_{\geq 1}$.

We study this spin by spin. We first note that we do not get any additional constraint from the case $p=q$, since $\Iq^{(2p-1)}$ trivially commutes with itself. Moreover, the spin-2 density satisfies this condition independently of the choice of other densities, since it is the stress-energy tensor. The first non-trivial constraint then arises from $\lbrace \Wq^{(4)}\, \Wq^{(6)} \rbrace^{(0)} = \p(\dots)$. As it turns out, this is enough to completely fix the parameters $\alpha$ (or $\alpha_{4,2}$) and $\alpha_{6,i}$ entering the ansatz for $\Wq^{(4)}$ and $\Wq^{(6)}$. More precisely, we get two different solutions, which we describe below.

\paragraph{The KdV solution.} The first solution corresponds to
\begin{equation}
    \alpha^{\KdV}=0\,, \qquad \alpha^{\KdV}_{6,2} = \frac{5k+4}{12(k+2)}\,, \qquad \alpha^{\KdV}_{6,3} = \alpha^{\KdV}_{6,4} = \alpha^{\KdV}_{6,5} = 0\,.
\end{equation}
To explain the superscript KdV, we recall our explicit choice \eqref{eq:BasesC246} of basis of $\Cas_4$ and $\Cas_6$. In particular, we note that the many parameters $\alpha_{2p,i}^{\KdV}$ which vanish are exactly the coefficients of the fields in these bases which cannot be written in terms of the stress-energy tensor only. The corresponding densities are then expressed purely in terms of $\Tq$ and more precisely read
\begin{equation}
   \Wq^{(4)}_{\KdV}=(\Tq\Tq)\qquad\text{ and }\qquad \Wq^{(6)}_{\KdV} = (\Tq\Tq\Tq) + \frac{5k+4}{12(k+2)} (\p\Tq\p\Tq)\,.
\end{equation}
Recalling the expression \eqref{eq:cWZW} of the central charge for the WZNW model, we see that these are nothing but the KdV densities \eqref{KdV_densities}, now written in terms of WZNW parameters.

This is to be expected and serves as a consistency check of our method. Indeed, the Sugawara construction $\Tq=\frac{1}{k+2}(\Jq^a\Jq^a)$ provides an embedding of the Virasoro algebra inside of the Casimir algebra $\Cas$. Since the former admits an integrable structure with our desired spin pattern, it is normal that we find these IMs among the solutions to our commutativity constraints.

\paragraph{The WZNW integrable structure.} What makes this work non-trivial is the existence of a second solution to the constraint $\lbrace \Wq^{(4)}\, \Wq^{(6)} \rbrace^{(0)} = \p(\dots)$, which we now describe. Explicitly, it reads
\begin{equation}
    \alpha = 1\,, \qquad \alpha_{6,2} = \frac{172 + 111 k}{100 (k+2)}\,,\qquad \alpha_{6,3}=\frac{603 + 380 k}{375 (k+2)^2}\,,
\end{equation}
\begin{equation*}
    \alpha_{6,4}=\frac{64}{25(k+2)^2}\,, \qquad \alpha_{6,5}=\frac{304}{25(k+2)^2}\,.
\end{equation*}
Accordingly, the spin-4 density
\begin{equation}\label{eq:W4WZW}
    \Wq^{(4)} = \bigl(\Tq\Tq\bigr) + \frac{2}{3} \frac{1}{k+2}\, \bigl( \p\Jq^a\p\Jq^a \bigr)
\end{equation}
and the spin-6 one (which we will not write for conciseness) contain terms which are not expressible in terms of the stress-energy tensor only and thus are quite distinct from the KdV solution above.

Having fixed the first two non-trivial densities, one can continue to explore the commutativity constraint \eqref{eq:KeyCond2} for higher-spins, which we have pushed up to spin-8. Interestingly, we found that the condition $\lbrace \Wq^{(4)}\,\Wq^{(8)}\rbrace^{(0)}=\p(\dots)$ is in fact enough to completely fix the density $\Wq^{(8)}$. We have further checked that the obtained solution then also satisfies $\lbrace \Wq^{(6)}\,\Wq^{(8)}\rbrace^{(0)}=\p(\dots)$, as required. For simplicity, we will not report on the explicit expression of $\Wq^{(8)}$ here, as it is rather heavy. These computations were done with the Mathematica package {\ttfamily OPEdefs}~\cite{mathematica} and already take some significant amount of time to run on a personal computer, showing that the next steps would probably require the use of more powerful machines. We are however quite confident in the existence of a solution to the conditions \eqref{eq:KeyCond2} at all spins, having found no obstacles so far. In what follows, we will call this the WZNW integrable structure (see the discussion below).

\paragraph{\label{par:Discussion}Discussion.} Let us summarise and analyse the results of this subsection. We have found strong evidence that the WZNW chiral algebra admits two different towers of commuting SU(2)-invariant local IMs of odd spins: the KdV ones $\Iq^{(2p-1)}_{\KdV}$ inside the Virasoro subalgebra and new ones, simply denoted by $\Iq^{(2p-1)}$, which are built from the current $\Jq^a$ but cannot be rewritten in terms of the stress-energy tensor only. It is interesting to note that this last property only holds at the quantum level. Indeed, the classical limit of these IMs can be computed by replacing $\Jq^a=J^a_\cl/\hbar$ and $k=k_\cl/\hbar$ and taking $\hbar\to 0$. For the explicit IMs found above up to spin-7, one then finds that $\Iq^{(2p-1)}_{\KdV}$ and $\Iq^{(2p-1)}$ have the same very simple classical limit, which coincides with the local charges \eqref{eq:ClassicalIM} discussed earlier. In particular, these charges only involve the classical stress-energy tensor $T_\cl$, and not even its derivatives. This is a consequence of the simplicity of the classical model, as also showcased by the absence of $\delta^{(3)}$-term in its Virasoro Poisson algebra \eqref{eq:PBT}. The two towers of commuting IMs $\Iq^{(2p-1)}_{\KdV}$ and $\Iq^{(2p-1)}$ thus constitute two different quantisations of our classical setup.

We expect the IMs $\Iq^{(2p-1)}$ to form the correct integrable structure for the quantum WZNW model. Indeed, they make use of the full extended conformal symmetry of the theory (\textit{i.e.} the affine algebra, in opposition to the Virasoro modes alone) and are thus inherent to the WZNW model, contrarily to the KdV IMs which are present in any CFT. In particular, as we will show in the next section, the fact that the KdV integrable structure  is built purely from the Virasoro subalgebra implies that it has degeneracies on the Hilbert space of the WZNW model, while the new IMs $\Iq^{(2p-1)}$ seem to have simple spectrum and are thus better suited to organise the states. In Subsection \ref{subsec:affine_bethe_ansatz}, we will also show that the operators $\Iq^{(2p-1)}$ can be diagonalised using an ``affine Bethe ansatz'', relying once again on the full algebraic structure underlying the WZNW model and thus making these operators quite natural. More conceptually, we expect these results to be part of a quite general quantisation program involving affine Gaudin models, an affinisation of the Geometric Langlands Correspondence and the so-called ODE/IQFT correspondence: in Subsection \ref{subsec:odeim}, we will show preliminary evidence that the IMs $\Iq^{(2p-1)}$ fit in this bigger perspective. Beyond these, the most convincing argument in favour of $\Iq^{(2p-1)}$ has to do with the non-local IMs of the theory. Indeed, recall that the classical local charges also Poisson-commute with the non-local quantities \eqref{eq:Kcl}. The latter have been quantised as so-called Kondo defects in~\cite{Bachas:2004sy,Gaiotto:2020fdr,Gaiotto:2020dhf}: as we shall discuss in Section \ref{sec:Kondo}, preliminary results indicate that these Kondo defects only commute with $\Iq^{(2p-1)}$, and not with the KdV charges $\Iq^{(2p-1)}_{\KdV}$. This suggests $\Iq^{(2p-1)}$ as the unique consistent quantisation for the WZNW integrable structure \eqref{eq:Kcl}--\eqref{eq:ClassicalIM}.

\section{Diagonalization of the integrable structure}
\label{sec:diagonalization}

Our goal in this section is to discuss the diagonalization of the new integrable structure found in Section \ref{sec:structure} on the Hilbert space of the WZNW model. We will start by recalling some well-known facts about this Hilbert space and the representation theory of the affine algebra $\widehat{\su(2)}_k$. We will then discuss the diagonalization of the integrable structure, focusing on the simplest higher-spin local IM $\Iq^{(3)}$. Finally, we will explain how these results fit in the so-called ODE/IQFT correspondence.

\subsection{The WZNW Hilbert space and $\widehat{\su(2)}_k$ Verma modules}
\label{subsec:module}

\paragraph{Verma modules.} By construction, the Hilbert space of the SU(2) WZNW model should be a unitary representation of the left-moving and right-moving copies of the affine algebra $\widehat{\su(2)}_k$. This naturally leads us to study the representation theory of this algebra, focusing here on the left-moving sector, formed by the Kac-Moody modes $\Jq^a_n$. Among those, the modes $\Jq^+_0$ and $\lbrace\Jq^a_n \rbrace_{n>0}$ will play the role of raising operators, while $\Jq^-_0$ and $\lbrace\Jq^a_{-n} \rbrace_{n>0}$ will serve as lowering operators, where we defined
\begin{equation}
 \Jq^\pm_n=\Jq^1_n\pm \ri\,\Jq^2_n\,.   
\end{equation}
We will explain later which quantum numbers these operators raise and lower.

The simplest representations of $\widehat{\su(2)}_k$ are the so-called Verma modules $\mathcal{V}_{k,\ell}$, depending on the level $k$ and an additional parameter $\ell$. Those can be defined for all values of $(k,\ell)$, although one often considers specific ones to ensure unitarity. Crucially, the module $\mathcal{V}_{k,\ell}$ contains a highest-weight vector $\ket{\ell}$, satisfying
\begin{equation}\label{eq:highest_weight}
    \Jq^3_0\ket{\ell}=\ell\ket{\ell},\qquad \Jq^+_0\ket{\ell}=0,\qquad \Jq^a_n\ket{\ell}=0\quad\forall n>0\,.
\end{equation}
In other words, this vector is annihilated by all raising operators and is an eigenstate of the Cartan generator $\Jq^3_0$, with weight $\ell$. The rest of the Verma module is then built by applying lowering operators on this highest-weight state:
\begin{equation}
    \mathcal{V}_{k,\ell}=\text{span}\bigl(\cdots(\Jq^-_{-2})^{m^-_2}(\Jq^3_{-2})^{m^3_2}(\Jq^+_{-2})^{m^+_2}(\Jq^-_{-1})^{m^-_1}(\Jq^3_{-1})^{m^3_1}(\Jq^+_{-1})^{m^+_1}(\Jq^-_0)^{m^-_0}\ket{\ell},\,m^a_i\in\Z_{\geq 0}\bigr). \label{eq:verma_module}
\end{equation}
In this equation, we restrict to states obtained by applying a finite number of lowering operators, \textit{i.e.} we consider the case where only finitely many of the integers $m^a_i$ are non-zero. 

\paragraph{Unitary representations and Hilbert space.} Unitary representations of $\widehat{\su(2)}_k$ are those equipped with a positive scalar product such that $(\Jq^a_n)^\dagger=\Jq^a_{-n}$. With this definition, the Verma module $\mathcal{V}_{k,\ell}$ is not unitary. To obtain unitarity, we first need to restrict to parameters $(k,\ell)$ with $k\in\Z_{\geq 1}$ and $\ell\in \bigl\lbrace 0,\frac{1}{2},1,\dots,\frac{k}{2}\bigr\rbrace$. Furthermore, we need to quotient out $\mathcal{V}_{k,\ell}$ by the two null vectors $(\Jq^-_0)^{2\ell+1}\ket{\ell}$ and $(\Jq^+_{-1})^{k+1-2\ell}\ket{\ell}$ (and their descendants). We will denote the resulting quotient by $\widetilde{\mathcal{V}}_{k,\ell}$, which is now a unitary representation of $\widehat{\su(2)}_k$. The Hilbert space of the SU(2) WZNW model at level $k\in\Z_{\geq 1}$ (and with diagonal modular invariance~\cite{Cappelli:1986hf,Cappelli:1987xt}) is then defined as 
\begin{equation}
    \mathcal{H}_k = \bigoplus_{\ell\,\in\, \lbrace 0,\frac{1}{2},1,\dots,\frac{k}{2}\rbrace} \widetilde{\mathcal{V}}_{k,\ell} \otimes \widetilde{\overline{\mathcal{V}}}_{k,\ell}\,,
\end{equation}
where the bar in the second tensor factor indicates that it is a representation of the right-moving Kac-Moody modes $\overline{\Jq}\null^{\hspace{1pt}a}_{\hspace{1pt}n}$. In the rest of this section, we will focus on the left-moving sector and will thus never have to consider those.

Note that the unitarity of the theory imposes the quantisation of the level $k\in\Z_{\geq 1}$ (which we already argued in equation \eqref{eq:QuantumLevel} from a path-integral point of view) and the restriction to a finite number of representations $\widetilde{\mathcal{V}}_{k,\ell}$, with weights $\ell\in \bigl\lbrace 0,\frac{1}{2},1,\dots,\frac{k}{2}\bigr\rbrace$. However, to keep the discussion general and to avoid having to deal with null vectors, we will mostly keep $k$ and $\ell$ generic and work with unquotiented Verma modules $\mathcal{V}_{k,\ell}$ throughout this section (with one exception in Subsection \ref{subsec:hints_simple_spectrum}).\\

We now wish to describe the finer structure of the Verma module $\mathcal{V}_{k,\ell}$ with respect to the operator $\Lq_0$ and the finite $\su(2)$-symmetry of the model. These commute with the integrable structure and will thus play an important role in its diagonalization in the next subsections.

\paragraph{$\bm{\Lq_0}$ and the conformal dimension.} The operator $\Lq_0$ measures the (left-moving contribution of the) \textit{conformal dimension} of states. We will then say that $\ket{\psi}\in\mathcal{V}_{k,\ell}$ has conformal dimension $h$ if $\Lq_0\ket{\psi}=h\,\ket{\psi}$. The Sugawara construction \eqref{eq:SugModes} of $\Lq_0$ and the properties \eqref{eq:highest_weight} of the highest-weight vector $\ket{\ell}$ imply that the latter has dimension\vspace{-3pt}
\begin{equation}\label{eq:conf_dim_hw}
    h_\ell := \frac{\ell(\ell+1)}{k+2}\,.\vspace{-2pt}
\end{equation}
Moreover, the commutation relation $[\Lq_0,\Jq^a_{-n}]=n\,\Jq^a_{-n}$ allows us to compute the conformal dimension of all the other states in the module:\vspace{-3pt}
\begin{equation}
    h\bigl(\Jq^{a_1}_{-n_1}...\Jq^{a_p}_{-n_p}\ket{\ell} \bigr) = h_\ell+\sum_{i=1}^p n_i\,.\label{eq:conf_dim_module}\vspace{-2pt}
\end{equation}
We will often call the sum $\sum_{i=1}^p n_i$ the \textit{depth} of the state $\Jq^{a_1}_{-n_1}...\Jq^{a_p}_{-n_p}\ket{\ell}$, which then corresponds to the eigenvalue of $\Lq_0 -h_\ell$. By construction, this depth is always a non-negative integer and it thus defines a natural $\Z_{\geq 0}$-grading on the Verma module $\mathcal{V}_{k,\ell}$.

The positive Kac-Moody modes $\lbrace \Jq^a_{n} \rbrace_{n>0}$ play the role of raising operators for minus the depth, while the negative modes $\lbrace \Jq^a_{-n} \rbrace_{n>0}$ act as lowering operators, as anticipated earlier. We recall from equation \eqref{eq:HP} that the Hamiltonian of the WZNW model on the cylinder is $\Hq=\frac{2\pi}{L}\left(\Lq_0+\overline{\Lq}_0-\frac{c}{12}\right)$. The highest-weight state $\ket{\ell}$ thus has minimal energy $\frac{2\pi}{L}\left(h_\ell-\frac{c}{12}\right)$ inside the Verma module $\mathcal{V}_{k,\ell}$ and the negative modes $\Jq^a_{-n}$ can be interpreted as creating quanta of energy $\frac{2\pi}{L}n$ (while the positive modes destroy these quanta\footnote{In particular, this explains why we consider the positive modes as annihilation operators, to ensure that the spectrum of the Hamiltonian stays bounded below.}). The $\Z_{\geq 0}$-grading defined by the depth then essentially measures the energy spectrum of the theory (in finite volume). 

\paragraph{Finite $\bm{\su(2)}$ symmetry and isospin.} The WZNW model also possesses an internal SU(2)-symmetry generated by the zero-modes $\Jq^a_0$, forming the finite $\su(2)$ algebra inside $\widehat{\su(2)}_k$. This symmetry is naturally associated with a quantum number, measured by the Cartan operator $\Jq^3_0$ and which we call the \textit{isospin}\footnote{We refer to it as the isospin to avoid confusion with the space-time Lorentz spin (see foornote \ref{fn:Spin}), which is associated with the relativistic symmetry of the theory and appeared for instance as a label for our local IMs $\Iq^{(s)}$.}. If $\ket{\psi}$ has isospin $j$, \textit{i.e.} $\Jq^3_0\ket{\psi}=j\ket{\psi}$, the standard $\su(2)$ commutation relations imply that $(\Jq_0^\pm)^p\ket{\psi}$ has isospin $j\pm p$. The modes $\Jq^+_0$ and $\Jq^-_0$ are thus respectively raising and lowering operators for the isospin.

This allows us to organise the states in $\mathcal{V}_{k,\ell}$ into representations of this finite $\su(2)$ symmetry. To this effect, it is useful to consider the distinguished set of $\su(2)$-highest-weight vectors, \textit{i.e.} states $\ket{\psi}$ such that $\Jq^+_0\ket{\psi}=0$ and $\Jq^3_0\ket{\psi}=j\ket{\psi}$ for some isospin $j$. The space 
\begin{equation}
    \text{span}\bigl((\Jq^-_0)^p\ket{\psi},\,p\in\Z_{\geq 0} \bigr)
\end{equation}
spanned by such a vector and its $\su(2)$-descendants forms a representation of $\su(2)$, which is isomorphic to the $\su(2)$-Verma module $\mathcal{V}^{\su(2)}_j$ of isospin $j$. Note that this representation is non-unitary and infinite dimensional, since we work in the unquotiented Verma module $\mathcal{V}_{k,\ell}$ and the label $p$ then runs over the whole set $\Z_{\geq 0}$. In particular, this representation contains states of all isospins in the set $\lbrace j,j-1,j-2,\dots\rbrace$. In contrast, all these vectors share the same eigenvalue $j(j+1)$ under the quadratic Casimir operator
\begin{equation}
    (\vec{\Jq}_0)^2:=(\Jq^3_0)^2+\frac{1}{2}(\Jq^+_0\Jq^-_0+\Jq^-_0\Jq^+_0)\,.
\end{equation}

\paragraph{The depth-isospin decomposition.} Since $\Lq_0$ and $\Jq^3_0$ commute, we can measure the conformal dimension / depth and the isospin simultaneously, \textit{i.e.} find states with both $\Lq_0\ket{\psi}=(h_\ell+n)\ket{\psi}$ and $\Jq_0^3\ket{\psi}=j\,\ket{\psi}$. The Kac-Moody mode $\Jq^\pm_0$ preserves the depth but raises/lowers the isospin, $\Jq^3_{m\neq 0}$ preserves the isospin but shifts the depth and $\Jq^\pm_{m\neq 0}$ shifts both the depth and the isospin. One checks that at a given depth $n$, only isospins $j \in \lbrace \ell+n,\ell+n-1,\ell+n-2,\dots\rbrace$ can appear: it is then natural to parametrise these isospins as $j=\ell+n-m$, with $m\in\Z_{\geq 0}$ a non-negative integer.\footnote{This parametrisation is quite natural from the point of view of the affine algebra $\widehat{\su(2)}_k$, whose lowering operators along simple roots are $\Jq_{-1}^+$ and $\Jq^-_0$. Indeed, a vector in $\mathcal{V}_{k,\ell}$ obtained by applying $n$ times $\Jq_{-1}^+$ and $m$ times $\Jq^-_0$ to $\ket{\ell}$ (in any given order) will have depth $n$ and isospin $\ell+n-m$. This will be useful in Subsection \ref{subsec:affine_bethe_ansatz}.}

To understand the structure of the Verma module, it is quite useful to introduce the space of all $\su(2)$-highest-weight vectors of depth $n$ and isospin $\ell+n-m$ inside $\mathcal{V}_{k,\ell}$:
\begin{equation}\label{eq:Upsilon}
    \Upsilon_{k,\ell}^{(n,m)}:=\Big\{ \ket{\psi}\in\mathcal{V}_{k,\ell} \; \Big| \; \Lq_0\ket{\psi}=(h_\ell+n)\ket{\psi}\,,\;\,\Jq^3_0\ket{\psi}=(\ell+n-m)\ket{\psi}\,, \;\, \Jq_0^+\ket{\psi}=0 \Big\}.
\end{equation}
Equivalently, the last condition can be rephrased as $(\vec{\Jq}_0)^2\ket{\psi}=(\ell+n-m)(\ell+n-m+1)\ket{\psi}$. One can check that this space is non-trivial if and only if $m\in\lbrace 0,\dots,2n\rbrace$, so that there are only a finite number of subspaces $\Upsilon_{k,\ell}^{(n,m)}$ at a given depth $n\in\Z_{\geq 0}$. Moreover, for any choice of $(n,m)$, the space $\Upsilon_{k,\ell}^{(n,m)}$ is always finite-dimensional and its dimension does not depend on the values of $(k,\ell)$.\footnote{Using characters, this dimension is the coefficient of $q^ny^{n-m}$ in $1/\prod_{r\in\Z_{\geq 1}}(1-q^r)(1-y q^r)(1-y^{-1}q^r)$.}  The $\su(2)$-descendants $(\Jq^-_0)^p\ket\psi$ of a vector $\ket{\psi}\in\Upsilon_{k,\ell}^{(n,m)}$ all have the same depth $n$ but do not belong to $\Upsilon_{k,\ell}^{(n,m)}$: together, they form a $\su(2)$-representation isomorphic to $\mathcal{V}^{\su(2)}_{\ell+n-m}$. Reciprocally, any state of depth $n$ is either in some $\Upsilon_{k,\ell}^{(n,m)}$ or a $\su(2)$-descendant of a vector in $\Upsilon_{k,\ell}^{(n,m)}$. We then obtain the depth-isospin decomposition
\begin{equation}\label{eq:DepthIso}
    \mathcal{V}_{k,\ell} = \bigoplus_{(n,p)\in\Z_{\geq0}^2}\, \bigoplus_{m\in\lbrace 0,\dots,2n\rbrace}\; (\Jq^-_0)^p\,\Upsilon_{k,\ell}^{(n,m)}\,. 
\end{equation}
This decomposition organises states according to their eigenvalues under the commuting operators $(\Lq_0,(\vec{\Jq}_0)^2,\Jq_0^3)$. More precisely, $(\Jq^-_0)^p\,\Upsilon_{k,\ell}^{(n,m)}$ is the eigenspace of these operators with eigenvalues
\begin{equation}
     \bigl(\,h_\ell+n\,,\, j(j+1)\,,\,j-p\,\bigr)\, \quad \text{ with } \quad j=\ell+n-m\,.
\end{equation}
Equivalently, this gives us the decomposition of $\mathcal{V}_{k,\ell}$ into representations of the abelian algebra $\text{span}(\Lq_0)$ times the finite $\su(2)$ algebra. More precisely, we have
\begin{equation}\label{eq:DecoL0su2}
    \mathcal{V}_{k,\ell} \simeq \bigoplus_{n\in\Z_{\geq0}}\,\bigoplus_{m\in\lbrace 0,\dots,2n\rbrace}\; \bigl(\dim \Upsilon_{k,\ell}^{(n,m)}\bigr)\, \mathcal{W}_{h_\ell+n} \otimes \mathcal{V}_{\ell+n-m}^{\text{su}(2)}\,,\vspace{-6pt}
\end{equation}
where $\mathcal{W}_h$ is the one-dimensional representation of $\text{span}(\Lq_0)$ of weight $h$. Note that the multiplicity of this decomposition is given by the dimension $\dim \Upsilon_{k,\ell}^{(n,m)}$, which is independent of $(k,\ell)$.

\paragraph{Decomposition of the quotient.} For $k\in\Z_{\geq 1}$ and $\ell\in \bigl\lbrace 0,\frac{1}{2},1,\dots,\frac{k}{2}\bigr\rbrace$, recall that the quotient $\widetilde{\mathcal{V}}_{k,\ell}$ of the Verma module by null vectors forms a unitary representation of $\widehat{\su(2)}_k$. Similarly to the construction above, we define $\widetilde{\Upsilon}_{k,\ell}^{(n,m)}$ as the space of all $\su(2)$-highest-weight vectors of depth $n$ and isospin $\ell+n-m$ within $\widetilde{\mathcal{V}}_{k,\ell}$. As before the quotient, all the states of depth $n$ in $\widetilde{\mathcal{V}}_{k,\ell}$ can be obtained as vectors in some $\widetilde{\Upsilon}_{k,\ell}^{(n,m)}$ or $\su(2)$-descendants thereof. The main difference is that these descendants now form $(2j+1)$-dimensional unitary representations $\widetilde{\mathcal{V}}^{\,\su(2)}_j$ of $\su(2)$, where $j=\ell+n-m$ (instead of infinite-dimensional Verma modules $\mathcal{V}_j^{\su(2)}$). We then replace the decomposition \eqref{eq:DepthIso} by
\begin{equation}
    \widetilde{\mathcal{V}}_{k,\ell} = \bigoplus_{n\in\Z_{\geq0}}\,  \bigoplus_{m\in\lbrace 0,\dots,2n\rbrace}\,\bigoplus_{p\in\lbrace 0,\dots  ,2(\ell+n-m)\rbrace}\; (\Jq^-_0)^p\,\widetilde{\Upsilon}_{k,\ell}^{(n,m)}\,,
\end{equation}
or equivalently \eqref{eq:DecoL0su2} by
\begin{equation}
    \widetilde{\mathcal{V}}_{k,\ell} \simeq \bigoplus_{n\in\Z_{\geq0}}\,\bigoplus_{m\in\lbrace 0,\dots,2n\rbrace}\; \bigl(\dim \widetilde\Upsilon_{k,\ell}^{(n,m)}\bigr)\, \mathcal{W}_{h_\ell+n} \otimes \widetilde{\mathcal{V}}_{\ell+n-m}^{\,\text{su}(2)}\,.\vspace{-8pt}
\end{equation}
The multiplicity $\dim \widetilde{\Upsilon}_{k,\ell}^{(n,m)}$ is always smaller or equal than the unquotiented one $\dim \Upsilon_{k,\ell}^{(n,m)}$ and now depends on the choice of $(k,\ell)$. 

\paragraph{Examples.} Let us illustrate the depth-isospin decomposition concretely, working in the unquotiented module $\mathcal{V}_{k,\ell}$. The lowest lying example is simply given by the $\widehat{\su(2)}_k$ highest weight $\ket{\ell}$, which trivially belongs to $\Upsilon_{k,\ell}^{(0,0)}$ and in fact spans it. Similarly, $\Jq^+_{-1}\ket{\ell}$ spans $\Upsilon_{k,\ell}^{(1,0)}$. However, note that generically $\Upsilon_{k,\ell}^{(n,m)}$ will be more than one-dimensional. For example, at depth $2$, the states
\begin{subequations}\label{eq:eigensystem_degeneracies}
\begin{align}
    \ket{\psi_1}&=\Jq^+_{-2}\ket{\ell}, \\[2pt]
    \ket{\psi_2}&=(\Jq^+_{-1})^2\Jq^-_0\ket{\ell}+2\ell\Jq^3_{-1}\Jq^+_{-1}\ket{\ell},
\end{align}
\end{subequations}
span the space $\Upsilon_{k,\ell}^{(2,1)}$, \textit{i.e.} their eigenvalues under  $(\Lq_0,(\vec{\Jq}_0)^2,\Jq_0^3)$ are $(h_\ell+2,(\ell+1)(\ell+2),\ell+1)$. Later on, we will also need the example of  $\Upsilon_{k,\ell}^{(2,2)}$, which is 3-dimensional and with eigenvalues $(h_\ell+2,\ell(\ell+1),\ell)$. A basis of this space is given by the states
\begin{subequations}\label{eq:Chi}
\begin{align}
    \ket{\chi_1} &= \Jq^3_{-2}\ket{\ell} + \Jq^3_{-1} \, \Jq^3_{-1} \ket{\ell} + \Jq^-_{-1} \, \Jq^+_{-1} \ket{\ell}\,, \\[2pt]
    \ket{\chi_2} &= \Jq^+_{-2}\,\Jq^-_0\ket{\ell} - 2\ell\, \Jq^-_{-1}\,\Jq^+_{-1}\ket{\ell} - 2\ell\, \Jq^3_{-1} \, \Jq^3_{-1} \ket{\ell}\,, \\[2pt]
    \ket{\chi_3} &= \Jq^+_{-1}\,\Jq^+_{-1}\,\Jq^-_{0}\,\Jq^-_{0} \ket{\ell} + 2(2\ell-1) \Jq^3_{-1}\,\Jq^+_{-1}\,\Jq^-_{0}\ket{\ell} - 2\ell(2\ell-1) \Jq^-_{-1}\,  \Jq^+_{-1} \ket{\ell}\,.
\end{align}
\end{subequations}

\paragraph{Degeneracies.} The fact that the spaces $\Upsilon^{(n,m)}_{k,\ell}$ are generically more that one-dimensional means that the commuting operators $(\Lq_0,(\vec{\Jq}_0)^2,\Jq_0^3)$ have degenerate spectrum: their eigenvalues are not enough to characterise a single state in $\mathcal{V}_{k,\ell}$. As we shall see in the next subsections, we conjecture that adding the new local IM $\Iq^{(3)}$ will lift these degeneracies and make the spectrum simple.\footnote{We note that the degeneracy of $(\Lq_0,(\vec{\Jq}_0)^2,\Jq_0^3)$ stays true in the quotient $\widetilde{\mathcal{V}}_{k,\ell}$ and thus is not just an accidental consequence of keeping non-physical states in the representation. Independently of any ``integrability'' considerations, \textit{i.e.} from a pure CFT point of view, we note that there is a slightly finer structure of the Hilbert space than the present decomposition in eigenspaces of $(\Lq_0,(\vec{\Jq}_0)^2,\Jq_0^3)$, which is to organise the states into representations of the $\su(2)$ algebra times the full Virasoro algebra (rather than $\Lq_0$ alone). As we will explain in Subsection \ref{subsec:hints_simple_spectrum}, this description is also degenerate. In particular, this will tell us that adding the KdV IM $\Iq^{(3)}_\KdV$ rather than $\Iq^{(3)}$ does not lift the degeneracies of $(\Lq_0,(\vec{\Jq}_0)^2,\Jq_0^3)$. We refer to Subsection \ref{subsec:hints_simple_spectrum} for details and a discussion.}

\subsection{Diagonalization of the local IMs}
\label{subsec:SpecI3}

\paragraph{Generalities.} In Section \ref{sec:structure}, we have discussed a new integrable structure for the SU(2) WZNW model: we now want to study the diagonalization of its commuting local IMs $\Iq^{(2p-1)}$ on the Verma module $\mathcal{V}_{k,\ell}$. We first argue that it is sufficient to consider the action of these operators on the subspaces $\Upsilon_{k,\ell}^{(n,m)}$ to understand their full spectrum.

To start with, notice that the first local IM in the integrable structure is $\Iq^{(1)}=\Lq_0 - \frac{c}{24}$. Up to a constant shift by $h_\ell-\frac{c}{24}$, its eigenvalue is simply the depth (\textit{i.e.} $n$ on $\Upsilon_{k,\ell}^{(n,m)}$) and its diagonalization was already performed in equation \eqref{eq:conf_dim_module}. Since the higher-spin IMs $\Iq^{(2p-1)}$ commute with $\Iq^{(1)}$, they will crucially preserve this depth. The second key point is that by construction, the integrable structure lies in the Casimir subalgebra, so it is built from fields invariant under the finite $\su(2)$ symmetry. As such, we have $[\Iq^{(2p-1)},\Jq^3_0]=[\Iq^{(2p-1)},(\vec{\Jq}_0)^2]=0$. Combining these two observations, we conclude that $\Iq^{(2p-1)}$ preserves the depth-isospin decomposition \eqref{eq:DepthIso} of $\mathcal{V}_{k,\ell}$ and in particular stabilises the subspaces $\Upsilon_{k,\ell}^{(n,m)}$ defined in equation \eqref{eq:Upsilon}. Since these are always finite dimensional, the action of $\Iq^{(2p-1)}$ on $\Upsilon_{k,\ell}^{(n,m)}$ is simply given by a finite size matrix (in a given basis), so that the eigenvalues and eigenvectors can be computed from standard linear algebra (at least in principle). Moreover, we recall that such an eigenvector $\ket{\psi}\in\Upsilon_{k,\ell}^{(n,m)}$ is a $\su(2)$-highest-weight state. Since the IMs commute with $\Jq^-_0$, all the $\su(2)$-descendants $(\Jq^-_0)^q\ket{\psi}$ are also eigenvectors of $\Iq^{(2p-1)}$ with the same eigenvalue, but lower isospins. This exhausts all the eigenstates of $\Iq^{(2p-1)}$ in $\mathcal{V}_{k,\ell}$.

In conclusion, it is completely sufficient to consider the action of the local IMs on the subspaces $\Upsilon_{k,\ell}^{(n,m)}$ and the spectral problem is reduced to diagonalising finite size matrices at each depth and isospin. We now want to study this spectrum in more details. For simplicity, we will focus on the first non-trivial IM $\Iq^{(3)}$. We give its expression in terms of modes:
\begin{equation}\label{eq:I3Modes}
   \Iq^{(3)}:=\Wq^{(4)}_0= \Lq_0^2-\frac{c+2}{12}\Lq_0+\frac{c(c+6)}{576}+2\sum_{n=1}^{+\infty}\Lq_{-n}\Lq_n+\frac{4}{3}\frac{1}{k+2}\sum_{n=1}^\infty n^2 \Jq^a_{-n}\Jq^a_{n}\;,
\end{equation}
which follows from the formulae \eqref{NO_cyl_modes} and \eqref{eq:W4WZW} and is needed to compute its action on the states (note that this action is always well-defined thanks to the normal-ordering of modes). In this subsection, we will study the diagonalization of $\Iq^{(3)}$ on the simplest subspaces $\Upsilon_{k,\ell}^{(n,m)}$. For more general, but conjectural, statements on the diagonalization of the whole integrable structure, in any subspace $\Upsilon_{k,\ell}^{(n,m)}$, we refer to the Subsections \ref{subsec:affine_bethe_ansatz} and \ref{subsec:odeim}. Also, we focus here on local IMs: non-local ones are diagonalised by the same basis, as briefly discussed at the end of Section \ref{sec:Kondo}.

Finally, for $(k,l)$ satisfying the unitarity constraints, we note that the null vectors of $\mathcal{V}_{k,l}$ are stable under the IMs $\Iq^{(2p-1)}$. These operators thus descend to the unitary quotient and one can use the diagonalization of $\Iq^{(2p-1)}$ on $\mathcal{V}_{k,l}$ to find the spectrum in $\widetilde{\mathcal{V}}_{k,l}$. For this, one has to check whether the different eigenvectors found in $\mathcal{V}_{k,l}$ survive the quotient or not. Although this can be done on examples, a systematic understanding of this point is beyond the scope of the present paper and constitutes a natural perspective to determine the fully physical integrable spectrum.

\paragraph{Action on $\Upsilon_{k,\ell}^{(0,0)}$ and $\Upsilon_{k,\ell}^{(1,0)}$.} We first compute the action on $\Upsilon_{k,\ell}^{(0,0)}$, that is on the $\widehat{\su(2)}_k$-highest-weight state $\ket{\ell}$. We start with the mode expansion \eqref{eq:I3Modes} of $\Iq^{(3)}$. Since the positive modes $\Lq_n$ and $\Jq^a_n$ ($n>0$) appearing in the last two terms annihilate $\ket{\ell}$, the only non-trivial contributions come from the first three terms, which involve constants and $\Lq_0$. The latter acts on $\ket{\ell}$ by the conformal dimension $h_\ell$ found in \eqref{eq:conf_dim_hw}, thus yielding
\begin{equation}
    \Iq^{(3)}\ket{\ell} = \Ec^{(3)}(h_\ell)\ket{\ell}\,, \qquad \text{ with } \qquad  \Ec^{(3)}(h):=\left(h^2 -\frac{c+2}{12} h +\frac{c(c+6)}{576}\right) \,. \label{eq:Q2_on_HW}
\end{equation}
This gives the eigenvalue of $\Iq^{(3)}$ on the reference state $\ket{\ell}$.\\

We can now proceed with excited states, starting with the vector $\Jq^+_{-1}\ket{\ell}$ spanning $\Upsilon_{k,\ell}^{(1,0)}$. As it is a typical illustration of the type of manipulations performed in this section, let us detail this computation. We start once again with the mode expansion \eqref{eq:I3Modes} of $\Iq^{(3)}$. The first three terms act on $\Jq^+_{-1}\ket{\ell}$ by multiplication by $\Ec^{(3)}(h_\ell+1)$, with $\Ec^{(3)}(h)$ as in equation \eqref{eq:Q2_on_HW} (taking into account the shift by 1 in the conformal dimension of $\Jq^+_{-1}\ket{\ell}$). We are thus left with computing the action of the last two terms in \eqref{eq:I3Modes}. To do so, we commute the positive modes $\Lq_n$ and $\Jq^a_n$ ($n>0$) through the lowering operator $\Jq^+_{-1}$, so that they act on the highest-weight state $\ket{\ell}$ and thus give zero. This leaves only terms involving $[\Lq_n,\Jq^+_{-1}]\ket{\psi}$ and $[\Jq^a_n,\Jq^+_{-1}]\ket{\psi}$, which are easily computed using the modes' algebra. In the end, we find
\begin{equation}
    \Iq^{(3)}\Jq^+_{-1}\ket{\ell}= \left(\Ec^{(3)}(h_\ell+1) +{\frac{2}{3}}\frac{k-2\ell}{k+2} \right)\Jq^+_{-1}\ket{\ell}\label{eq:Q2_on_first_excited}.
\end{equation}
The expressions \eqref{eq:Q2_on_HW} and \eqref{eq:Q2_on_first_excited} provide nice simple examples of eigenvectors and eigenvalues of $\Iq^{(3)}$ (which will be useful in Subsection \ref{subsec:odeim} to check the ODE/IQFT conjecture). However, they are a bit trivial, as the subspaces $\Upsilon_{k,\ell}^{(0,0)}$ and $\Upsilon_{k,\ell}^{(1,0)}$ are 1-dimensional. Indeed, since $\Iq^{(3)}$ stabilises these spaces, it is clear that $\ket{\ell}$ and $\Jq^+_{-1}\ket{\ell}$ have to be eigenvectors of $\Iq^{(3)}$.

\paragraph{Action on $\bm{\Upsilon_{k,\ell}^{(2,1)}}$.} More interesting examples arise in the presence of degeneracies, \textit{i.e.} when the space $\Upsilon_{k,\ell}^{(n,m)}$ is multi-dimensional. The lowest lying case is $\Upsilon_{k,\ell}^{(2,1)}$, which is spanned by the two states $\ket{\psi_{1,2}}$ in equation \eqref{eq:eigensystem_degeneracies}.

One can compute the action of $\Iq^{(3)}$ on $\ket{\psi_{i}}$ using the same techniques as above. The first three terms in the mode expansion \eqref{eq:I3Modes} simply act by multiplication by $\Ec^{(3)}(h_\ell+2)$. We are thus left with the action of the last two terms, which is computed by commuting positive modes with lowering operators. The main difference with the previous examples is that this computation will not produce a vector proportional to $\ket{\psi_{i}}$ itself, but will also involve terms proportional to the other vector in the basis. Explicitly, we find
\begin{align}
\hspace{-6pt}   \Iq^{(3)}\ket{\psi_1}&=  \frac{2409 k^2\hspace{-1pt}+\hspace{-1pt}4 k \left(172 \ell ^2\hspace{-1pt}-\hspace{-1pt}468 \ell \hspace{-1pt}+\hspace{-1pt}2003\right)\hspace{-1pt}+\hspace{-1pt}64 \left(3 \ell ^4\hspace{-1pt}+\hspace{-1pt}6 \ell ^3\hspace{-1pt}+\hspace{-1pt}26 \ell ^2\hspace{-1pt}-\hspace{-1pt}57 \ell \hspace{-1pt}+\hspace{-1pt}100\right)}{192 (k+2)^2}\ket{\psi_1} + \frac{4}{k+2}\ket{\psi_2}, \nonumber\\[8pt]
\hspace{-6pt} \label{eq:I3n2j1}   \Iq^{(3)}\ket{\psi_2}&=\frac{4 \ell  (k (\ell +4)-2 \ell  (\ell +3)+4)}{k+2} \ket{\psi_1}\\[2pt]
 &\hspace{25pt} +\frac{873 k^2+4 k \left(364 \ell ^2+684 \ell +851\right)+64 \left(3 \ell ^4+6 \ell ^3+50 \ell ^2+87 \ell +52\right)}{192 (k+2)^2}\ket{\psi_2}. \nonumber
\end{align}
This is simply a $2\times2$ mixing matrix, which we can diagonalise to find the eigenvalues
\begin{align}
&\frac{1641 k^2+4 k (4 \ell  (67 \ell +27)+1427)+64 (\ell  (\ell  (3 \ell  (\ell +2)+38)+15)+76)}{192 (k+2)^2}\nonumber \\
&\hspace{15pt} \mp\frac{2 \sqrt{ 4 (k+1)^2 -8 k \ell  +\ell ^4+4 \ell ^3+8 \ell ^2-8\ell}}{(k+2)}\,, \label{eq:Eigen21}
\end{align}
with eigenvectors
\begin{equation}
\ket{\psi_\pm}:=\frac{1}{2} \left(2 k-\ell (\ell+6)+2 \pm\sqrt{4 (k+1)^2 -8 k \ell  +\ell ^4+4 \ell ^3+8 \ell ^2-8\ell}\right)\ket{\psi_1} + \ket{\psi_2}.\label{eq:explicit_eigenvectors}
\end{equation}

Crucially, notice that these eigenvalues are distinct (except for the very specific and non-unitary cases $(k,\ell)=(-1,0)$ and $(-3,-2)$): we have thus lifted the degeneracy of this subsystem. More precisely, although the set of commuting operators $(\Lq_0,(\vec{\Jq}_0)^2,\Jq_0^3)$ has degenerate spectrum on the subspace $\Upsilon_{k,\ell}^{(2,1)}$, adding the operator $\Iq^{(3)}$ makes the spectrum simple. We also observe that these eigenvalues are real (for all real values of $k$ and $\ell$). Finally, we stress again that diagonalising this matrix provides not only eigenvalues of $\Iq^{(3)}$ on $\ket{\psi_\pm}$, but also on all of their descendants $(\Jq_0^-)^p\ket{\psi_\pm}$.

\paragraph{Discussion.} We can then proceed analogously for other spaces $\Upsilon_{k,\ell}^{(n,m)}$. Their dimension grows with $n$ (and as $m$ approaches $n$), so that the computations become more and more involved. It will be useful for Section \ref{sec:Kondo} to look at the next simplest case, namely the 3-dimensional space $\Upsilon_{k,\ell}^{(2,2)}$, whose basis is given by equation \eqref{eq:Chi}. The computations being heavier, we report on these results in Appendix \ref{App:n2j0} (see in particular the equation \eqref{eq:I3_chi} for the action of $\Iq^{(3)}$ in this basis).

The examples $\Upsilon_{k,\ell}^{(2,1)}$ and $\Upsilon_{k,\ell}^{(2,2)}$ are both multi-dimensional and our computations show that $\Iq^{(3)}$ has simple spectrum on these spaces in both cases (excluding specific non-unitary values of $(k,\ell)$). Based on these observations, we conjecture that the action of $\Iq^{(3)}$ generically lifts the degeneracy of the spaces $\Upsilon_{k,\ell}^{(n,m)}$, or in other words that we can uniquely label states in $\mathcal{V}_{k,\ell}$ by their eigenvalues under $(\Lq_0,(\vec{\Jq}_0)^2,\Jq_0^3,\Iq^{(3)})$. We will provide a further check of this conjecture in the next subsection and will show that this property is not shared by the KdV charges $\Iq^{(2p-1)}_\KdV$, making the new IMs $\Iq^{(2p-1)}$ quite natural. Note that, if this conjecture is true, the higher-spin IMs $\Iq^{(2p-1)}$ ($p>2$) would automatically be diagonalised in the eigenspaces $(\Lq_0,(\vec{\Jq}_0)^2,\Jq_0^3,\Iq^{(3)})$, as these spaces would then be one-dimensional and stable under $\Iq^{(2p-1)}$.

So far, we have taken a quite brute-force approach to the problem of diagonalising $\Iq^{(3)}$. Although this provided explicit expressions for the eigenvalues and eigenvectors in the simplest examples -- for low $(n,m)$, the next cases would quickly become quite involved or even impossible, as one would have to solve higher-degree polynomial equations. Moreover, although the same method can in principle be used to diagonalise the higher-spin IMs $\Iq^{(2p-1)}$, this endeavour also rapidly grows in complexity as we increase the spin, making it technically challenging. This shows the need for more conceptual and structural approaches. Such developments include the affine Bethe ansatz, which gives a (conjectural) construction for all eigenvectors, and the ODE/IQFT correspondence, which allows (once again conjecturally) to access the spectrum of all higher-spin charges $\Iq^{(2p-1)}$. We will discuss these in Subsections \ref{subsec:affine_bethe_ansatz} and \ref{subsec:odeim} and in particular check the validity of these conjectures on the explicit examples presented above.


\subsection{Simple spectrum and comparison with KdV charges}\label{subsec:hints_simple_spectrum}

\paragraph{Goal.} In the previous subsection, we have discussed the spectrum of the local IM $\Iq^{(3)}$ on the simplest states in the Verma module $\mathcal{V}_{k,\ell}$. In particular, we showed that including this additional commuting operator lifts the degeneracies of $(\Lq_0,(\vec{\Jq}_0)^2,\Jq_0^3)$ on these states. We thus conjectured that this extends to the whole Verma module, \textit{i.e.} that $(\Lq_0,(\vec{\Jq}_0)^2,\Jq_0^3,\Iq^{(3)})$ has simple spectrum on $\mathcal{V}_{k,\ell}$ (for generic values of $(k,\ell))$. Here, we present another example in favour of this conjecture.

The main difference with our previous examples is due to the following observation. So far, we have only considered the diagonalization of the new integrable structure $\Iq^{(2p-1)}$ introduced in Section \ref{sec:structure} (focusing on $\Iq^{(3)}$). However, recall that the analysis of Section \ref{sec:structure} also recovered another tower of commuting local IMs, the KdV charges $\Iq^{(2p-1)}_\KdV$, built only from the stress-energy tensor and which differ from $\Iq^{(2p-1)}$ except for the first one $\Iq^{(1)}_\KdV=\Iq^{(1)}=\Lq_0-\frac{c}{24}$. Since the KdV charges also commute with $(\Lq_0,(\vec{\Jq}_0)^2,\Jq_0^3)$, it is natural to wonder if they could also lift the degeneracies of these operators on the spaces $\Upsilon_{k,\ell}^{(n,m)}$. For the examples $\Upsilon_{k,\ell}^{(2,1)}$ and $\Upsilon_{k,\ell}^{(2,2)}$ considered in the previous subsection, one can check that $\Iq^{(3)}_\KdV$ also has simple spectrum, as did $\Iq^{(3)}$ earlier. In this subsection, we will exhibit an explicit example where the KdV IMs do not lift the degeneracy, whereas the new IM $\Iq^{(3)}$ does. This shows that the new integrable structure $\Iq^{(2p-1)}$ is more natural and powerful than the KdV one for the SU(2) WZNW-model and provides a new argument in favour of the simple spectrum of $(\Lq_0,(\vec{\Jq}_0)^2,\Jq_0^3,\Iq^{(3)})$ (by now comparing with other sets of commuting operators).

\paragraph{The Virasoro-isospin decomposition.} The KdV charges $\Iq^{(2p-1)}_\KdV$ are built from the stress-energy tensor only and are thus expressed purely in terms of Virasoro modes $\Lq_p$. To discuss their spectrum, it will therefore be quite useful to consider the decomposition of $\mathcal{V}_{k,\ell}$ with respect to the Virasoro algebra $\text{Vir}=\text{span}(\Lq_p,p\in\Z)$. As we shall see, this is a finer structure than the depth-grading, which only took into account the zero-mode $\Lq_0$. Since all Virasoro generators $\Lq_p$ commute with the finite $\su(2)$ generators $\Jq^a_0$, we can perform this decomposition simultaneously with the isospin one.

Recall the space $\Upsilon_{k,\ell}^{(n,m)}$ of $\su(2)$-highest-weight vectors of isospin $\ell+n-m$ and depth $n$ in $\mathcal{V}_{k,\ell}$. It contains further distinguished states, which are also Virasoro-primaries\footnote{We could also have used the term Virasoro-highest-weight. To facilitate their distinction, we will keep the terminology ``highest-weight'' for the affine and finite $\su(2)$ algebras and the terminology ``primary'' for the Virasoro algebra.}, \textit{i.e.} are annihilated by the positive modes $\Lq_{p>0}$. They form the subspace
\begin{equation}
    \Xi_{k,\ell}^{(n,m)} = \Bigl\lbrace \ket{\psi} \in \mathcal{V}_{k,\ell}  \; \Big| \; \Lq_0\ket{\psi}=(h_\ell+n)\ket{\psi}\,,\;\,\Jq^3_0\ket{\psi}=(\ell+n-m)\ket{\psi}\,, \;\, \Jq^+_0\ket{\psi}=\Lq_{p>0}\ket{\psi}=0 \Bigr\rbrace
\end{equation}
of $\Upsilon_{k,\ell}^{(n,m)}$. Given a vector in $\Xi_{k,\ell}^{(n,m)}$, we create new states by applying negative Virasoro modes $\Lq_{m<0}$ to raise the depth and $\su(2)$-lowering operators $\Jq^-_0$ to shift the isospin. This provides the Virasoro-isospin decomposition of the affine Verma module:
\begin{equation}\label{eq:VirIsospin}
    \mathcal{V}_{k,\ell} \simeq \bigoplus_{n\in\Z_{\geq0}}\,\bigoplus_{m\in\lbrace 0,\dots,2n\rbrace}\;  \alpha^{(n,m)}\,\mathcal{V}^{\text{Vir}}_{h_\ell+n} \otimes \mathcal{V}^{\su(2)}_{\ell+n-m}\,,\vspace{-6pt}
\end{equation}
where $\mathcal{V}^{\su(2)}_j$ is the $\su(2)$-Verma module of isospin $j$, as earlier, and $\mathcal{V}^{\text{Vir}}_{h}$ is the Virasoro-Verma module of conformal dimension $h$.\footnote{In this decomposition, the $\su(2)$ generators $\Jq^a_0$ then move us along $\mathcal{V}^{\su(2)}_{\ell+n-m}$ but stays at the same state of $\mathcal{V}^{\text{Vir}}_{h_\ell+n}$, while the Virasoro modes $\Lq_p$ do the opposite.} Crucially, we have included multiplicities $\alpha^{(n,m)}\in\Z_{\geq 0}$ in equation \eqref{eq:VirIsospin}, to take into account that the same tensor product $\mathcal{V}^{\text{Vir}}_{h_\ell+n} \otimes \mathcal{V}^{\su(2)}_{\ell+n-m}$ can appear several times in the decomposition. In the notations above, this multiplicity is nothing but
\begin{equation}
    \alpha^{(n,m)} = \dim\Xi_{k,\ell}^{(n,m)}\,,
\end{equation}
which turns out to be independent of the values of $(k,\ell)$.

The equation \eqref{eq:VirIsospin} gives the decomposition of $\mathcal{V}_{k,\ell}$ into representations of $\text{Vir}\oplus\su(2)$ and is thus a refinement of \eqref{eq:DecoL0su2}, which concerned only $\text{span}(\Lq_0) \oplus\su(2)$. In particular, the multiplicities $\alpha^{(n,m)}$ of the former are always smaller or equal than the ones $\dim\Upsilon_{k,\ell}^{(n,m)}$ of the latter.\\

For $k\in\Z_{\geq 1}$ and $\ell\in\lbrace 0,\frac{1}{2},1,\dots,\frac{k}{2}\rbrace$, we can transfer all of the above construction to the unitary quotient $\widetilde{\mathcal{V}}_{k,\ell}$. In particular, we define subspaces $\widetilde{\Xi}_{k,\ell}^{(n,m)}$ formed by Virasoro-primaries inside $\widetilde{\Upsilon}_{k,\ell}^{(n,m)}$ and the Virasoro-isospin decomposition
\begin{equation}
    \widetilde{\mathcal{V}}_{k,\ell} \simeq \bigoplus_{n\in\Z_{\geq0}}\,\bigoplus_{m\in\lbrace 0,\dots,2n\rbrace}\; \widetilde{\alpha}^{\,(n,m)}_{k,\ell}\,\widetilde{\mathcal{V}}^{\hspace{1pt}\text{Vir}}_{h_\ell+n} \otimes \widetilde{\mathcal{V}}^{\,\su(2)}_{\ell+n-m}\,,\vspace{-6pt}
\end{equation}
where $\widetilde{\mathcal{V}}^{\hspace{1pt}\text{Vir}}_{h}$ is the infinite-dimensional unitary representation of Virasoro of weight $h$ and $\widetilde{\mathcal{V}}^{\,\su(2)}_j$ is the $(2j+1)$-dimensional unitary representation of $\su(2)$. The multiplicity $    \widetilde{\alpha}^{\hspace{1pt}(n,m)}_{k,\ell} = \dim\widetilde{\Xi}_{k,\ell}^{\hspace{1pt}(n,m)}$ is always smaller or equal than the unquotiented one $\alpha^{(n,m)}$ and now depends on the values of $(k,\ell)$. These multiplicities essentially define the branching rules from the affine algebra $\widehat{\su(2)}_k$ to the subalgebra  $\text{Vir} \oplus\su(2)$. They are most easily computed with the help of characters (as the coefficients in the decomposition of $\widehat{\su(2)}_k$-characters as sums of products of Virasoro and $\su(2)$-characters).

\paragraph{Virasoro/KdV degeneracies.} Before illustrating the Virasoro-isospin decomposition \eqref{eq:VirIsospin} with examples, let us explain its relation with our initial problem. Since the KdV charges $\Iq^{(2p-1)}_\KdV$ are built only from Virasoro modes, they stabilise the Virasoro modules $\mathcal{V}^{\text{Vir}}_h$ embedded in $\mathcal{V}_{k,l}$ and their spectrum on such a module depends only on the weight $h$ of the primary. Therefore, the presence of a non-trivial multiplicity $\alpha^{(n,m)}>1$ in the decomposition \eqref{eq:VirIsospin} would by construction imply a degeneracy of the commuting operators $(\Lq_0,(\vec{\Jq}_0)^2,\Jq^3_0,\Iq^{(2p-1)}_\KdV)$ on $\mathcal{V}_{k,\ell}$.\footnote{Let us be a bit more precise. It is expected that the spectrum of KdV charges on a single Virasoro-module $\mathcal{V}_h$ is generically simple and more precisely that the eigenvalues of $\Iq^{(1)}_\KdV=\Lq_0-\frac{c}{24}$ and $\Iq^{(3)}_\KdV$ are enough to distinguish all the states in $\mathcal{V}^{\text{Vir}}_h$. Adding the operator $\Jq_0^3$ then allows to separate the states on a tensor product $\mathcal{V}^{\text{Vir}}_{h}\otimes\mathcal{V}^{\su(2)}_j$ and adding $(\vec{\Jq}_0)^2$ distinguishes between different values of $j$. The degeneracies of the whole set $\bigl(\Lq_0,(\vec{\Jq}_0)^2,\Jq^3_0,\Iq^{(2p-1)}_\KdV\bigr)$ on $\mathcal{V}_{k,\ell}$ thus exactly correspond to multiple apparitions of the same product $\mathcal{V}^{\text{Vir}}_{h}\otimes\mathcal{V}^{\su(2)}_j$ in the decomposition \eqref{eq:VirIsospin}, which are precisely tracked by the multiplicities $\alpha^{(n,m)}>1$.} To answer the question raised at the beginning of this subsection, we thus want to find a case with such a multiplicity and check whether the new IM $\Iq^{(3)}$ is enough to lift the degeneracy that would not be lifted by the KdV charges. To make sure that this degeneracy is not an accident of keeping non-physical states in $\mathcal{V}_{k,\ell}$, we should also check that this statement still holds in the quotient $\widetilde{\mathcal{V}}_{k,\ell}$ and thus need to look for non-trivial multiplicities $\widetilde{\alpha}^{\,(n,m)}_{k,\ell}>1$. Below, we exhibit a simple example of such a phenomenon.

\paragraph{Example with non-trivial multiplicity.} For depth $n=0,1,2$, we find that all the subspaces $\Xi_{k,\ell}^{(n,m)}$ are one-dimensional. The lowest lying example with a non-trivial multiplicity is $\Xi_{k,\ell}^{(3,2)}$, which we now describe. To illustrate the structure, we start with the larger space $\Upsilon_{k,\ell}^{(3,2)}$ formed by $\su(2)$-highest-weight vectors of isospin $\ell+1$ and depth $3$. One finds that this space is of dimension 5. Among these vectors, only 2 independent ones are also Virasoro-primaries, thus spanning $\Xi_{k,\ell}^{(3,2)}$. Those can be chosen as
\begin{subequations}\label{eq:Vir_prim}
\begin{align}
    \ket{\kappa_1}=& \ell (2\ell -1) (8 + k (-2 + \ell) -  4 \ell - 4 \ell^2 (4 + \ell)) \Jq^+_{-3}\ket{\ell} + 3 (2\ell -1) \Big(\ell (k + 2 \ell (2 + \ell)) \Jq^3_{-2}\Jq^+_{-1}\ket{\ell}\nonumber\\
    &+ \ell (12 - k + 2 \ell (9 + 2 \ell)) \Jq^+_{-2}\Jq^3_{-1}\ket{\ell} + (2 + 3 \ell) (-2 \ell (\Jq^3_{-1})^2\Jq^+_{-1}\ket{\ell}
    + (4 + \ell) \Jq^+_{-2}\Jq^+_{-1}\Jq^-_{0}\ket{\ell})\Big)\nonumber \\
    &-  6 (-2 + \ell + 6 \ell^2) \Jq^3_{-1}(\Jq^+_{-1})^2\Jq^-_{0}\ket{\ell}
    -  3 (2 + 3 \ell) (\Jq^+_{-1})^3(\Jq^-_{0})^2\ket{\ell} \,,\\[4pt]
    \ket{\kappa_2}=& -4 \ell (2\ell -1) (20 + 34 \ell + 8 \ell^2- k (11 + 2 \ell)) \Jq^+_{-3}\ket{\ell}- 3 (-5 k +  2 (2 + \ell) (1 + 2 \ell)) (\Jq^+_{-1})^3(\Jq^-_{0})^2\ket{\ell}\nonumber \\
    &+ 6 (2\ell -1) \Big(4 \ell (6 - k (4 + \ell) + 2 \ell (5 + \ell)) \Jq^3_{-2}\Jq^+_{-1}\ket{\ell}\nonumber\\
    &+  4 \ell (2 + k + 2 \ell) (\Jq^3_{-1})^2\Jq^+_{-1}\ket{\ell} + 
    (8- 10 k  +  16 \ell -  2 k \ell  +  4 \ell^2 )\Jq^+_{-2}\Jq^+_{-1}\Jq^-_0\ket{\ell}\nonumber \\
    &+ (5 k -  2 (2 + \ell) (1 + 2 \ell)) \Jq^3_{-1}(\Jq^+_{-1})^2\Jq^-_{0}\ket{\ell} + 
     \ell (12 - k + 2 \ell (9 + 2 \ell)) \Jq^-_{-1}(\Jq^+_{-1})^2\ket{\ell}\Big) \, .
\end{align}
\end{subequations}
This means that $\mathcal{V}^{\text{Vir}}_{h_\ell+3}\otimes\mathcal{V}^{\su(2)}_{\ell+1}$ appears twice in the Virasoro-isospin decomposition \eqref{eq:VirIsospin}. As explained earlier, the action of $\bigl(\Lq_0,(\vec{\Jq}_0)^2,\Jq^3_0,\Iq^{(2p-1)}_\KdV\bigr)$ on $\Upsilon_{k,\ell}^{(3,2)}$ is thus degenerate, having the same eigenvalues on both $\ket{\kappa_1}$ and $\ket{\kappa_2}$. For example, they share the eigenvalue $(h_\ell+3)^2-\frac{c+2}{12}(h_\ell+3)+\frac{c(5c+22)}{2880}$ under the action of $\Iq^{(3)}_\KdV$.

We now want to make sure that this degeneracy survives when taking the unitary quotient to $\widetilde{\mathcal{V}}_{k,\ell}$. This depends on the values of $(k,\ell)$, since these parameters enter the definition of the null vectors. We will not try to keep track of these null vectors for general $(k,\ell)$ and will therefore restrict to a simple example where the degeneracy survives, namely $(k,\ell)=(4,1)$. This will be enough to illustrate our point and will drastically simplify some of our explicit computations below. For this choice, all the vectors in $\Upsilon_{4,1}^{(3,2)}$ survive the quotient, so that $\widetilde{\Upsilon}_{4,1}^{(3,2)}$ and $\widetilde{\Xi}_{4,1}^{(3,2)}$ stay 5-dimensional and 2-dimensional respectively. The latter is spanned by the image of $\ket{\kappa_{1,2}}$ in the quotient and the operators $\bigl(\Lq_0,(\vec{\Jq}_0)^2,\Jq^3_0,\Iq^{(2p-1)}_\KdV\bigr)$ are degenerate on these vectors, as desired.

Finally, we want to check if this degeneracy is lifted by using the new local IMs $\Iq^{(2p-1)}$ instead of the KdV charges $\Iq^{(2p-1)}_\KdV$. We thus consider the action of $\Iq^{(3)}$ on $\widetilde{\Upsilon}_{4,1}^{(3,2)}$. Since it does not commute with the Virasoro modes $\Lq_n$ (except for $\Lq_0$), there are no reasons for $\Iq^{(3)}$ to stabilise the Virasoro-primaries $\widetilde{\Xi}_{4,1}^{(3,2)}$ inside $\widetilde{\Upsilon}_{4,1}^{(3,2)}$. The action of $\Iq^{(3)}$ on $\widetilde{\Upsilon}_{4,1}^{(3,2)}$ thus mixes all the states and results in a non-trivial $5\!\times\!5$ matrix in a choice of basis. Since we have specialised to $(k,\ell)=(4,1)$, the entries of this matrix are very explicit rational numbers (which we will not write here for simplicity). Although it would be difficult to find an explicit expression for the eigenvalues of this matrix, one can easily evaluate them numerically. Doing so, we check that the five eigenvalues are all distinct, so that $\Iq^{(3)}$ has simple spectrum on $\widetilde{\Upsilon}_{4,1}^{(3,2)}$, contrarily to the KdV charges. We have performed various other explicit checks for different values of $(k,\ell)$ (satisfying the unitarity constraints), always finding a simple spectrum. This provides a new argument in favour of our conjecture that $\Iq^{(3)}$ lifts the degeneracies of $\bigl(\Lq_0,(\vec{\Jq}_0)^2,\Jq^3_0\bigr)$, making it more powerful than the KdV charges.


\subsection{Affine Bethe Ansatz}\label{subsec:affine_bethe_ansatz}

\paragraph{Generalities.} So far, we have followed a quite naive and brute force approach to the diagonalization of our integrable structure, by explicitly diagonalising the operator $\Iq^{(3)}$ on the first few subspaces $\Upsilon_{k,\ell}^{(n,m)}$. It is clear that this approach has limits and would become more and more inapplicable as we raise the depth $n$. Fortunately for us, there exists a very powerful conjectural construction of \textit{all} the eigenstates of the WZNW integrable structure, which we will refer to as the \textit{affine Bethe ansatz}. This conjecture was initially developed in~\cite{Feigin:2007mr,Lacroix:2018fhf} for the more general framework of affine Gaudin models. It is based on a similar construction~\cite{gaudin2014bethe,Babujian:1993ts,Feigin:1994in} for the so-called finite Gaudin models\footnote{These finite Gaudin models are associated with finite-dimensional Lie algebras and define systems of interacting spins, with finitely many degrees of freedom. In contrast, the affine Gaudin models are associated with infinite-dimensional Lie algebras and take the form of field theories.} and makes extensive use of a remarkable work~\cite{Schechtman:1991hgd} by Schechtman and Varchenko on solutions of the Knizhnik–Zamolodchikov equation associated to any symmetrisable Kac-Moody algebra (thus including affine ones). For the case of the SU(2) WZNW model, it was further studied in~\cite{Gaiotto:2020dhf} as a method to diagonalise Kondo defects, \textit{i.e.} the non-local IMs of the WZNW integrable structure (which share the same eigenvectors as the local ones, since they mutually commute -- see Section \ref{sec:Kondo}). This construction strongly relies on the affine algebra $\widehat{\su(2)}_k$ underlying the model and shows again the naturality of the IMs $\Iq^{(2p-1)}$. In this subsection, we recall this construction and show that it reproduces the eigenvectors in $\Upsilon_{k,\ell}^{(0,0)}$, $\Upsilon_{k,\ell}^{(1,0)}$ and $\Upsilon_{k,\ell}^{(2,1)}$ found earlier by the brute-force approach. This then also serves as a check of the affine Bethe ansatz, which is only conjectural at the moment.\\

The $\widehat{\su(2)}_k$ Bethe ansatz depends on complex variables $\lbrace w_1,\dots,w_m \rbrace$ and $\lbrace w'_1,\dots,w'_n\rbrace$ respectively called the \textit{finite and affine Bethe roots} and which we collectively denote as $\{t_i\}=\{w_i\}\cup\{w_i'\}$. To each Bethe root $t_i$ is attached an operator $\Fq_{t_i}$, given by the standard lowering-operator $\Fq_{w_i}=\Jq^-_0$ of the finite $\su(2)$ algebra for a finite root and by the affine lowering-operator $\Fq_{w'_i}=\Jq_{-1}^+$ for an affine root. We then define the associated \textit{off-shell Bethe vector} as
\begin{equation}\label{eq:BA}
\ket{\{t_i\}}_{\text{BA}}=\sum_{\sigma\in S_{m+n}}\frac{\Fq_{t_{\sigma (1)}}\Fq_{t_{\sigma(2)}}\cdots \Fq_{t_{\sigma(m+n)}}}{(t_{\sigma(1)}-t_{\sigma(2)})(t_{\sigma(2)}-t_{\sigma(3)})\cdots(t_{\sigma(m+n-1)}-t_{\sigma(m+n)})t_{\sigma(m+n)}}\,\ket{\ell}\,,
\end{equation}
where $S_{m+n}$ is the symmetric group in $m+n$ elements (which permutes the indices of Bethe roots $\{t_i\}$). The main conjectural claim of the Bethe ansatz is that the state \eqref{eq:BA} is an eigenvector of the WZNW integrable structure if and only if the roots satisfy the Bethe ansatz equations
\begin{subequations}\label{eq:BAE}
    \begin{equation}
-\frac{\ell}{w_i}+\sum_{j\neq i}\frac{1}{w_i-w_j}-\sum_j\frac{1}{w_i-w_j'}=0\,,\label{eq:BAE1}
\end{equation}
\begin{equation}
\frac{1}{2}-\frac{\frac{k}{2}-\ell}{w_i'}+\sum_{j\neq i}\frac{1}{w_i'-w_j'}-\sum_{j}\frac{1}{w_i'-w_j}=0\,.\label{eq:BAE2}
\end{equation}
\end{subequations}
Under this constraint, we call $\ket{\{t_i\}}_{\text{BA}}$ an \textit{on-shell Bethe vector}. This conjecture was checked in~\cite{Gaiotto:2020dhf} for the first states and the first few non-local IMs in the integrable structure (see Section \ref{sec:Kondo}). The higher-spin local IMs had not been discussed explicitly in the literature so far and were the main motivations for the present work. We are now in position to extend the conjecture, by stating that $\ket{\{t_i\}}_{\text{BA}}$ should be an eigenvector of the local IMs $\Iq^{(2p-1)}$ defined in Section \ref{sec:structure} (and not of the KdV charges $\Iq^{(2p-1)}_\KdV$, showing that $\Iq^{(2p-1)}$ are also quite natural from this point of view).

Before performing simple checks of this conjecture, let us analyse further the Bethe vector \eqref{eq:BA}. Recalling the definition $\Fq_{w_i}=\Jq^-_0$ and $\Fq_{w'_i}=\Jq_{-1}^+$ of the lowering operators, it is clear that $\ket{\{t_i\}}_{\text{BA}}$ has conformal dimension $h_\ell+n$, where $n$ is the number of affine Bethe roots, and isospin $\ell+n-m$, where $m$ is the number of finite Bethe roots. Moreover, the conjecture in fact includes an additional statement, namely that the on-shell Bethe vector is a $\su(2)$-highest-weight state. In other words, if the roots $\{t_i\}$ are solutions of the Bethe ansatz equation \eqref{eq:BAE}, the vector $\ket{\{t_i\}}_{\text{BA}}$ is expected to be annihilated by $\Jq^+_0$. In the notation of equation \eqref{eq:Upsilon}, this simply means that
\begin{equation}\label{eq:UpsilonBA}
    \ket{\{t_i\}}_{\text{BA}} \,\in\, \Upsilon_{k,\ell}^{(n,m)} \; \text{ if \eqref{eq:BAE} hold} \,.
\end{equation}
Finally, one conjectures that the Bethe ansatz is (generically) complete, \textit{i.e.} that it provides all the eigenstates of the integrable structure in the spaces $\Upsilon_{k,\ell}^{(n,m)}$, at least for $(k,\ell)$ generic. In that case, we recall that all the other eigenstates in $\mathcal{V}_{k,\ell}$ would then be obtained as $\su(2)$-descendants of these on-shell Bethe vectors (see the discussion at the beginning of Subsection \ref{subsec:SpecI3}). 

The brute-force diagonalization of $\Iq^{(3)}$ on the first few spaces $\Upsilon_{k,\ell}^{(n,m)}$ was the main topic of Subsection \ref{subsec:SpecI3}. Therefore, we now want to check that the Bethe ansatz reproduces these results.

\paragraph{Trivial checks: $\Upsilon_{k,\ell}^{(0,0)}$ and $\Upsilon_{k,\ell}^{(1,0)}$.} The first examples considered in Subsection \ref{subsec:SpecI3} were $\Upsilon_{k,\ell}^{(0,0)}$ and $\Upsilon_{k,\ell}^{(1,0)}$. The former is spanned by the highest-weight vector $\ket{\ell}$, which trivially corresponds to the Bethe vector $\ket{\emptyset}_{\text{BA}}$ with no roots.

Due to the property \eqref{eq:UpsilonBA}, the second example $\Upsilon_{k,\ell}^{(1,0)}$ should correspond to the Bethe ansatz with $n=1$ affine root and no finite roots ($m=0)$). Denoting by $w'$ the corresponding unique affine root, the off-shell Bethe vector simply reads $\ket{\{w'\}}_{\text{BA}}=\frac{1}{w'}\Jq_{-1}^+\ket{\ell}$. This vector already spans $\Upsilon_{k,\ell}^{(1,0)}$, as one could expect since this space is only one-dimensional. It is thus automatically an eigenvector, even before putting it on-shell, \textit{i.e.} before solving the Bethe ansatz equation for $w'$. This is a low-dimensional accident, which will not happen for more complicated setups. For later comparison with the ODE/IQFT correspondence in Subsection \ref{subsec:odeim}, it will however be useful to still solve the corresponding Bethe ansatz equation, which yields
\begin{equation}\label{eq:aff_BAE_1st}
    w'=k-2\ell \,.
\end{equation}

\paragraph{Non-trivial check: $\Upsilon_{k,\ell}^{(2,1)}$.} The main non-trivial example analysed in Subsection \ref{subsec:SpecI3} is $\Upsilon_{k,\ell}^{(2,1)}$. We recall that it is 2-dimensional and that the direct diagonalization of $\Iq^{(3)}$ on this space resulted in the eigenvectors \eqref{eq:explicit_eigenvectors}. To recover those from the affine Bethe ansatz, we need to consider $m=1$ finite root $w$ and $n=2$ affine roots $w'_1,w'_2$. We then have to solve the corresponding BAEs \eqref{eq:BAE}. We find two independent solutions, which both have the same affine roots\vspace{-2pt}
\begin{subequations}\label{eq:bethe_roots_explicit}
\begin{align}
   w_1'&= -\frac{1}{2} \left(\sqrt{2 \ell  (\ell +2)-2 \sqrt{4 (k+1)^2 -8 k \ell  +\ell ^4+4 \ell ^3+8 \ell ^2-8\ell}}-2 k+2 \ell -2\right)\,,\\
   w_2'&= -\frac{1}{2} \left(-\sqrt{2 \ell  (\ell +2)-2 \sqrt{4 (k+1)^2 -8 k \ell  +\ell ^4+4 \ell ^3+8 \ell ^2-8\ell}}-2 k+2 \ell -2\right)\,,
\end{align}
but correspond to two different choices for the finite root:
\begin{equation}
w_\pm = \frac{\mp\sqrt{4 (k+1)^2 -8 k \ell  +\ell ^4+4 \ell ^3+8 \ell ^2-8\ell}+2 k (\ell    +1)+(\ell -2) \ell -2}{4 (\ell +2)}\,.
\end{equation}
\end{subequations}
Following equation \eqref{eq:BA}, the corresponding Bethe vectors are
\begin{align}\label{eq:bethe_vector_explicit}
\ket{\{w_\pm, w'_1, w'_2 \}}_{\text{BA}} &= \bigg( \frac{1}{w'_1 - w'_2} \frac{1}{w'_2 - w_\pm} \frac{1}{w_\pm} + \frac{1}{w'_2 - w'_1} \frac{1}{w'_1 - w_\pm} \frac{1}{w_\pm} \bigg) \Jq^+_{-1} \Jq^+_{-1} \Jq^-_0 \ket{\ell}\nonumber\\
&\quad + \bigg( \frac{1}{w'_1 - w_\pm} \frac{1}{w_\pm - w'_2} \frac{1}{w'_2} + \frac{1}{w'_2 - w_\pm} \frac{1}{w_\pm - w'_1} \frac{1}{w'_1} \bigg) \Jq^+_{-1} \Jq^-_0 \Jq^+_{-1} \ket{\ell}\nonumber\\
&\qquad + \bigg( \frac{1}{w_\pm - w'_1} \frac{1}{w'_1 - w'_2} \frac{1}{w'_2} + \frac{1}{w_\pm - w'_2} \frac{1}{w'_2 - w'_1} \frac{1}{w'_1} \bigg) \Jq^-_0 \Jq^+_{-1} \Jq^+_{-1} \ket{\ell}.
\end{align}
Using the explicit form \eqref{eq:bethe_roots_explicit} of the Bethe roots and re-arranging the lowering operators, we then express this vector in terms of the states $\ket{\psi_{1,2}}$ of equation \eqref{eq:eigensystem_degeneracies}. Up to a proportionality factor (which rescales the coefficient of $\ket{\psi_2}$ to $1$), we simply find\vspace{-2pt}
\begin{equation}
    \ket{\{w_\pm, w'_1, w'_2 \}}_{\text{BA}} \propto \ket{\psi_\pm}\,,
\end{equation}
with $\ket{\psi_\pm}$ the eigenvectors found in equation \eqref{eq:explicit_eigenvectors}. We thus showed that the affine Bethe ansatz correctly reproduces our brute-force diagonalization of $\Iq^{(3)}$, as expected.\vspace{-2pt}

\paragraph{Discussion.} The above example provides a first non-trivial check of the affine Bethe ansatz. More generally, the latter provides a conjectural construction for the eigenstates of $\Iq^{(2p-1)}$ on any space $\Upsilon_{k,\ell}^{(n,m)}$. We note that the corresponding Bethe ansatz equations \eqref{eq:BAE} become more involved as $(n,m)$ grow and generally define quite complicated coupled algebraic equations. We do not expect those to be exactly solvable by radicals, except for a few low-lying cases similar to the one above. The affine Bethe ansatz thus does not provide a completely closed solution to the diagonalization problem. However, it significantly reduces its complexity by relating it to explicit algebraic equations, exhibiting the underlying structure and making it quite more amenable to numerical computations.

Our discussion so far concerned the eigenstates in the Verma module $\mathcal{V}_{k,\ell}$. As mentioned in the previous subsection, a natural follow-up is to determine which of the Bethe eigenvectors survive the quotient to the unitary representation $\widetilde{\mathcal{V}}_{k,\ell}$. This question is well-understood for the finite Gaudin model, where it led to the rich theory of ``reproductions'' of Bethe ansatz solutions~\cite{mukhin2004critical,Frenkel:2003qx}. It would be interesting to extend these results to affine Gaudin models and the WZNW model.

We note that the affine Bethe ansatz constructs the eigenvectors of $\Iq^{(2p-1)}$ but does not provide the corresponding eigenvalues (except for $\Iq^{(1)}$). Still, admitting the conjecture, we can already observe that the eigenvalues of $\Iq^{(2p-1)}$ can be expressed as rational functions of the parameters $(k,\ell)$ and of the Bethe roots $(t_i)$. Indeed, the Bethe vector \eqref{eq:BA} is defined as a linear combination of descendant states in the Verma module, with rational functions of the Bethe roots as coefficients (see \eqref{eq:bethe_vector_explicit} for an explicit example). The action of the operator $\Iq^{(2p-1)}$ mixes these descendent states, with coefficients which do not involve the Bethe roots but depend rationally on $(k,\ell)$. The eigenvalue is extracted as the proportionality coefficient between $\Iq^{(2p-1)}\ket{\{t_i\}}_{\text{BA}}$ and $\ket{\{t_i\}}_{\text{BA}}$ and should thus be a rational function of $(k,\ell,t_i)$.\footnote{Of course, one should not forget that the Bethe roots $\{t_i\}=\{w_i\}\cup\{w'_i\}$ are themselves constrained by the Bethe ansatz equations \eqref{eq:BAE} and are thus complicated algebraic functions of $(k,\ell)$.} Remarkably, there is in fact another conjecture which describes this eigenvalue  more explicitly and which we discuss in the next subsection. We also refer to the end of Section \ref{sec:Kondo} for a brief discussion of the eigenvalues of non-local Kondo operators.\footnote{In particular, these non-local eigenvalues take the form of ``supersymmetric'' polynomials in the Bethe roots $(t_i)$~\cite{Gaiotto:2020dhf}, contrarily to the local eigenvalues discussed here, which are only rational functions of these quantities. This is in contrast with the Bethe ansatz proposed in~\cite{Bazhanov:2003ni} for the KdV integrable structure, in which it is the local eigenvalues which are symmetric polynomials of the Bethe roots, rather than the non-local ones. In fact, it is known that the KdV integrable structure admits several Bethe formulations~\cite{Feigin:2007mr,Litvinov:2013zda,feigin2017integrals,Prochazka:2023zdb,masoero2025q}, with different algebraic origins and different types of functional dependence on the Bethe roots. We expect the affine Bethe ansatz for the WZNW model to be closer to the formulation of KdV as an affine Gaudin model on the coset $\widehat{\mathfrak{sl}(2)}_{k} \times \widehat{\mathfrak{sl}(2)}_{1} / \widehat{\mathfrak{sl}(2)}_{k+1}$~\cite{feigin2017integrals,masoero2025q}.}

\subsection{ODE/IQFT correspondence and Langlands duality}\label{subsec:odeim}

\paragraph{Motivation.} As mentioned earlier, the SU(2) WZNW model can be seen as one of the simplest example in the class of affine Gaudin models, which are themselves field-theoretic generalisations of the finite Gaudin models~\cite{gaudin2014bethe}. The latter have been extensively studied in the literature and their spectral problem is now very well-understood. In particular, the works~\cite{Feigin:1994in,BD,Frenkel:2003qx} showed that the spectrum of finite Gaudin models can be encoded in differential operators called opers, as part of a deeper mathematical framework named the Geometric Langlands correspondence. In their seminal paper~\cite{Feigin:2007mr}, Feigin and Frenkel defined affine Gaudin models (associated with infinite-dimensional affine Lie algebras) and conjectured that the Geometric Langlands correspondence generalises to these models, so that their spectrum should be encoded in so-called affine opers. In particular, they proposed that this relation could explain the deeper origin of the ODE/IQFT correspondence\footnote{It is often also called ODE/IM correspondence, with IM standing for either Integrable Models or Integrals of Motion.}, which relates the spectrum of Integrable Quantum Field Theories with Ordinary Differential Equations and which was initially discovered for the KdV integrable structure~\cite{Dorey:1998pt,Bazhanov:1998wj}. This affine Geometric Langlands correspondence  has been further developed and checked in~\cite{Frenkel:2016gxg,Lacroix:2018fhf,Lacroix:2018itd,Kotousov:2021vih,Kotousov:2022azm,Gaiotto:2020dhf}, although it is still largely conjectural at the time. In particular, it can be applied to the case of the SU(2) WZNW model to produce a conjectural ODE/IQFT correspondence computing the eigenvalues of the local IMs $\Iq^{(2p-1)}$ on the Bethe vectors discussed earlier. This is the subject of the present subsection. For brevity, we will not re-explain the origin of this construction here and will simply give a self-contained description of its concrete consequences.\vspace{-2pt}

\paragraph{The ODE.} Let us consider a set of finite and affine Bethe roots $\{t_i\}=\{w_1,\dots,w_m\}\cup\{w_1',\dots,w_n'\}$, in the context of the Bethe ansatz \eqref{eq:BA}. We associate to it the ODE~\cite{Feigin:2007mr,Lacroix:2018fhf,Gaiotto:2020dhf}\footnote{The ODE associated with the vacuum state for level $k=1$ and other related ones already appeared in~\cite{Bazhanov:2003ua,Lukyanov:2003rt,Lukyanov:2006cu}.}
\begin{equation}\label{eq:ODE}
    \Bigl(\p_z^2 - V(z) + \Lambda^{-2}\,\Pc(z) \Bigr)\, \psi(z) = 0\,, 
\end{equation}
where $z$ is a complex coordinate, $\Lambda$ is a complex parameter and
\begin{equation}\label{eq:PV}
    \Pc(z) = z^k\,e^{-z}\,, \qquad V(z) = s(z)^2-\p_z s(z)\,, \qquad s(z) = \frac{\ell}{z} - \sum_{i=1}^m \frac{1}{z-w_i} + \sum_{i=1}^{n} \frac{1}{z-w'_i}\,.
\end{equation}
We note that the function $\Pc(z)$ depends only on the underlying WZNW model, through the level $k$, whereas $V(z)$, called the \textit{potential}, depends on the choice of representation (through $\ell$) and on the Bethe roots $\{t_i\}=\{w_i\}\cup\{w_i'\}$. The equation \eqref{eq:ODE} more precisely defines a family of ODEs, depending on the auxiliary parameter $\Lambda$. This family is essentially what is called an affine oper, or $\Lambda$-oper, in the literature. It is worth noting that $\p_z^2 - V(z)$ can be factorised as $(\p_z - s(z))(\p_z + s(z))$: we call that the Miura form of the oper, which reflects the fundamental nature of the function $s(z)$ in the definition of the ODE. Finally, it will be useful to introduce the \textit{twist function} $\vp(z)$, defined as the logarithmic derivative of $\Pc(z)$:
\begin{equation}\label{eq:Twist}
    \vp(z) = \p_z\log\Pc(z) = \frac{k}{z} - 1\,.
\end{equation}

\paragraph{Finite Bethe equations.} The function $s(z)$ has poles at the Bethe roots $w_i$ and $w_i'$. A straightforward computation shows that the double pole at a finite root $w_i$ cancels in the specific combination $V(z) = s(z)^2-\p_z s(z)$. Remarkably, the residue of the remaining simple pole is proportional to the left-hand side of the Bethe ansatz equation \eqref{eq:BAE1}. In other words, the ODE \eqref{eq:ODE} is regular at $z=w_i$ if and only if the finite Bethe root $w_i$ is on-shell. This shows that the ODE/IQFT correspondence also naturally encodes the Bethe equations, at least for finite roots so far. In contrast, one can check that the potential $V(z)$ has a pole at the affine roots $w_i'$, even if they are on-shell (in fact a double pole). The ODE/IQFT origin of the affine Bethe equations is therefore a bit more subtle and will be discussed at a later stage in this subsection.

\paragraph{Exact WKB expansion.} We now search for solutions to the ODE \eqref{eq:ODE} through an expansion at small $\Lambda$, using the \textit{exact WKB method}. Roughly speaking, it states that the ansatz
\begin{subequations}\label{eq:WKB}
\begin{equation}\label{eq:WKB1}
    \psi(z) = \frac{1}{\sqrt{\Pi(z,\Lambda)}} \exp\left( \int^z \Pi(z,\Lambda)\,\dd z \right) \,,
\end{equation}
\begin{equation}\label{eq:WKB2}
    \Pi(z,\Lambda) = \frac{\sqrt{\Pc(z)}}{\Lambda} + \sum_{p=1}^{+\infty}  \left(\frac{\Lambda}{\sqrt{\Pc(z)}}\right)^{2p-1}\,\frac{(-1)^{p+1}(2p)!}{4^p (p!)^2 (2p-1)} \,\Pi_{2p-1}(z)\,,
\end{equation}
\end{subequations}
is compatible with the ODE\footnote{The normalisation of $\Pi_{2p-1}(z)$ in equation \eqref{eq:WKB2} has been chosen for future convenience. We note that this expansion is only an asymptotic series in $\Lambda$, which generally does not converge. The true solutions of the ODE are then obtained from this asymptotic result by taking into account non-perturbative effects (such as Borel resummation and resurgence). We will not need any of this more advanced technology here and will care only about the algebraic form of the coefficients $\Pi_{2p-1}(z)$ in the asymptotic series. We thus treat \eqref{eq:WKB} as a formal power series in $\Lambda$, which we reinsert in the ODE \eqref{eq:ODE}. Extracting the coefficients of different powers of $\Lambda$ in this equation then results in the recursion relation for $\Pi_{2p-1}(z)$.} \eqref{eq:ODE} and results in a recursion relation expressing the coefficient $\Pi_{2p-1}(z)$ in terms of the lower ones $\Pi_{2q-1}(z)$ ($q<p$) and their derivatives, sourced by the functions $\Pc(z)$ and $V(z)$. Crucially, this recursion relation does not depend on the derivatives of $\Pi_{2p-1}(z)$ (only the ones of lower coefficients) and can thus be solved purely algebraically, order by order.

The object $\Pi(z,\Lambda)$ is called the \textit{WKB momentum} and will play a crucial role in the ODE/IQFT correspondence. Its first two coefficients (computed from the recursion relation) are given by
\begin{equation}\label{eq:Pip}
\Pi_1(z) =  V(z) - \frac{\vp(z)^2-4\vp'(z)}{16} \,, \vspace{-6pt}
\end{equation}
\begin{align}
\Pi_3(z) =& \;   V(z)^2 - \frac{13\vp(z)^2-12 \vp'(z)}{8}\,V(z) + \frac{5}{2}\vp(z)V'(z) - V''(z) \\
   & \hspace{10pt} +\frac{25 \varphi
   (z)^4 -200
   \varphi (z)^2 \varphi '(z) +112 \varphi '(z)^2+192 \varphi (z) \varphi ''(z)-64 \varphi ^{(3)}(z)}{256}\,, \notag
\end{align}
where we recall from equation \eqref{eq:Twist} that $\vp(z)=\p_z\log\Pc(z)$. More generally, $\Pi_{2p-1}(z)$ is a differential polynomial in $V(z)$ and $\vp(z)$, of the form $\Pi_{2p-1}(z) =   V(z)^p + \dots \,,$ where the dots represent terms of lower degree in $V$. Reinserting the expressions \eqref{eq:PV} and \eqref{eq:Twist} of $V(z)$ and $\vp(z)$ relevant to our case, we find that $\Pi_{2p-1}(z)$ is a rational function of $z$ with poles at $0$, $w_i'$ and $w_i$, with the latter disappearing when the Bethe ansatz equations are enforced.

\paragraph{Eigenvalues of the local IMs and monodromy.} Let $\mathcal{C}$ be the Hankel contour in the $z$-plane which starts at $-\infty$, runs below the horizontal axis towards the origin, loops around it counter-clockwise and goes back to $-\infty$ above the horizontal axis (see Figure \ref{fig:Hankel}). 

\begin{center}

    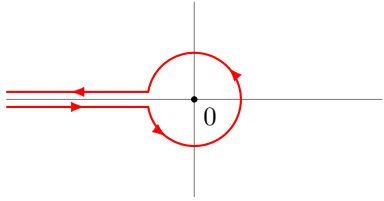
\begin{figure}[H]
            \centering
   \begin{tikzpicture}[scale=1,>=latex]

\draw[gray,thin] (-2.5,0) -- (2.5,0);
\draw[gray,thin] (0,-1.3) -- (0,1.3);

\filldraw (0,0) circle (1pt) node[below right] {$0$};

 \draw[thick,red,decoration={ markings,  
      mark=at position 0.55 with {\arrow{latex}}}, 
      postaction={decorate}]
  (-0.6,0.1) -- (-2.5,0.1) ;

 \draw[thick,red,decoration={ markings,  
      mark=at position 0.55 with {\arrow{latex}}}, 
      postaction={decorate}]
  (-2.5,-0.1) -- (-0.6,-0.1) ;

 \draw[thick,red,decoration={ markings,  
      mark=at position 0.125 with {\arrow{latex}}, 
      mark=at position 0.625 with {\arrow{latex}}}, 
      postaction={decorate}]
  (0,0) ++(-170:.62) arc (-170:170:.62);

\end{tikzpicture}

    \caption{The Hankel contour $\mathcal{C}$. }
    \label{fig:Hankel}  \vspace{-10pt}
    \end{figure}
    
\end{center}

The first ODE/IQFT conjecture relevant for this paper is then the following.\\

\noindent \textbf{Conjecture 1:} For $k\notin\lbrace -2,-4,-6,\dots\rbrace$, the eigenvalue $\mathcal{I}^{(2p-1)}_{\ell,\lbrace t_i \rbrace}$ of the local IM $\Iq^{(2p-1)}$ on the on-shell Bethe vector $\ket{\lbrace t_i\rbrace}_{\text{BA}}$ is given by the integral
\begin{equation}\label{eq:EigenODE}
    \mathcal{I}^{(2p-1)}_{\ell,\lbrace t_i \rbrace} = a_{2p-1}(k) \int_{\mathcal{C}} \Pc(z)^{-(2p-1)/2}\, \Pi_{2p-1}(z)\,\dd z\,,
\end{equation}
where
\begin{equation}\label{eq:ap}
a_{2p-1}(k)=   \frac{\,\Gamma\Big(\frac{k}{2}(2p-1)+2p\Big)}{2\ri \pi (k+2)^p} \left(\frac{2p-1}{2}\right)^{-\frac{k}{2}(2p-1)-2p-1}\,.
\end{equation}

The presence of the exponential term in $\Pc(z)=z^k e^{-z}$ ensures the convergence of the integral \eqref{eq:EigenODE} for $p\geq 1$, as the integrand then decays exponentially when approaching $-\infty$. We also note that this integrand is generically multi-valued due to the $\Pc(z)$ term,  with a branch cut that we put along the negative horizontal axis. Then, the choice of Hankel loop guarantees that the integrand admits a single-valued branch along the integration contour (specified by $z^\alpha=|z|^\alpha e^{\ri\alpha\arg(z)}$ with $-\pi < \arg(z) < \pi$) and thus that the integral \eqref{eq:EigenODE} is well-defined. In the specific case where $k\in 2\Z$ is an even integer, the integrand becomes single-valued and one can pinch off the Hankel loop into a simple circle around the origin: the integral then becomes a standard residue, which is non-vanishing only for $k$ non-negative (hence the restriction on $k$ in the conjecture).

This conjecture shows that the ODE \eqref{eq:ODE} encodes the eigenvalues of all the local IMs on the Bethe vector, through its WKB expansion \eqref{eq:WKB}. More compactly, the formal $\Lambda$-expansion of the integral $\int_{\mathcal{C}} \Pi(z,\Lambda)\, \dd z$ plays the role of generating function for these eigenvalues.\footnote{Technically, the $\Lambda^{-1}$-term of this expansion is $\int_{\mathcal{C}} \sqrt{\Pc(z)}\, \dd z$, which does not converge. Fortunately for us, this is the only coefficient which does not encode the eigenvalue of an IM, so that we can simply ignore it and more precisely focus on the regular part of $\int_{\mathcal{C}} \Pi(z,\Lambda)\,\dd z$.} This integral is essentially computing the monodromy of the WKB solution \eqref{eq:WKB} around the path $\cal C$: the ODE/IQFT correspondence then relates the eigenvalues of our IQFT with the monodromy properties of well-chosen ODEs. The above conjecture is a simple adaptation and refinement of the works~\cite{Lacroix:2018fhf,Lacroix:2018itd}, which were themselves guided by the affine Langlands duality of~\cite{Feigin:2007mr} and the ODE/IQFT interpretation of local IMs proposed in~\cite{Lukyanov:2010rn}. Below, we will provide checks of this conjecture and discuss a slightly different but equivalent formulation, which will be more algebraic and will not require the use of integrals.

\paragraph{Gauge transformations.} The key ingredient in this alternative conjecture is the notion of gauge transformations of the WKB momentum. We will only briefly explain the origin of these transformations here and refer to~\cite{Lacroix:2018fhf} for details. Essentially, this is based on the reformulation of the 2nd-order ODE \eqref{eq:ODE} as a 1st-order ODE $(\p_z+A(z)\bigr)\Psi(z)=0$ on a 2-vector $\Psi(z)$, with $A(z)$ a $2\times 2$ matrix meromorphic in $z$. Under a $z$-dependent linear redefinition $\Psi(z) \mapsto H(z)^{-1}\Psi(z)$, this matrix transforms as a gauge connection $A(z) \mapsto H(z)^{-1}A(z)H(z)+H(z)^{-1}\p_z H(z)$. This leads to a different but equivalent ODE. Of importance for us is the existence of specific gauge transformations which preserve the WKB ansatz \eqref{eq:WKB1} but shift the WKB momentum by a $z$-derivative:
\begin{equation}\label{eq:TransfPi}
    \Pi(z,\Lambda) \;\longmapsto \; \Pi^h(z,\Lambda):= \Pi(z,\lambda) + \p_z h(z,\Lambda)\,,
\end{equation}
where $h(z,\Lambda)$ is the local gauge parameter, whose form we will specify shorty. To understand the relevance of these transformations to our problem, we note that
\begin{equation}\label{eq:GaugeInv}
    \int_{\mathcal{C}} \Pi(z,\Lambda)\,\dd z = \int_{\mathcal{C}} \Pi^h(z,\Lambda)\,\dd z\,,
\end{equation}
since the integrands differ by a $z$-derivative, assuming appropriate boundary conditions which we discuss below. Our conjecture above states that this integral is the generating function for the eigenvalues $\mathcal{I}^{(2p-1)}_{\ell,\lbrace t_i \rbrace}$, which are thus left unchanged by these transformations.\footnote{The integral $\int_{\mathcal{C}} \Pi(z,\Lambda)\,\dd z$ essentially encodes monodromy properties for the matricial ODE $(\p_z+A(z)\bigr)\Psi(z)=0$: it is therefore natural that this quantity is invariant under gauge transformations.} In other words, the WKB momentum is subject to gauge transformations \eqref{eq:TransfPi} and the eigenvalues $\mathcal{I}^{(2p-1)}_{\ell,\lbrace t_i \rbrace}$ are the gauge-invariant quantities encoded in this function, which can be extracted by taking a well-chosen integral. As we shall see, there are others ways to extract these gauge-invariant quantities, which will lead us to the second formulation of the ODE/IQFT conjecture.

For this, we first have to specify the form of the allowed gauge transformations \eqref{eq:TransfPi}. Recall that $\Pi(z,\Lambda)$ is defined as the $\Lambda$-expansion \eqref{eq:WKB2}. To preserve this form, we will also consider the gauge parameter $h(z,\Lambda)$ to be a formal power series in $\Lambda$, of the form
\begin{equation}
 h(z,\Lambda) = \sum_{p=1}^{+\infty} \left(\frac{\Lambda}{\sqrt{\Pc(z)}}\right)^{2p-1} \,\frac{(-1)^{p+1}(2p)!}{4^p (p!)^2 (2p-1)}\, h_{2p-1}(z)\,.
\end{equation}
The transformation \eqref{eq:TransfPi} then acts on the coefficients of the WKB momentum by
\begin{equation} \label{eq:GaugePip}
 \Pi_{2p-1}(z) \; \longmapsto \; \Pi_{2p-1}^h(z):= \Pi_{2p-1}(z)+ \p_z h_{2p-1}(z) - \frac{2p-1}{2} \vp(z)\,h_{2p-1}(z)\,,
\end{equation} 
where we recall that $\vp(z)=\p_z\log \Pc(z)$. In the language of~\cite{Lacroix:2018fhf}, we say that $\Pi_{2p-1}(z)$ is shifted by the $(2p-1)$-twisted derivative of $h_{2p-1}(z)$. Due to the functional form of $\Pi_{2p-1}(z)$ (see the discussion below equation \eqref{eq:Pip}), we further restrict ourselves to gauge parameters $h_{2p-1}(z)$ which are rational functions of $z$, with poles at $0$ and the Bethe roots, such that the transformations \eqref{eq:GaugePip} preserve this form. This specifies the set of allowed gauge transformations of the WKB momentum. Under these conditions, $h(z,\Lambda)$ is single-valued along the Hankel contour $\mathcal{C}$ and decays exponentially at $-\infty$, ensuring that the boundary term discarded to establish the gauge invariance \eqref{eq:GaugeInv} indeed vanished. Alternatively, focusing on the $\Lambda^{2p-1}$-term, these choices of $h$ are such that
\begin{equation}\label{eq:GaugeInvp}
    \int_{\mathcal{C}} \Pc(z)^{-(2p-1)/2}\, \Pi_{2p-1}(z)\,\dd z = \int_{\mathcal{C}} \Pc(z)^{-(2p-1)/2}\, \Pi^h_{2p-1}(z)\,\dd z\,,
\end{equation}
so that the integral \eqref{eq:EigenODE} encoding the eigenvalue $\mathcal{I}^{(2p-1)}_{\ell,\lbrace t_i \rbrace}$ is truly gauge-invariant, as desired. 

\paragraph{Affine Bethe equations.} We are now in position to come back to the question of the ODE/IQFT interpretation of the affine Bethe equations \eqref{eq:BAE2}, left unanswered earlier. Indeed, recall that the finite Bethe equations coincide with the conditions under which the WKB momentum $\Pi(z,\Lambda)$ is regular at the finite roots $z=w_i$. However, this function is not regular at the affine roots $z=w_i'$. Remarkably, this can be bypassed by considering the gauge transformations $\Pi(z,\lambda) \mapsto \Pi^h(z,\lambda)$ introduced above. Namely, there exists a gauge $\Pi^h(z,\lambda)$ which is regular at $z=w_i'$ if and only if the affine Bethe equation \eqref{eq:BAE2} holds. This claim can be proven using the same type of reasoning as in~\cite[Subsection 4.3]{Lacroix:2018fhf}. Since the WKB momentum can be used to measure monodromy properties of the ODE \eqref{eq:ODE}, we see that this ODE has no monodromy around the Bethe roots $w_i$ and $w_i'$ (both finite and affine) if and only if these roots are on-shell. This gives the origin of the full set of Bethe ansatz equations in the ODE/IQFT correspondence.

\paragraph{Simple representative and eigenvalues of local IMs.} Following the above discussion, we now restrict ourselves to gauges where the WKB momentum $\Pi^h(z,\Lambda)$ is regular at the Bethe roots. In that case, we are only left with poles at the origin in the coefficients $\Pi^h_{2p-1}(z)$. We note that there are still allowed gauge transformations \eqref{eq:GaugePip} which preserve this form, namely those whose gauge parameters also only has poles at $z=0$. It is a little algebraic exercise to convince oneself that these residual gauge transformations can always be used to eliminate all terms in $\Pi^h_{2p-1}(z)$ except for a pole of order $2p$ at the origin. In other words, there exists a gauge transformation $\Pi(z,\Lambda) \mapsto \Pi^{\text{simp}}(z,\Lambda)$ that brings all coefficients of the WKB momentum to the form $\Pi^{\text{simp}}_{2p-1}(z) \propto \frac{1}{z^{2p}}$. This gauge is unique and we will refer to it as the \textit{simple representative}. We are now in position to state the second formulation of our ODE/IQFT conjecture.\\

\noindent \textbf{Conjecture 2:} The simple representative $\Pi^{\text{simp}}(z,\Lambda)$ of the WKB momentum is a generating function for the local IM eigenvalues $\mathcal{I}^{(2p-1)}_{\ell,\lbrace t_i \rbrace}$. More precisely:
\begin{equation}\label{eq:EigenODE2}
    \Pi^{\text{simp}}_{2p-1}(z) = \frac{(k+2)^p}{z^{2p}}\, \mathcal{I}^{(2p-1)}_{\ell,\lbrace t_i \rbrace}\,.
\end{equation}

This conjecture is equivalent to our previous one \eqref{eq:EigenODE}, formulated in terms of integrals. This is easily proven using the gauge invariance \eqref{eq:GaugeInvp} of the WKB integral and the identity
\begin{equation}
    \int_{\mathcal{C}} \frac{\dd z}{\Pc(z)^{(2p-1)/2}} \frac{(k+2)^p}{z^{2p}} =\frac{ 2 \ri\pi (k+2)^p}{\Gamma(\frac{k}{2}(2p-1)+2p)} \left(\frac{2p-1}{2}\right)^{\frac{k}{2}(2p-1)+2p+1} = \frac{1}{a_{2p-1}(k)}\,,
\end{equation}
with $a_{2p-1}(k)$ chosen as in equation \eqref{eq:ap}. This identity is found using the representation of $1/\Gamma(u)$ as an integral over the Hankel  loop -- see for instance~\cite[(5.9.2)]{NIST:DLMF}.

To summarise, the ODE/IQFT conjectures extract the eigenvalues $\mathcal{I}^{(2p-1)}_{\ell,\lbrace t_i \rbrace}$ as the gauge-invariant quantities encoded in the WKB momentum $\Pi(z,\Lambda)$ of the ODE. The first one \eqref{eq:EigenODE} does so by taking a well-chosen integral of this momentum, while the second one \eqref{eq:EigenODE2} passes through a gauge-fixing to its simple representative $\Pi^{\text{simp}}(z,\Lambda)$, which always exists and is unique. The main advantage of this second formulation is that it gives a completely algebraic algorithm to determine the eigenvalue $\mathcal{I}^{(2p-1)}_{\ell,\lbrace t_i \rbrace}$. Namely, we start with the expression of $\Pi_{2p-1}(z)$ arising from the WKB expansion, perform its partial fraction decomposition, act on it by a gauge transformation \eqref{eq:GaugePip} and fine-tune the gauge parameter such that we eliminate all the other terms than $1/z^{2p}$. This provides the simple representative $\Pi^{\text{simp}}_{2p-1}(z)$, from which we extract $\mathcal{I}^{(2p-1)}_{\ell,\lbrace t_i \rbrace}$. In particular, we never have to compute any integral in this process, contrarily to our first conjecture \eqref{eq:EigenODE}. This avoids the apparition of transcendental prefactors like the Gamma function and makes it clear that the eigenvalue of $\Iq^{(2p-1)}$ can be expressed as a rational function of $(k,\ell)$ and the Bethe roots (see the checks and discussion below). On the other hand, this reformulation required introducing the notions of gauge transformation of the WKB momentum and of simple representative, whereas the integral considered in \eqref{eq:EigenODE} was a more direct way of extracting the gauge-invariant content in $\Pi_{2p-1}(z)$. 

\paragraph{Checks on simple examples.} We first checked the conjecture 2 against explicit calculations for $p=1,2$ and for the ground and first excited states, corresponding respectively to no Bethe roots and one affine root. For these four cases, we recovered directly the eigenvalues of $\Iq^{(1,3)}$  on $\ket{\ell}$ and $\Jq^+_{-1}\ket{\ell}$, that is, equations \eqref{eq:Q2_on_HW} and \eqref{eq:Q2_on_first_excited}. In particular, the eigenvalues of our new charge $\Iq^{(3)}$ on these low lying states are correctly predicted by conjecture 2, which we consider a valuable check of both $\Iq^{(3)}$ and of the ODE/IQFT conjecture. As they are rather straightforward to obtain following the above prescription, we will not report the explicit expressions of the corresponding $\Pi^{\text{simp}}_{2p-1}(z)$, with the exception of $\Pi^{\text{simp}}_1(z)$ for the first excited state, to highlight how the affine Bethe equation enters through gauge transformations.

Let us then consider the setup with a single affine root $w'$, so that the function $s(z)$ in equation \eqref{eq:PV} takes the form $s(z)=\frac{\ell}{z} +\frac{1}{z-w'}$. The WKB coefficient $\Pi_1(z)$ of equation \eqref{eq:Pip} then reads
\begin{equation}
\Pi_1(z)=\frac{16 \ell(\ell+1)-k(k+4) }{16 z^2}+\frac{k w'-16 \ell }{8 w' z}+\frac{2\ell }{w' (z-w')}+\frac{2}{(z-w')^2}-\frac{1}{16}\,.
\end{equation}
Following the above procedure, we then want to perform a gauge transformation \eqref{eq:GaugePip} on $\Pi_1(z)$ to remove the simple pole at $z=0$, the double pole at $z=w'$ and the constant term. This is done by taking $h_1(z)=\frac{32 \ell -k (w'+16)}{8 w' z}+\frac{2}{z-w'}+\frac{1}{8}$, yielding
\begin{equation}
\Pi^h_1(z)=\frac{8 k^2-k w'-16 k \ell +16 k+8 w' \ell ^2+8 w' \ell -32 \ell }{8 w' z^2}+\frac{w'+2 \ell-k }{ w' (z-w')}\,.
\end{equation}

At this stage, no gauge transformation can remove the $1/(z-w')$ term without introducing new singularities at $z=w'$: we must then demand that it vanishes by setting $w'=k-2\ell$. This coincides precisely with equation \eqref{eq:aff_BAE_1st}, which is the solution of the affine Bethe ansatz equation for a single affine root.
Putting $w'$ to its on-shell value in $\Pi^h_1(z)$ finally gives $\Pi^{\text{simp}}_1(z)$, which we can match against the eigenvalue of $\Iq^{(1)}$ on $\Jq^+_{-1}\ket{\ell}$:
\begin{equation}
\Pi^{\text{simp}}_1(z)=\frac{\ell(\ell+1)+7k/8+2}{z^2}=\frac{k+2}{z^2}\left(h_\ell+1-\frac{c}{24}\right)=\frac{k+2}{z^2}\, \mathcal{I}^{(1)}_{\ell,\{w'\}}\,,
\end{equation}
in agreement with conjecture 2. 

\paragraph{Off-shell eigenvalues.} The computation above showed the extraction of the eigenvalue $\mathcal{I}^{(1)}_{\ell,\{w'\}}$ from the ODE/IQFT correspondence. One can in principle proceed through analogous steps for all other $\mathcal{I}^{(2p-1)}_{\ell,\{t_i\}}$, with higher spin and an arbitrary number of Bethe roots, although expressions get increasingly involved. One complication which arises in this computation is the fact that the Bethe ansatz equations quickly become non explicitly solvable when the number of roots increases. As such, we cannot easily set the roots to their on-shell values in the computation of $\Pi_{2p-1}^{\text{simp}}(z)$. This leads us to introduce the notion of ``off-shell'' WKB representatives and eigenvalues. In the rest of this paragraph, we will thus think of the Bethe roots $\{t_i\}=\{w_1,\dots,w_m\}\cup\{w_1',\dots,w_n'\}$ as being free parameters, and not as solutions of the Bethe ansatz equations.

Starting from the WKB coefficient $\Pi_{2p-1}(z)$ of the ODE \eqref{eq:ODE}--\eqref{eq:PV}, we perform gauge transformations to remove all terms except from a fraction $1/z^{2p}$ and simple poles at the Bethe roots. This gauge always exists, is unique and takes the form
\begin{equation}\label{eq:off_shell_eigenvalues}
   \Pi_{2p-1}^{\text{off}}(z) = \frac{(k+2)^p}{z^{2p}} \, \mathcal{I}^{(2p-1)}_{\ell,\{t_i\},\text{off}} + \sum_{i=1}^m \frac{a_i\,\BAE_i}{z-w_i} + \sum_{i=1}^n \frac{a_j'\,\BAE'_j}{z-w'_j}
\end{equation}
for some coefficients $\bigl(\mathcal{I}^{(2p-1)}_{\ell,\{t_i\},\text{off}},\,a_i,\,a_j'\bigr)$, and where $\BAE_i$ and $\BAE'_i$ denote the left-hand side of the Bethe ansatz equations \eqref{eq:BAE1} and \eqref{eq:BAE2}. This off-shell form makes it clear that the poles at the Bethe roots disappear exactly when we impose the Bethe ansatz equations $\BAE_i=\BAE_j'=0$. In that case, $\Pi_{2p-1}^{\text{off}}(z)$ then becomes the simple representative \eqref{eq:EigenODE2} of Conjecture 2. In particular, the coefficient $\mathcal{I}^{(2p-1)}_{\ell,\{t_i\},\text{off}}$ can be seen as an  ``off-shell'' form of the eigenvalue of $\Iq^{(2p-1)}$ on $\ket{\{t_i\}}_{\text{BA}}$.

Crucially, all the steps performed in the above procedure only involve rational functions and never require the resolution of polynomial or differential equations. The resulting off-shell expression for the eigenvalue $\mathcal{I}^{(2p-1)}_{\ell,\{t_i\},\text{off}}$ is then a rational function of $(k,\ell,w_i,w_j')$.
Let us illustrate this with the example of the spin-1 eigenvalue. 
Starting from the WKB coefficient $\Pi_1(z)$ in equation \eqref{eq:Pip}, it is a direct computation to show that the gauge transformation by
\begin{equation}
    h_1(z) = \sum_{j=1}^n \frac{2}{z-w'_j} + \frac{1}{8} - \frac{1}{z} \Bigl( \frac{k}{8} + \sum_{i=1}^m \frac{4\ell}{w_i} + \sum_{j=1}^n \frac{2k-4\ell}{w_j'} \Bigr)
\end{equation}
converts $\Pi_1(z)$ to the form \eqref{eq:off_shell_eigenvalues}, with $a_i=a_j'=2$ and
\begin{equation} \label{eq:I1t2}
    \mathcal{I}^{(1)}_{\ell,\{t_i\},\text{off}} = \frac{\ell(\ell+1)-k/8}{k+2} + \sum_{i=1}^m \frac{2\ell}{w_i} + \sum_{j=1}^n \frac{k-2\ell}{w_j'}\,.
\end{equation} 
As anticipated, the off-shell eigenvalue $ \mathcal{I}^{(1)}_{\ell,\{t_i\},\text{off}}$ is rational in $(k,\ell,w_i,w_j')$. As it turns out, this expression can be drastically simplified by using the Bethe ansatz equations. Indeed, adding $2\sum_{i=1}^m \BAE_i + 2\sum_{j=1}^n \BAE'_j$ to the right-hand side of equation \eqref{eq:I1t2} removes all the dependence on the Bethe roots and reduces $ \mathcal{I}^{(1)}_{\ell,\{t_i\},\text{off}}$ to its on-shell value 
\begin{equation}\label{eq:I1t}
\mathcal{I}^{(1)}_{\ell,\{t_i\}} = \frac{\ell(\ell+1)}{k+2} - \frac{c}{24} + n = \mathcal{I}^{(1)}_{\ell,\emptyset} + n \,.   
\end{equation}
This is the eigenvalue expected by acting directly with $\Iq^{(1)}$ on $\ket{\{t_i\}}_{\text{BA}}$, \textit{i.e.} the energy of the reference state $\ket{\ell}$ plus the depth of the Bethe vector (equal to the number $n$ of affine roots). This constitutes a further check of the ODE/IQFT conjecture. Moreover, it illustrates how this correspondence can be used to extract, at least in principle, explicit off-shell expressions $\mathcal{I}^{(2p-1)}_{\ell,\{t_i\},\text{off}}$ for the local eigenvalues.\footnote{Of course, there are infinitely many possible off-shell expressions for such an eigenvalue, since one can add to it any multiple of the Bethe equations. The ODE/IQFT procedure produces a unique choice among those. However, as illustrated in the spin-1 example, this choice might generally not be the simplest (nonetheless, we note that in the higher-spin case, it should not be possible to completely eliminate the dependence on the Bethe roots in this expression, contrarily to the case of spin-1).} However, even the computation of the next non-trivial one $\mathcal{I}^{(3)}_{\ell,\{t_i\},\text{off}}$ is rather involved and we thus leave it for future developments.


\section{Non-local integrals of motion and Kondo defects}
\label{sec:Kondo}

\subsection{The classical non-local integrals of motion}

\paragraph{Classical Kondo defect.} This section is devoted to the non-local charges in the integrable structure of the WZNW model. At the classical level, they are defined through equation \eqref{eq:Kcl}. This definition depends on the choice of a representation of SU(2): for simplicity, we will focus here on the fundamental one\footnote{We make this choice because we only aim at presenting the simplest examples of non-local IMs, rather than a systematic analysis. We note however that considering different representations allows one to build a larger class of non-local charges and to study interesting structural properties such as fusion rules / Hirota equations. We refer to~\cite{Gaiotto:2020fdr,Gaiotto:2020dhf} for a discussion of these aspects.} (hence dropping the label $V$ in the notations). We are thus considering
\begin{equation}
    K_{\cl}(\rho) = \Tr\left[\Pexp\left( \ri\rho\int_0^L J_{\cl}(x)\,\dd x \right)\right]\,,
\end{equation}
where $J_\cl(x)$ is the classical left-moving Kac-Moody current, $\rho$ is the spectral parameter and the trace is taken in the fundamental representation. This object is built from the path-ordered exponential of a left-moving field and thus takes the form of what is called a \textit{chiral Kondo defect} in the underlying 2d CFT (here at the classical level). It satisfies
\begin{equation}\label{eq:InvolNL}
    \bigl\lbrace K_\cl(\rho), K_\cl(\s) \bigr\rbrace = 0 \qquad \text{ and } \qquad \Bigl\lbrace K_\cl(\rho), I^{(2p-1)}_\cl \Bigr\rbrace = \Bigl\lbrace K_\cl(\rho), \overline{I}^{(2p-1)}_\cl \Bigr\rbrace = 0
\end{equation}
for all choices of spectral parameters $\rho,\s$, where $I^{(2p-1)}_\cl$ are the classical local IMs \eqref{eq:ClassicalIM} and $\overline{I}^{(2p-1)}_\cl$ are their analogues in the right-moving sector. Since $I^{(1)}_\cl+\overline{I}^{(1)}_\cl$ is proportional to the Hamiltonian of the theory, the Kondo defect $K_\cl(\rho)$ is conserved along the dynamics. As usual, there also exists a Poisson-commuting Kondo defect $\overline{K}_\cl(\rho)$ built from the right-moving Kac-Moody current $\overline{J}_\cl(x)$. In this section, we will focus only on the left-moving sector.

\paragraph{$\bm\rho$-expansion.} We extract an infinite number of non-local IMs by expanding the Kondo defect in powers of $\rho$:
\begin{equation}\label{eq:KLambda}
      K_\cl(\rho) = 2 + \sum_{p=1}^{+\infty} K^{(p)}_\cl\,\left( \ri\rho\right)^p\,.
\end{equation}
The standard perturbative definition of path ordered-exponentials allows us to express $K^{(p)}_\cl$ as nested integrals of the Kac-Moody current:
\begin{equation}\label{eq:KpClass}
    K^{(p)}_\cl = \int_0^L \dd x_1 \int_0^{x_1} \dd x_2 \cdots \int_0^{x_{p-1}} \dd x_p\,\Tr\bigl( J_\cl(x_1) \cdots J_\cl(x_p) \bigr)\,.
\end{equation}
Below, we list a few general properties of these non-local IMs.
\begin{enumerate}
    \item The index $p$ essentially measures the ``level of non-locality'' of $K^{(p)}_\cl$, \textit{i.e.} the number of nested integrals in its expression \eqref{eq:KpClass}.
    \item By construction, the equation \eqref{eq:InvolNL} implies
\begin{equation}\label{eq:PbK}
    \Bigl\lbrace K^{(p)}_\cl, K^{(q)}_\cl \Bigr\rbrace = \Bigl\lbrace K^{(p)}_\cl,  I^{(2q-1)}_\cl \Bigr\rbrace  = 0 \,.
\end{equation}
    \item The presence of the trace in equation \eqref{eq:KpClass} ensures the invariance of $K^{(p)}_\cl$ under the global SU(2)-symmetry $J_\cl \mapsto U_L^{-1} J_\cl \,U_L$, generated by the 0-modes $J^a_{\cl,0}$.
\end{enumerate}

\paragraph{Mode expansion.} Reinserting the Fourier expansion $J_\cl(x)= \frac{2\ri\pi}{L}\sum_{n\in\Z} J^a_{\cl,n} e^{-2\ri\pi\,nx/L}\,T^a$, we rewrite equation \eqref{eq:KpClass} in the form
\begin{equation}\label{eq:KpClass2}
    K^{(p)}_\cl = \ri^p\,\Tr\bigl(T^{a_1}\cdots T^{a_p}\bigr)\,\sum_{n_1,\dots,n_p\in\Z \vspace{2pt} \atop n_1+\cdots+n_p=0} \omega_{n_1,\dots,n_p}\,J^{a_1}_{\cl,n_1}\cdots J^{a_p}_{\cl,n_p}\,,
\end{equation}
where a summation is implied over the indices $a_i$ and where we note that the invariant $p$-tensor $\Tr\bigl(T^{a_1}\cdots T^{a_p}\bigr)$ is not symmetrised in its indices. The main new ingredient in this formula is the coefficient
\begin{equation}\label{eq:omega}
    \omega_{n_1,\dots,n_p}     =  \int_0^{2\pi} \dd u_1 \int_0^{u_1} \dd u_2 \cdots \int_0^{u_{p-1}} \dd u_p\,e^{-\ri(n_1u_1+\cdots+n_pu_p)}\,,
\end{equation}
controlling the mode expansion of $K^{(p)}_\cl$, and which can straightforwardly be obtained from \eqref{eq:KpClass} (with $u_i=2 \pi x_i/L$). For a given $p$, this integral is straightforward to compute, at least in principle. We will not try to find a general expression here and will simply determine the first few $K_\cl^{(p)}$ below.

Before that, we first comment on a feature of the mode expansion \eqref{eq:KpClass2} of $K^{(p)}_\cl$, namely the restriction of the sum to modes with $n_1+\dots+n_p=0$. This constraint is initially not present in the computation of equation \eqref{eq:KpClass2}: in particular, the coefficients $\omega_{n_1,...,n_p}$ with $n_1+\dots+n_p\neq 0$ are generally non-zero. However, various cancellations occur between these different contributions, leaving only terms with $n_1+\dots+n_p= 0$ in the final expression. Note that this is equivalent to the expected Poisson-commutativity of $K^{(p)}_\cl$ with $I^{(1)}_\cl=-\frac{L}{4\pi^2}\int_0^L T_\cl(x)\,\dd x$. Indeed, the latter (being the classical equivalent of $\Lq_0$) acts on Kac-Moody modes by $\bigl\lbrace I^{(1)}_\cl, J^a_{\cl,n} \bigr\rbrace = \ri n\,J^a_{\cl,n}$.

\paragraph{Expression for $\bm{K_\cl^{(1)}}$ and $\bm{K_\cl^{(2)}}$.} We first note that $\su(2)$-matrices are traceless, so that the 1-tensor $\Tr(T^a)$ vanishes. Therefore, we simply have $K_\cl^{(1)}=0$.

We then turn our attention to the next case, \textit{i.e.} $K_\cl^{(2)}$. The invariant 2-tensor appearing in this quantity is the symmetric bilinear form $\Tr(T^a T^b) = \frac{1}{2}\, \delta^{ab}$. Moreover, as explained above, its mode expansion is restricted to $(n_1,n_2)=(n,-n)$, with $n\in\Z$. A straightforward computation yields the corresponding coefficient:
\begin{equation}
    \omega_{n,-n} = \left\lbrace \begin{array}{cl}
        2\pi^2 & \text{ if } n=0  \vspace{4pt} \\
        -\dfrac{2\ri\pi}{n} & \text{ if } n\neq 0
    \end{array}  \right.
\end{equation}
Reinserting these expressions for $\Tr(T^a T^b)$ and $\omega_{n,-n}$ in equation \eqref{eq:KpClass2}, we see that the term in the sum with a given $n\neq 0$ cancels with the term arising from $-n$. Therefore, we are only left with the zero-modes contribution:
\begin{equation}
    K^{(2)}_\cl = -\pi^2 J^a_{\cl,0}\,J^a_{\cl,0} \,.
\end{equation}
Up to a prefactor, this is nothing but the (classical) Casimir of the finite $\su(2)$ generators $J^a_{\cl,0}$, which we expect in our integrable structure.

\paragraph{Expression for $\bm{K_\cl^{(3)}}$.} To obtain a genuinely new IM, we have to consider the next case $K_\cl^{(3)}$. For $\su(2)$, the invariant 3-tensor $\Tr(T^aT^bT^c)$ is simply proportional to the structure constants: $\Tr(T^aT^bT^c)=\frac{\ri}{4}\varepsilon^{abc}$. To pursue, we have to determine the mode coefficients $\omega_{n_1,n_2,n_3}$, under the constraint $n_1+n_2+n_3=0$. In the case where $n_1,n_2,n_1+n_2\neq 0$, we find
\begin{equation}
    \omega_{n_1,n_2,-n_1-n_2} = -\frac{2\pi}{n_1(n_1+n_2)}\,.
\end{equation}
Moreover, we have the following non-generic cases (with $n\neq 0$):
\begin{equation}
    \omega_{n,0,-n} = -\frac{4\pi}{n^2}\,, \qquad \omega_{n,-n,0}=\omega_{0,n,-n} = \frac{2\pi(1-\ri\pi n)}{n^2}\,, \qquad \omega_{0,0,0} = \frac{4\pi^3}{3}\,.
\end{equation}
Taking into account the specific form of $\omega_{n_1,n_2,n_3}$ and the skew-symmetry of $\Tr(T^aT^bT^c)$, we find many cancellations in the computation of $K^{(3)}_\cl$ from equation \eqref{eq:KpClass2}. In the end, we are left with the following rather simple expression:
\begin{equation}\label{eq:K3Class}
    K^{(3)}_\cl = -2\pi^2\,\sum_{n> 0} \frac{\ri}{n}\,\varepsilon^{abc}\,J^a_{\cl,-n}\,J^b_{\cl,0}\,J^c_{\cl,n}\,.
\end{equation}
We note that the quantities $\frac{1}{n}J^a_{\cl,n}$ are essentially the Fourier modes of a primitive of $J^a_\cl(x)$, exhibiting the non-local nature of $K^{(3)}_\cl$. We will not discuss the higher non-local IMs $K^{(p)}_\cl$ here and refer to~\cite{Bachas:2004sy,Gaiotto:2020fdr,Gaiotto:2020dhf} for more explicit examples.

\subsection{The quantum non-local integrals of motion}\label{subsec:non_loc_quantum_IMs}

\paragraph{Generalities.} We now come to the quantisation of the non-local IMs extracted from the Kondo defect. A perturbative approach to this problem, based on the $\rho$-expansion \eqref{eq:KLambda} of $K_\cl(\rho)$, was proposed by Bachas and Gaberdiel~\cite{Bachas:2004sy} and later used by Gaiotto, Lee, Vicedo and Wu~\cite{Gaiotto:2020fdr,Gaiotto:2020dhf} in the context of integrability. This construction requires the regularisation of certain divergences and the renormalisation of the spectral parameter $\rho$, which plays the role of coupling constant in this Kondo problem. It produces quantum operators $\Kq^{(p)}$, which we expect to satisfy the following properties:
\begin{enumerate}
    \item $\Kq^{(p)}$ are well-defined operators on Verma modules of $\widehat{\su(2)}_k$ built from the Kac-Moody modes $\Jq^a_n$ and reproducing the non-local IMs \eqref{eq:KpClass2} in the classical limit (with $\Jq^a_n=\frac{1}{\hbar}J^a_{\cl,n}$):
    \begin{equation}\label{eq:KClassLimit}
        \Kq^{(p)} = \frac{K^{(p)}_\cl+O(\hbar)}{\hbar^p}\,.
    \end{equation}
    \item The operators $\Kq^{(p)}$ commute one with another and with the local IMs:
    \begin{equation}\label{eq:IntK}
        \bigl[ \Kq^{(p)}, \Kq^{(q)} \bigr] = \bigl[ \Kq^{(p)}, \Iq^{(2q-1)} \bigr] = 0\,.
    \end{equation}
    \item The operators $\Kq^{(p)}$ are invariant under the global SU(2)-symmetry of the WZNW model, \textit{i.e.}
    \begin{equation}\label{eq:Ksu2}
        \bigl[ \Jq^a_0, \Kq^{(p)} \bigr] = 0\,.
    \end{equation}
\end{enumerate}

Let us further comment on these desired properties and how they are concretely realised. Recall first the expansion \eqref{eq:KpClass2} of $K^{(p)}_\cl$ in terms of classical Kac-Moody modes $J^a_{\cl,n}$. At the quantum level, we expect a similar (albeit more involved) expansion, which then requires a prescription for the ordering of modes $\Jq^a_n$. Here, we will work with normal-ordered products $:\!\Jq^{a_1}_{n_1}\cdots\Jq^{a_q}_{n_q}\!:$, with mode numbers $n_i$ sorted out in increasing order from left to right\footnote{For instance, ${:\!\Jq^a_1\Jq^a_{3}\Jq^c_{-4}\!:}\null=\Jq^c_{-4}\Jq^a_1\Jq^a_{3}$. Note that this is a slightly different notion of normal-ordering than the one for fields $(\p^{m_1}\Jq^{a_1}\cdots\p^{m_p}\Jq^{a_p})$, as it is defined directly in terms of Fourier modes and sorts them out completely by increasing mode number. The two procedures are however related by the formula \eqref{NO_cyl_modes} and potential re-arrangements.}. With these notations, the renormalisation procedure of~\cite{Bachas:2004sy,Gaiotto:2020fdr,Gaiotto:2020dhf} produces operators of the form
\begin{equation}\label{eq:KpQuant}
    \Kq^{(p)} = \ri^p\,\Tr\bigl(T^{a_1}\cdots T^{a_p}\bigr)\,\sum_{n_1,\dots,n_p\in\Z \vspace{2pt} \atop n_1+\cdots+n_p=0} \omega_{n_1,\dots,n_p}\,:\hspace{-1pt}\Jq^{a_1}_{n_1}\cdots \Jq^{a_p}_{n_p}\hspace{-1pt}: \, + \, \dots\,,
\end{equation}
where the dots involve products $:\!\Jq^{a_1}_{n_1}\cdots \Jq^{a_q}_{n_q}\!:$ with a lower number of modes $q< p$, summed over $(n_1,\dots,n_q)\in\Z^q$ with the condition $n_1+\dots+n_q=0$. This last constraint, together with the normal-ordering, ensures that these infinite sums have growing positive modes placed on the right, so that they finitely truncate on affine Verma modules and thus yield well-defined operators.

Recall moreover that the classical limit of quantum operators is obtained by taking $\hbar\to 0$ with $\Jq^a_n=J^a_{\cl,n}/\hbar$. With this in mind, we see that the top term in equation \eqref{eq:KpQuant} dominates in this limit and recreates the expected semi-classical behaviour \eqref{eq:KClassLimit}. In other words, the additional terms in equation \eqref{eq:KpQuant}, contained in the dots, can be interpreted as quantum corrections to $\Kq^{(p)}$, arising from ordering issues and the renormalisation procedure. 

We further expect the products $:\!\Jq^{a_1}_{n_1}\cdots \Jq^{a_q}_{n_q}\!:$ in these quantum corrections to be contracted with ad-invariant $q$-tensors $\tau^{a_1\dots a_q}$ on the $\su(2)$ algebra. This property translates the invariance of $\Kq^{(p)}$ under the global SU(2)-symmetry, \textit{i.e.} equation \eqref{eq:Ksu2}.

Last but not least, the desired commutativity property \eqref{eq:IntK} is the quantum analogue of equation \eqref{eq:PbK} and encodes integrability. It is quite less trivial than the other conditions and requires a very fine-tuned choice of quantum corrections. There lies the remarkable power of the procedure of~\cite{Bachas:2004sy,Gaiotto:2020fdr,Gaiotto:2020dhf}, which (conjecturally) produces such commuting operators $\Kq^{(p)}$, as has been checked for various cases in these works. In contrast, the commutativity with the higher-spin local IMs $\Iq^{(2p-1)}$ has not been investigated so far in the literature. As we shall see below on simple examples, it seems to also completely fix the quantum corrections in $\Kq^{(p)}$ and to single out the local IMs $\Iq^{(2p-1)}$ found in Subsection \ref{subsec:quant-Hierarchy} (compared to the KdV charges).

\paragraph{An ansatz for $\bm{\Kq^{(3)}}$.} It will be enough for our discussion to focus on the first non-trivial Kondo IM $\Kq^{(3)}$. Its classical version was given in terms of Kac-Moody modes in equation \eqref{eq:K3Class}. A proposal for the quantisation of this IM is the one produced by the renormalisation procedure of~\cite{Bachas:2004sy,Gaiotto:2020fdr,Gaiotto:2020dhf}. Here, we will stay agnostic and ignore this result for the moment, trying to follow a more naive approach. Following the general discussion above, a natural ansatz for this quantum IM is
\begin{equation}\label{eq:K3Quant}
    \Kq^{(3)} = -2\pi^2\,\left(\sum_{n> 0} \frac{\ri}{n}\,\varepsilon^{abc}\,\Jq^a_{-n}\,\Jq^b_{0}\,\Jq^c_{n} + \beta \sum_{n>0} \frac{2}{n} \Jq^a_{-n}\,\Jq^{a}_{n} \right)\,,
\end{equation}
where $\beta$ is a parameter (free for the moment). The first term is simply the naive normal-ordered quantisation of the classical expression \eqref{eq:K3Class}, as in the general equation \eqref{eq:KpQuant}. To motivate the second term, it is useful to study other possible choices of ordering. For instance, one could consider the ordering $\varepsilon^{abc}\,\Jq^a_{-n}\,\Jq^c_{n}\,\Jq^b_{0}$. Putting this back in normal-order creates a correction proportional to $\varepsilon^{abc}\,\Jq^a_{-n}\,[\Jq^b_{0},\Jq^c_{n}]=\ri\,\varepsilon^{abc}\varepsilon^{bcd}\,\Jq^a_{-n}\Jq^d_n=2\ri\, \Jq^a_{-n}\Jq^a_n$, of the same form as the second term in equation \eqref{eq:K3Quant}. We therefore decide to include this term in our ansatz with an arbitrary coefficient $\beta$, interpreting it as a quantum correction (indeed, being formed by 2 Kac-Moody modes and not 3, this term will be subdominant in the classical limit compared to the first one). Note that there are other possible choices of ordering: a careful analysis shows that putting those back in normal-order only creates the same type of corrections as above, thus justifying our ansatz.\footnote{One could also imagine that the quantisation creates completely new terms, which are not simply interpreted as ordering ambiguities. However, taking into account our expectation that these terms should involve only 1 or 2 Kac-Moody modes, contracted with invariant $\su(2)$-tensors and commuting with $\Lq_0$, we are naturally led to corrections of the form $\Jq^a_{-n}\Jq^a_n$. In our ansatz, we make the choice of including such a term with a coefficient proportional to $1/n$: this is quite natural so as to keep only expressions with the same ``level of non-locality'' than our starting point (recalling that divisions by the mode number $n$ heuristically amount to integrating a field).}

The renormalisation procedure of~\cite{Bachas:2004sy,Gaiotto:2020fdr,Gaiotto:2020dhf} produces an expression for $\Kq^{(3)}$ exactly of the form \eqref{eq:K3Quant}, with $\beta=1$. As we shall see below, we recover the same result by requiring the commutativity of $\Kq^{(3)}$ with the higher-spin local IM $\Iq^{(3)}$ (we note that $\Kq^{(3)}$ automatically commutes with $\Iq^{(1)}=\Lq_0-\frac{c}{24}$ due to the form of the ansatz \eqref{eq:K3Quant}, with mode numbers summing to zero).

\paragraph{Commutativity with $\bm{\Iq^{(3)}}$.} Recall the expression \eqref{eq:I3Modes} of $\Iq^{(3)}$ in terms of modes. As explained in Section \ref{sec:diagonalization}, the action of this operator on the Verma module $\mathcal{V}_{k,\ell}$ stabilises the subspaces $\Upsilon_{k,\ell}^{(n,m)}$ formed by $\su(2)$-highest-weight states of isospin $\ell+n-m$ and conformal dimension $h_\ell + n$. Since the Kondo operator $\Kq^{(3)}$ commutes with $\Lq_0$ and $\Jq_0^a$, it is clear that it also stabilises $\Upsilon_{k,\ell}^{(n,m)}$. In particular, we recall that these subspaces are of finite dimension, so that the action of $\Iq^{(3)}$ and $\Kq^{(3)}$ on it can be written in terms of finite matrices. This gives a particularly convenient way to check the commutativity of these operators.

For this check to be non-trivial, we need to consider subspaces $\Upsilon_{k,\ell}^{(n,m)}$ of dimension strictly greater than 1. The simplest case is that of $\Upsilon_{k,\ell}^{(2,1)}$, which we recall is spanned by the two vectors $\ket{\psi_{1,2}}$ defined in \eqref{eq:eigensystem_degeneracies}. In particular, we wrote the action of $\Iq^{(3)}$ on these vectors in equation \eqref{eq:I3n2j1}. It is a simple exercise to also compute the action of the Kondo operator \eqref{eq:K3Quant}, yielding
\begin{subequations}
\begin{align}
   -(2\pi^2)^{-1}\Kq^{(3)}\ket{\psi_1}&= \bigl(\ell(2+k-2\beta) + 
 2(2+k) \beta - 2 -\ell^2\bigr) \ket{\psi_1} -\ket{\psi_2}\,, \\[5pt]
 -(2\pi^2)^{-1}\Kq^{(3)}\ket{\psi_2}&=\bigl( 2 \ell^3 + \ell^2 (2-k+4\beta) -  2\ell (4+k-2\beta+k\beta) \bigr)\ket{\psi_1}\nonumber\\[2pt]
 & \hspace{30pt} + \bigl( \ell (k-2-4\beta) - 2\ell^2 + 
 4(1+k)\beta \bigr) \ket{\psi_2}\,.
\end{align}
\end{subequations}
It is then straightforward to compute the commutator $\bigl[\Iq^{(3)},\Kq^{(3)}\bigr]$ on this space. Remarkably, we find that this commutator vanishes if and only if $\beta=1$. As mentioned above, this is the value predicted by the renormalisation procedure of~\cite{Bachas:2004sy,Gaiotto:2020fdr,Gaiotto:2020dhf}. We thus recover their construction by requiring commutativity between local and non-local IMs.

\paragraph{Fixing $\bm{\Iq^{(3)}}$.} Recall from Subsection \ref{subsec:quant-Hierarchy} that the first local IMs themselves were found by starting with a general ansatz and imposing their commutativity spin by spin. We then found two independent solutions: the KdV charges, built only in terms of the stress-energy tensor, and a new integrable structure which we claimed is the most natural one for the WZNW model. A natural question at this point is whether this choice of local charges is also forced by commutativity with Kondo-like non-local IMs. To explore this, we take a step back and forget about the results of Subsection \ref{subsec:quant-Hierarchy} for a while, going back to the general ansatz \eqref{eq:W4} for (the density of) $\Iq^{(3)}$. For convenience, we recall it here and also give its corresponding mode expansion:
\begin{align}\label{eq:kondo_local_mode_general}
    \Iq^{(3)} &= \frac{L^3}{16\pi^4} \int_0^L \left( \bigl(\Tq(x)^2\bigr) + \frac{2\alpha}{3(k+2)} \bigl( \p\Jq^a(x)\,\p\Jq^a(x)\bigr) \right)\dd x \\ &  = 2\sum_{n=1}^{+\infty}\Lq_{-n}\Lq_n+\Lq_0^2-\frac{c+2}{12}\Lq_0+\frac{c(5c+22)}{2880} +\frac{4}{3}\frac{\alpha}{k+2}\sum_{n=1}^\infty n^2 \Jq^a_{-n}\Jq^a_{n} + \frac{\alpha\,c}{360}\,, \nonumber
\end{align}
depending on a free parameter $\alpha$. With these notations, the KdV charge corresponds to $\alpha=0$ while the new ``WZNW charge'' of Subsection \ref{subsec:quant-Hierarchy} is obtained for $\alpha=1$. 

We can now recompute the action of $\Iq^{(3)}$ on $\Upsilon_{k,\ell}^{(2,1)}$ for this more general operator and impose the commutativity with $\Kq^{(3)}$. This turns out to impose only one relation between the parameters $\alpha$ and $\beta$ entering the ansatz of $\Iq^{(3)}$ and $\Kq^{(3)}$, namely $\alpha=2\beta-1$. In particular, this is not enough to completely fix our IMs. We thus push to the next non-trivial subspace $\Upsilon_{k,\ell}^{(2,2)}$, which turns out to be 3-dimensional. The results being rather heavy, we gather them in Appendix \ref{App:n2j0}. In the present discussion, suffices to say that the commutativity $\bigl[\Iq^{(3)},\Kq^{(3)}\bigr]=0$ on this subspace forces both $\alpha=1$ and $\beta=1$. Therefore, we find that the existence of one non-trivial Kondo IM $\Kq^{(3)}$ inside the integrable structure also fixes the local charge $\Iq^{(3)}$ to be the one we advocated for in Subsection \ref{subsec:quant-Hierarchy} (in particular, this condition excludes the KdV charges). This is one of the main reason why we expect this new local integrable structure to be quite natural for the WZNW model.

We have performed similar checks for the case $\widetilde{\Upsilon}_{4,1}^{(3,2)}$ considered in Subsection \ref{subsec:hints_simple_spectrum}. In particular, commutativity of $\Iq^{(3)}$ and $\Kq^{(3)}$  again fixes $\alpha=\beta=1$. Moreover, $\Kq^{(3)}$ also has simple spectrum on this subspace, hence lifting its Virasoro degeneracies.

\paragraph{Discussion.} Note that the only checks of $\bigl[\Iq^{(3)},\Kq^{(3)}\bigr]=0$ that we have performed so far were restricted to simple subspaces $\Upsilon_{k,\ell}^{(n,m)}$, so that we have not yet proved this commutativity on the whole Hilbert space (contrarily to that among local IMs, which we could study directly in the chiral algebra). However, based on our preliminary results and those of~\cite{Gaiotto:2020fdr,Gaiotto:2020dhf}, we are confident that the integrability condition \eqref{eq:IntK} holds on the whole Hilbert space, for well-chosen operators $\Iq^{(2p-1)}$ and $\Kq^{(p)}$. Moreover, we conjecture that commutativity with either $\Iq^{(3)}$ or $\Kq^{(3)}$ is enough to completely fix all the other operators in the integrable structure. 

Once properly constructed, these operators should then be simultaneously diagonalisable on the Verma module $\mathcal{V}_{k,\ell}$. In particular, the eigenvectors of $\Iq^{(3)}$ discussed in Section \ref{sec:diagonalization}, either through the direct diagonalisation or through the affine Bethe ansatz, should automatically be eigenvectors of the Kondo operators $\Kq^{(p)}$. The authors of~\cite{Gaiotto:2020dhf} have proposed a quite simple conjecture for the eigenvalue of $\Kq^{(3)}$ on the Bethe vector $\ket{\{t_i\}}$, expressed in terms of a ``supersymmetric'' sum of the Bethe roots $t_i$. Moreover, they have shown evidence that the eigenvalues of $\Kq^{(p)}$ on this state are encoded into well-chosen Wronskians of the ODE \eqref{eq:ODE}, showing that the non-local Kondo IMs also naturally fit in the ODE/IQFT correspondence.

\section{Conclusion and perspectives}
\label{sec:Conc}

The goal of this paper was to investigate the first principle quantum integrability of the SU(2) WZNW model on the cylinder (motivated in the long term by the study of its $\lambda$-deformation). The first step in this program is to define the algebra of quantum field operators of the theory: as reviewed in Section \ref{sec:2D_CFT}, the conformal invariance provides a rigorous and already well-established answer to that question, in the form of the current chiral algebra generated by the Kac-Moody current $\Jq^a(x)$. This allowed us to proceed with the second and third steps of the program, that is to explicitly build and diagonalise the integrable structure of the model.

\paragraph{Quantum integrable structure.} Our first main achievement is the construction of the first few local higher-spin IMs $\Iq^{(2p-1)}$ within this quantum integrable structure. This was achieved in Section \ref{sec:structure} by starting with a quite general ansatz, guided by known classical results, and imposing the commutativity of these IMs spin by spin. This way, we found new local IMs inherent to the WZNW model, which are distinct from the well-known quantum KdV charges and which make use of the fully extended conformal algebra (rather than its Virasoro subalgebra only). More precisely, we have explicitly built the first four charges $\Iq^{(1)}$ to $\Iq^{(7)}$. A strong limitation of our analysis is the lack of a general construction for the full tower $\Iq^{(2p-1)}$. In view of the technicality of our direct approach, we expect that significant conceptual progress is needed to find such a general construction.

In Section \ref{sec:Kondo}, we have further reviewed the construction~\cite{Bachas:2004sy,Gaiotto:2020fdr,Gaiotto:2020dhf} of the non-local Kondo IMs $\Kq^{(p)}$ entering the quantum integrable structure and have investigated their commutativity with the local ones $\Iq^{(2p-1)}$. Our main observation was that commutativity of the first non-trivial elements $\Iq^{(3)}$ and $\Kq^{(3)}$ in these towers is a very strong constraint, which in particular singles out our specific choice of IM $\Iq^{(3)}$, while excluding the KdV one. This is a strong consistency check that the new local IMs proposed in Section \ref{sec:structure} form the correct quantised integrable structure of the SU(2) WZNW model, together with the non-local Kondo defects. Note that we have only checked the commutativity of $\Iq^{(3)}$ and $\Kq^{(3)}$ on specific subspaces of the affine Verma module, rather than on the full module. While this still gave very strong constraints on the form of the integrable structure, it would be interesting to show this commutativity on the full Hilbert space and for all the operators $\Iq^{(2p-1)}$ and $\Kq^{(p)}$.

\paragraph{Spectrum of the integrable structure.} 
In Section \ref{sec:diagonalization}, we then turned to step 3 of our first principle quantum integrability program, that is the diagonalisation of the integrable structure on the model's Hilbert space, which is built from $\widehat{\su(2)}_k$-Verma modules. The first local IM $\Iq^{(1)}$ in the integrable structure is the light-cone Hamiltonian, whose diagonalisation on the Verma module was already well-understood and provides the energy spectrum of the theory. However, this operator is quite degenerate, as there exist many states in the module with the same energy.  This is where the integrable structure, and in particular the higher-spin IMs $\Iq^{(2p-1)}$, enter the story. These operators stabilise the eigenspaces of $\Iq^{(1)}$ but generally mix the different states in these spaces. Their diagonalisation is therefore non-trivial and provides a finer decomposition of the energy eigenstates, associated with new quantum numbers. We have explicitly built the first few eigenstates of $\Iq^{(3)}$ and computed the corresponding eigenvalues. In all the examples that we have covered, this new quantum number was generically enough to lift the degeneracies in the energy spectrum, up to the finite $\su(2)$-symmetry of the WZNW model (with which the integrable structure commutes). Based on these preliminary results, we conjectured that this extends to the whole Verma module, \textit{i.e.} that the eigenstates are uniquely identified by their energy, their eigenvalue with respect to $\Iq^{(3)}$ and their $\su(2)$-quantum numbers. In contrast, we have exhibited explicit examples where the KdV charges $\Iq^{(2p-1)}_\KdV$ do not lift the energy degeneracies, providing another argument for the naturalness of the new IMs $\Iq^{(2p-1)}$ compared to the KdV ones.

\paragraph{Affine Bethe Ansatz, ODE/IQFT and Langlands duality. } The results summarised above concerned the direct diagonalisation of $\Iq^{(3)}$ on low-lying states in the Verma module. Remarkably, there exists a conjectural construction of all the eigenstates of the integrable structure, called the affine Bethe ansatz~\cite{Feigin:2007mr,Lacroix:2018fhf,Schechtman:1991hgd,Gaiotto:2020dhf}. In Section \ref{sec:diagonalization}, we have checked that this construction successfully reproduces the first few eigenstates found by direct diagonalisation of $\Iq^{(3)}$ (while similar checks were performed in~\cite{Gaiotto:2020dhf} for the non-local Kondo IM $\Kq^{(3)}$).

The affine Bethe ansatz provides the eigenstates of the integrable structure, but not its eigenvalues. Fortunately, there exists another set of powerful conjectures which describe this spectrum, based on an affine geometric Langlands duality~\cite{Feigin:2007mr,Lacroix:2018fhf,Gaiotto:2020dhf}. These conjectures fit into the so-called ODE/IQFT correspondence, which relates the spectrum of Integrable QFTs with properties of well-chosen ODEs. More precisely, this correspondence associates an explicit ODE to each Bethe eigenvector. The eigenvalues of the non-local Kondo IMs on this state are then expectedly encoded in specific Wronskians of this ODE, as checked on explicit examples in~\cite{Bazhanov:2003ua,Lukyanov:2003rt,Lukyanov:2006cu,Gaiotto:2020fdr,Gaiotto:2020dhf}. In this paper, we proposed an extension of this correspondence to the eigenvalues of the local IMs $\Iq^{(2p-1)}$, which we conjecture are encoded in the exact WKB expansion of the ODE. Again, we have checked this conjecture against the explicit diagonalisation of $\Iq^{(3)}$ on low-lying states.

The brute-force construction and diagonalisation of commuting IMs $\Iq^{(2p-1)}$ presented in this paper thus provides very strong support for these affine Bethe ansatz and ODE/IQFT conjectures. A very natural perspective for future developments is to find a complete proof of these statements. Moreover, one can hope that the existence of these general conjectures and their relations with deeper mathematical structures such as Langlands duality will provide us with useful hints on the construction of the full integrable structure itself.

\paragraph{Higher-rank WZNW.} Another obvious future direction is to extend the results and methods of the present paper to WZNW models on higher-rank Lie groups $G$ (we note that this could in principle also include non-compact Lie groups). We expect our direct approach to the construction of local IMs $\Iq^{(s)}$ to pull through and to yield similar results, albeit with higher technical difficulties and with the IM's spins following a different pattern (the affine exponents of the group $G$). At the classical level, Poisson-commuting local IMs of the $G$-WZNW model can be built following~\cite{Evans:1999mj,Evans:2000hx,Lacroix:2017isl} and we expect them to coincide with the (dispersion-less) $G$-Drinfeld-Sokolov hierarchy, which is a higher-rank generalisation of KdV. As in the SU(2) case treated in this paper, we expect this coincidence to be a classical accident and the natural quantum local IMs to be different from those of the Drinfeld-Sokolov integrable structure.\footnote{In the SU(2) case, one can build the quantum KdV IMs $\Iq^{(2p-1)}_\KdV$ inside of the current algebra, using the Sugawara realisation of Virasoro, so that $\Iq^{(2p-1)}$ and $\Iq^{(2p-1)}_\KdV$ define two consistent quantisations of the same classical local IMs. For higher-rank groups, we expect that the principal $\mathcal{W}(\g)$-algebra, in which the Drinfeld-Sokolov IMs are built, cannot be embedded inside of the current algebra. In this case, there might thus be a unique consistent quantisation of the classical local IMs. This point deserves further investigations.} For non-local Kondo IMs, the construction of~\cite{Bachas:2004sy,Gaiotto:2020fdr,Gaiotto:2020dhf} also applies for higher-rank groups and is still expected to produce commuting operators.

Concerning the diagonalization of the integrable structure, we stress that the affine Bethe ansatz conjecture naturally extends to the higher-rank case, using the results of~\cite{Schechtman:1991hgd,Feigin:2007mr,Lacroix:2018fhf}. Similarly, an ODE/IQFT correspondence for these models can be formulated following~\cite{Feigin:2007mr,Lacroix:2018fhf,Gaiotto:2020dhf}, with the ODEs taking more complicated forms (for instance ODEs of order $N$ for $G=\;$SU(N)). It would be interesting to compare these conjectures with a direct diagonalization of the commuting IMs.

\paragraph{Level 1, spinons and Yangians.} WZNW models at level $k=1$ are simpler but still quite interesting theories. For instance, they are used to describe tensionless superstrings on AdS$_3$~\cite{Gaberdiel:2018rqv}. In condensed matter physics, they are also known to describe spinons and to admit a Yangian symmetry~\cite{Haldane:1992sj,Bernard:1994wg}. The latter is an infinite-dimensional algebra generated by the zero-modes of the Kac-Moody current and an additional non-local current.\footnote{See also~\cite{Bernard:1996bp} for a different action of the Yangian in WZNW models, this time for arbitrary level.} This formulation is also related to integrable many-body systems and long-range spin chains (such as the Calogero-Sutherland and Haldane-Shastry models). Furthermore, various results on the integrable structure and the ODE/IQFT correspondence of the WZNW model at level $k=1$ have been discussed in~\cite{Bazhanov:2003ua,Lukyanov:2003rt,Lukyanov:2006cu}. It would be interesting to investigate the relation between these different approaches and the $k=1$ specialisation of this paper and of~\cite{Gaiotto:2020fdr,Gaiotto:2020dhf}.

\paragraph{Quantum integrability for the $\bm\lambda$-model.} As mentioned in the introduction, the WZNW model is the UV fixed point of the so-called $\lambda$-model~\cite{Balog:1993es,Sfetsos:2013wia}. This is a non-conformal sigma-model, obtained as a relevant perturbation of the WZNW theory by the operator $\lambda\,\Jq^a\overline{\Jq}\null^a$, which breaks the chirality of the Kac-Moody currents and triggers an RG-flow of the parameter $\lambda$~\cite{Kutasov:1989dt,Itsios:2014lca}. Although this model is classically integrable, its first principle quantum integrability has not been established yet in the literature and is a challenging open question. The results of this paper and of~\cite{Gaiotto:2020fdr,Gaiotto:2020dhf} provide preliminary steps towards this objective at the conformal point $\lambda=0$. A very interesting (but difficult) perspective is therefore to extend these results to non-zero $\lambda$, by showing that the integrable structure of the WZNW CFT can be deformed to define commuting IMs of the $\lambda$-model, all along the RG-flow (it would then be interesting to compare this approach with the formalism of~\cite{Bernard:1990jw}, based on massive current algebras). As a first step, this question can be studied pertubatively in $\lambda$, using the techniques of~\cite{Zamolodchikov:1987jf,Zamolodchikov:1989hfa,Yurov:1989yu} (we note however that we expect the main non-trivial effects of the deformation to appear only at the 2nd order in $\lambda$, as is the case for instance for the RG-flow).

The approach advertised above is based on the microscopic field-theoretic description of the $\lambda$-model in finite volume, starting from its UV fixed-point. At the other end of the RG-flow, the $\lambda$-model is expected to become strongly-coupled in the IR and thus to develop a dynamical mass-gap. The large-distance physics of this model should then be described by massive asymptotic particles, obeying a factorised scattering theory (if we assume quantum integrability). However, it is very difficult to relate this formulation with the microscopic field-theoretic one, since the massive excitations will arise from complicated non-perturbative effects and should not be directly identified with quantum excitations of the sigma-model fields or of the Kac-Moody currents. Remarkably, and despite these difficulties, Ahn, Bernard and LeClair have proposed an exact description of the mass spectrum and of the S-matrix in~\cite{Ahn:1990gn} (see also~\cite{Hollowood:2015dpa}).
Moreover, these results have also been argued to arise from spin-chain discretisations in~\cite{Bazhanov:1989yk,Appadu:2017fff}. Furthermore, the Thermodynamic Bethe Ansatz of this model has been investigated in~\cite{Zamolodchikov:1991vg,Ravanini:1992fs,Hollowood:1993ac,Babichenko:2003rf}, providing a conjecture for its ground-state energy in finite volume.\footnote{See also~\cite{Evans:1994hi} for the energy in presence of a chemical potential. In particular, this result has been used to provide a non-trivial check of the conjectured S-matrix, by comparing it with a perturbative computation in the sigma-model. Finally, a non-linear integral equation for this model has been proposed in~\cite{Hegedus:2005bg}.} In principle, this result (and would-be generalisations to excited states and to eigenvalues of other IMs) provides a bridge to the UV physics and potential comparisons with the first-principle approach sketched above. These, however, constitute very long term perspectives.

\paragraph{Integrable sigma-models and Affine Gaudin Models.}
The $\lambda$-model is a prototypical example of the quite wider class of 2d integrable sigma-models (see for instance the review~\cite{Hoare:2021dix} and references therein). Classically, the framework of Affine Gaudin Models (AGMs)~\cite{Feigin:2007mr,Vicedo:2017cge} provides a unifying description of many theories in this class, and in fact has been used to systematically construct many new ones (as reviewed in the lectures~\cite{Lacroix:2023gig}).\footnote{The inclusion of coset sigma-models requires the use of gauged AGMs, where current algebras are essentially replaced by cosets thereof. The main ideas depicted in this conclusion still hold in this case, with appropriate technical adaptations.} These AGMs are integrable theories built from an arbitrary number of classical Kac-Moody currents (or their so-called Takiff generalisations~\cite{Vicedo:2017cge,Babichenko:2012uq,Quella:2020uhk}). In general, these currents are not chiral and the theory is not conformal at the quantum level, but the classical dynamics is known to be integrable. This construction forms a vast generalisation of the $\lambda$-model, which corresponds to the case with only two currents.

The quantum integrability of these theories is an important and vast open question (in fact, many of the methods used in the simplest examples of sigma-models, such as the integrable S-matrix bootstrap, have not been applied to these generalisations). Much in the spirit of the present paper, we can follow the same logic as for the $\lambda$-model and attack the problem of their first principle quantum integrability by their UV fixed points. This requires knowing their RG-flow, which has been derived in full generality in~\cite{Delduc:2020vxy,Hassler:2023xwn,Lacroix:2025ias}, at least at 1-loop. In well-behaved cases, the UV limit exists and is characterised by a chiral splitting of the theory: essentially, half of the Kac-Moody currents become left-moving, while the other half becomes right-moving. The classical integrable structure is then described by two independent AGMs, each built from half the number of currents, either left- or right-moving. We will refer to these as ``chiral AGMs''. This conformal setup can be quantised using the standard techniques of 2d CFTs, with the underlying chiral algebra being a multi-current generalisation of the current algebra of WZNW models (or a coset thereof in the gauged case). It was conjectured in~\cite{Feigin:2007mr,Lacroix:2018fhf} (see also~\cite{Frenkel:2016gxg,Lacroix:2018itd,Gaiotto:2020dhf,Kotousov:2021vih,Kotousov:2022azm,Franzini:2022duf}) that the integrable structure of these chiral AGMs can be quantised into commuting operators in the multi-current chiral algebra (although so far this has been explicitly done only for the simplest operators). This forms a quite natural generalisation of the WZNW integrable structure studied here, although there are a few differences which illustrate the richness of this generalised setup. For instance, we expect the local commuting IMs of these models to come into several infinite towers with the same spin pattern (as many towers as there are currents in the algebra), so that there exist more than one IM of each spin. More importantly, the integrable structure of general chiral AGMs should depend on external continuous parameters, which do not enter the definition of the underlying chiral algebra: consequently, the same algebra then contains an infinite number of different integrable structures (which we expect to be compatible with different integrable massive deformations of the same CFT).  

As well, the affine Bethe ansatz and ODE/IQFT conjectures studied in this paper for the WZNW model admit very natural generalisations to arbitrary chiral AGMs~\cite{Feigin:2007mr,Lacroix:2018fhf,Gaiotto:2020dhf,Schechtman:1991hgd}. In fact, the general AGM framework and its relation to affine Langlands duality were the main guides for the formulation of these conjectures in the WZNW case. Finally, the long term goal would be to extend these conformal integrable structures along the RG flow of well-chosen massive deformations to achieve the first principle quantum integrability of (most) 2d integrable sigma-models and develop a massive PDE/IQFT description of their spectrum (in the spirit of~\cite{Lukyanov:2010rn,Bazhanov:2013cua}). This clearly represents a very long term perspective and much work remains to be done.

\paragraph{The problem of non-utralocality.} Readers familiar with integrable sigma-models might wonder how the above program sits with respect to the problem of non-ultralocality~\cite{Maillet:1985fn,Maillet:1985ek}, which is well-known to hinder the proof of their first-principle quantum integrability. To answer that question, let us start by briefly recalling the nature of this problem. It originates from the presence of derivatives of the Dirac distribution in the Poisson bracket of the Lax connection. These terms create ambiguities in the Poisson algebra of the monodromy matrix and thus make its quantisation difficult. It is worth noting that taking the conformal limit of integrable sigma-models, for example by going from the $\lambda$-model to the WZNW model, does not cure this issue (for instance, the monodromy matrix of the WZNW model is the path-ordered exponential of the Kac-Moody current, whose Poisson bracket contains a non-ultralocal $\delta'$-term). Crucially, the quantisation program discussed in the present paper avoids the non-ultralocality problem in two ways.

First, we note that the non-ultralocality is an obstacle for computing the Poisson algebra of the full monodromy matrix, but not of its trace, from which we extract the Poisson-commuting non-local IMs. The Kondo defects considered in~\cite{Bachas:2004sy,Gaiotto:2020fdr,Gaiotto:2020dhf} are meant as a quantisation of these IMs and thus do not suffer from non-ultralocality: consistently, this construction allows us to build these defects as well-defined operators on the Hilbert space, which furthermore seem to be pairwise commuting, without ambiguities. We note however that this commutativity has only been checked on simple explicit examples: it is not obvious at this stage whether there could exist a general proof of this statement which would not require computing the algebra of the full monodromy first and which would then completely avoid the non-ultralocality problem.\footnote{As an alternative, it is worth mentioning that preliminary results for other models~\cite{Bazhanov:2018xzh,Kotousov:2022azm} suggest that the non-ultralocality problem might be cured by appropriate prescriptions for the ambiguities and that one could hope for an RTT-relation of the full quantum monodromy. Other approaches~\cite{Freidel:1991jx} suggest to widen the type of algebra expected for this object, by considering more general relations than the RTT one. At the moment, there seems to be no definitive and general answer to that problem, which remains a very interesting question for future developments.}

The second way in which our quantisation program can avoid the non-ultralocality problem is by focusing on the higher-spin local IMs. Indeed, non-ultralocality is really an obstacle to quantising the algebra of monodromy elements, yet in the classical integrable structure, the local IMs are not extracted from this monodromy but rather from an alternative method. Commutators of local charges do not suffer any ambiguity from the presence of derivatives of $\delta$-distributions, contrarily to non-local quantities, and one might thus hope for a direct quantisation of the local IMs without any problems with non-ultralocality. This is well illustrated in Section \ref{sec:structure} of this paper, where no issues of this sort arise in the direct construction of commuting local IMs in the WZNW model. Although these results concerned only the first few charges of a specific example, some preliminary developments~\cite{Lacroix:2018fhf,Lacroix:2018itd,Kotousov:2022azm,Franzini:2022duf} suggest that, in the long term, a general construction of all quantum local IMs of chiral affine Gaudin models might be possible without passing through monodromy-type constructions (thus avoiding the non-ultralocality problem).

\paragraph{Other integrable structures and massive deformations.} To end this conclusion, we note that other integrable structures for the WZNW model have been discussed in~\cite{Kobayashi:1989sg,Leblanc:1990fi,Bonora:1990hc,Durganandini:1991gma,Brazhnikov:1996fa}, motivated by different integrable massive deformations of this CFT. Contrarily to the $\Jq^a\overline{\Jq}\null^a$-perturbation underlying the present paper, these deformations do not preserve the full SU(2) symmetry of the theory and thus correspond to conformal local IMs which do not belong to the Casimir subalgebra. The integrable structures considered in~\cite{Kobayashi:1989sg,Leblanc:1990fi,Bonora:1990hc,Durganandini:1991gma,Brazhnikov:1996fa} are then different from the one built here (and are neither deformations or specialisations thereof). This illustrates the fact that the same underlying CFT can host several integrable structures, compatible with different massive deformations. It would be interesting to study the spectrum of these other structures and their potential inclusion in the ODE/IQFT and/or affine Gaudin formalisms.

\section*{Acknowledgements} 
We would like to thank Denis Bernard, Sibylle Driezen, Matthias Gaberdiel, Ji Hoon Lee, Beno\^it Vicedo, Jingxiang Wu and Jean-Bernard Zuber for useful and interesting discussions. We are particularly grateful to Sibylle Driezen and Jean-Bernard Zuber for helpful comments on the draft of this paper. This work is partly based on the Master thesis of A.M. prepared at ETH Z\"urich. We thank Niklas Beisert for acting as co-advisor for this thesis and for helpful interactions and comments during its preparation. A.M.~acknowledges support by the NCCR SwissMAP.

\appendix

\section{From the cylinder to the sphere}
\label{app:Sphere}

In the main text, we consider CFTs defined on the cylinder, with periodic spatial coordinate $x\in \R/L\Z$. However, it is often quite standard to study CFTs on the sphere. For completeness, we explain the link between these two formalisms in this appendix.

\paragraph{Coordinate change.} We begin by considering the change of coordinate $x\mapsto z=e^{2\ri\pi x/L}$. For $x\in \R/L\Z$, this $z$ belongs to the unit circle of the complex plane. Focusing on left-moving fields, we now re-establish the time-dependence, \textit{i.e.} replace $x$ by $x+t$, and further perform a Wick rotation to Euclidean time $\tau=\ri t$. We then get $z=e^{2\pi(\tau+\ri x)/L}$, which we interpret as a complex coordinate on the Riemann sphere $\CP$. As is usual in radial quantisation, the coordinate change $(x,\tau) \mapsto  z=e^{2\pi(\tau+\ri x)/L}$ then maps the Euclidean cylinder to the 2-punctured sphere $\CP\setminus\lbrace 0,\infty\rbrace$. Through these steps, left-moving (resp. right-moving) fields on the Lorentzian cylinder are mapped to holomorphic (resp. anti-holomorphic) fields on the sphere. In particular, the latter are then described by a Laurent expansion in $z$ (resp. $\overline{z}$), instead of a Fourier expansion in $x$.

\paragraph{Field transformation and primary fields.} The behaviour of a general  field $\Aq\in\Ac$ under an infinitesimal change of coordinate is controlled by its OPE with the stress-energy tensor (see for instance~\cite{Gaberdiel:1994fs}). This transformation rule can then be exponentiated to describe the behaviour under a finite change of variable, as for instance the one $x\mapsto z$ from the cylinder to the sphere. In this respect, the simplest fields are the so-called primary ones, whose OPE with $\Tq$ stops at order 2:
\begin{equation}
    \Tq(x)\Aq^{\text{pr}}(y) = \frac{\p\Aq^{\text{pr}}(y)}{x-y} + \frac{h(\Aq)\,\Aq^{\text{pr}}(y)}{(x-y)^2} + \reg\,.
\end{equation}
For such a primary field, the change of variable to the sphere simply reads
\begin{equation}
    \Aq^{\text{pr}}_{\text{sphere}}(z) = \left(\frac{\partial x}{\partial z}\right)^{h(\Aq^{\text{pr}})} \Aq^{\text{pr}}\bigl(x(z)\bigr)\,.
\end{equation}
Since $\frac{\partial x}{\partial z}=\frac{L}{2\ri\pi z}$, the Fourier expansion \eqref{mode_expansion_general} on the cylinder is then mapped to the Laurent expansion
\begin{equation}
    \Aq^{\text{pr}}_{\text{sphere}}(z) = \sum_{n\in\Z} \Aq^{\text{pr}}_n\,z^{-n-h(\Aq^{\text{pr}})}
\end{equation}
on the sphere. In particular, the Fourier modes and Laurent modes coincide (with the appropriate shift by $h(\Aq^{\text{pr}})$ in the Laurent expansion). This is a specificity of primary fields, which does not hold for more complicated ones.

\paragraph{Transformation of the stress-energy tensor.} The simplest example of a non-primary field is the stress-energy tensor $\Tq$. Indeed, its OPE \eqref{OPE_T} with itself contains a 4$^{\text{th}}$-order pole, encoding the central charge $c$. The behaviour of $\Tq$ under the change of coordinate $x\mapsto z$ can be derived from this OPE and reads
\begin{equation}
     \Tq_{\text{sphere}}(z)=\left(\frac{\partial x}{\partial z}\right)^{2}\Tq\bigl(x(z)\bigr) + \frac{c}{12}\{x;z\}\,,
\end{equation}
where we have introduced the Schwarzian derivative
\begin{equation}
    \{x;z\}=\frac{\partial_z^3x}{\partial_zx}-\frac{3}{2}\left(\frac{\partial^2_zx}{\partial_zx}\right)^2\,.
\end{equation}
For our specific change of variable, the latter is simply equal to $\frac{1}{2z^2}$ and the (shifted) Fourier expansion \eqref{modes_T_spher_vs_cylinder} is then mapped to the Laurent expansion
\begin{equation}\label{eq:Tsphere}
     \Tq_{\text{sphere}}(z)=\frac{1}{z^2}\left[\left(\frac{L}{2\ri\pi}\right)^{2}\Tq\bigl(x(z)\bigr) + \frac{c}{24} \right] = \sum_{n\in\mathbb{Z}}\Lq_n\, z^{-n-2} \,.
\end{equation}
This illustrates a general feature of non-primary fields: their Fourier modes on the cylinder and their Laurent modes on the sphere are related but not exactly the same. As anticipated in the main text, in the case of the stress-energy tensor, these Laurent modes are the standard Virasoro generators $\Lq_n$. For more complicated non-primary fields, the transformation rule is derived in a similar fashion but can become quite involved. In practice, we will not need it in this paper, as we will work exclusively on the cylinder.

\paragraph{Transformation of the Kac-Moody current.} Of particular interest for this paper is the chiral algebra of the WZNW model, generated by a Kac-Moody current $\Jq(x)=\Jq^a(x)\,T^a$. The OPE of this current with the stress-energy tensor is given by equation \eqref{eq:OpeTJ}, such that $\Jq^a(x)$ is a primary field of conformal dimension $h(\Aq)=1$. According to the discussion above, the Fourier and Laurent expansions of this field, on the cylinder and the sphere respectively, are then expressed directly in terms of the same modes:
\begin{equation}
    \Jq^a(x) = \left(\frac{2\ri\pi}{L}\right) \sum_{n\in\Z} \Jq^a_n \,e^{-2\ri\pi\,nx/L}\qquad \text{ and } \qquad \Jq^a_{\text{sphere}}(z) = \sum_{n\in\Z} \Jq^a_n z^{-n-1}\,.
\end{equation}
The WZNW stress-energy tensor is built as the normal ordered product $\Tq=(\Jq^a\Jq^a)/(k+2)$. On the sphere, the Laurent expansion of this field is then given by
\begin{equation}
    \Tq_{\text{sphere}}(z) = \sum_{n\in\mathbb{Z}}\Lq_n\, z^{-n-2}\,, \qquad \text{ with } \qquad \Lq_n=\frac{1}{k+2}\left(\sum_{m\geq0 }\Jq^a_{n-m}\Jq^a_m+\sum_{m<0 }\Jq^a_m \Jq^a_{n-m} \right)\,.
\end{equation}
Indeed, the normal ordering on the sphere simply consists of placing positive modes to the right and negative ones to the left. As illustrated by the third term in equation \eqref{NO_cyl_modes}, the normal ordering on the cylinder contains additional finite size corrections. Using this equation and the Kac-Moody OPE \eqref{kac_moody_OPE}, the Fourier expansion of the stress-energy tensor is then given by
\begin{equation}
    \Tq(x)= \left(\frac{2\ri\pi}{L}\right)^2\sum_{n\in\mathbb{Z}}\Tq_n\, e^{-2\ri\pi\,nx/L}
\end{equation}
with
\begin{align*}
    \Tq_n &= \frac{1}{k+2}\left(\sum_{m\geq0 }\Jq^a_{n-m}\Jq^a_m+\sum_{m<0 }\Jq^a_m \Jq^a_{n-m} + c_0 \lbrace \Jq^a \Jq^a \rbrace^{(0)}_n  + c_1 \lbrace \Jq^a \Jq^a \rbrace^{(1)}_n  \right) \\
    &= \Lq_n + \frac{1}{k+2} \left( \frac{1}{2}\ri\,\varepsilon^{aac}\Jq^c_n - \frac{k}{24}\delta^{aa}\, \delta_{n,0} \right) \\
    &= \Lq_n - \frac{1}{24} \frac{3k}{k+2}\delta_{n,0}\,.
\end{align*}
In the second equality, we have used $c_0=\frac{1}{2}$ and $c_1=-\frac{1}{12}$, while the third equality follows from $\varepsilon^{aac}=0$ by skew-symmetry and $\delta^{aa}=\dim\su(2)=3$. The second contribution in the last line can be rewritten as $-\frac{c}{24}\,\delta_{n,0}$, in terms of the central charge \eqref{eq:cWZW}: this is the standard finite-size shift in the Fourier mode $\Tq_0$ on the cylinder compared to the Laurent mode $\Lq_0$ on the sphere. The appearance of this shift was explained earlier in this appendix for a general CFT, by considering the behaviour of $\Tq$ under a change of coordinate. We recover it here by a different approach in the case of the WZNW model, viewing $\Tq$ as a composite field built from the Kac-Moody current and applying the rule \eqref{NO_cyl_modes} for the normal ordering on the cylinder.

\section{Action of $\Iq^{(3)}$ and $\Kq^{(3)}$ on $\Upsilon_{k,\ell}^{(2,2)}$}
\label{App:n2j0}

In this appendix, we describe the action of the IMs $\Iq^{(3)}$ and $\Kq^{(3)}$ on the degenerate subspace $\Upsilon_{k,\ell}^{(2,2)}$ of the Verma module. It lies at depth $2$ and consists of the $\su(2)$ highest-weight states of isospin $\ell$. A basis was given in \eqref{eq:Chi}, which we recall here for convenience:
\begin{subequations}\label{eq:Chi_app}
\begin{align}
    \ket{\chi_1} &= \Jq^3_{-2}\ket{\ell} + \Jq^3_{-1} \, \Jq^3_{-1} \ket{\ell} + \Jq^-_{-1} \, \Jq^+_{-1} \ket{\ell}\,, \\[2pt]
    \ket{\chi_2} &= \Jq^+_{-2}\,\Jq^-_0\ket{\ell} - 2\ell\, \Jq^-_{-1}\,\Jq^+_{-1}\ket{\ell} - 2\ell\, \Jq^3_{-1} \, \Jq^3_{-1} \ket{\ell}\,, \\[2pt]
    \ket{\chi_3} &= \Jq^+_{-1}\,\Jq^+_{-1}\,\Jq^-_{0}\,\Jq^-_{0} \ket{\ell} + 2(2\ell-1) \Jq^3_{-1}\,\Jq^+_{-1}\,\Jq^-_{0}\ket{\ell} - 2\ell(2\ell-1) \Jq^-_{-1}\,  \Jq^+_{-1} \ket{\ell}\,.
\end{align}
\end{subequations}
For the purpose of Subsection \ref{subsec:non_loc_quantum_IMs}, we initially add a free coefficient $\alpha$ in the definition of $\Iq^{(3)}$, as in equation \eqref{eq:kondo_local_mode_general} (we will later specialise to $\alpha=1$, corresponding to the true IM). The action of this operator on the above vectors reads
\begin{subequations}\label{eq:I3_chi}
\begin{align}
\Iq^{(3)} \ket{\chi_1}=&\,\frac{1}{960 (k+2)^2}\Bigg(k^2 (161 (8 \alpha +37)+9600 \ell )+4 k (1284 \alpha +20 \ell  (96 \alpha +43 \ell +475)+4731)\nonumber\\
   & \hspace{70pt} +320 (16 \alpha +\ell  (48
   \alpha +\ell  (3 \ell  (\ell +2)+26)+119)+44)\Bigg) \ket{\chi_1}\nonumber \\
 &+ \frac{2}{k+2} \ket{\chi_2} +\frac{ 4 \alpha +5 k-4 \ell +8}{k+2} \ket{\chi_3}\, ,
\end{align}
\begin{align}
 \Iq^{(3)} \ket{\chi_2}= &\,\frac{4\ell  (2 \ell -1) (k (2 \alpha -2 \ell +3)+2 \ell  (-2 \alpha +2 \ell +3)+18)}{k+2} \ket{\chi_1}\nonumber\\
 &+\frac{1}{960 (k+2)^2}\Bigg( (1288 \alpha +3077) k^2+4 k \left(1284 \alpha +860 \ell ^2+4700 \ell +1371\right)\nonumber\\
 &\hspace{80pt} +320 (16\alpha +\ell  (\ell  (3 \ell  (\ell +2)+26)+119)-4)\Bigg)\ket{\chi_2}\nonumber \\
 &+\frac{4 (2 \ell -1) \left(k (\alpha -\ell +2)+2 \ell ^2-2 (\alpha +1) \ell +8\right)}{k+2}\ket{\chi_3}\, ,\\[7pt]
   \Iq^{(3)} \ket{\chi_3}=&\,  \frac{2\ell  \left(4 \alpha  (k-2 \ell +1)-10 k \ell +k+8 \ell ^2-4 \ell +20\right)}{ (k+2)}\ket{\chi_1} + \frac{4-4 \ell }{k+2} \ket{\chi_2}\nonumber\\
   &+\frac{1}{960 (k+2)^2}\Bigg(k^2 (5128 \alpha -9600 \ell +6917)+4 k (4164 \alpha +20 \ell  (-96 \alpha +235 \ell -389) +9051)\nonumber\\
&\hspace{75pt} +320 (20 (2 \alpha +7)+\ell  (-48 \alpha +\ell  (3 \ell  (\ell +2)+122)-73))\Bigg)\ket{\chi_3} \,.
\end{align}
\end{subequations}
Although very lengthy, these expressions just form an explicit 3 $\times$ 3 mixing matrix. For $\alpha=1$, we get the action of the true local IM $\Iq^{(3)}$, which one can diagonalise to get its eigenvalues and eigenvectors. This involves solving a lengthy cubic equation, making the computation quite heavy analytically. We have checked that these eigenvalues are again all distinct for generic $(k,\ell)$, with some degeneracies appearing at specific low-lying values of these parameters.\\ 

We now play the same game for the non-local IM $\Kq^{(3)}$, depending on an arbitrary coefficient $\beta$ as in equation \eqref{eq:K3Quant}. Its action on $\ket{\chi_{1,2,3}}$ reads
\begin{subequations}\label{eq:K3_chi}
\begin{align}  
 -(2\pi^2)^{-1}\Kq^{(3)}\ket{\chi_1}=&\,  (-4 + 3 \ell + 4 \beta - 
      2 \ell (\ell + \beta) + k (-2 + \
\ell + 2 \beta))\ket{\chi_1} + 
  \frac{1}{2}\ket{\chi_2}\nonumber\\
  &+ \frac{1}{2}(5 + k - 2 \ell - 2 \beta)\ket{\chi_3}\,,\\[3pt]
-(2\pi^2)^{-1}\Kq^{(3)}\ket{\chi_2}=&\,\ell (2\ell -1) (-k + 2 \ell) (-1 + 
    2 \ell + 2 \beta)\ket{\chi_1} - 
 \frac{1}{2} (k + 2 \ell + 4 \ell^2 - 
    4 (2 + k) \beta)\ket{\chi_2}\nonumber\\
    &- 
 \frac{1}{2} (2\ell -1) (8 + 6 \ell - 
    4 \ell (\ell + \beta) + k (1 + 
       2 \ell + 2 \beta))\ket{\chi_3}\,, \\[3pt]
 -(2\pi^2)^{-1}\Kq^{(3)}\ket{\chi_3}=&\,\ell (5 + k (3 - 2 \ell - 2 \beta) - 
    2 \beta + 4 \ell (-1 + \ell + \beta))\ket{\chi_1} - 
 \frac{1}{2} (1 + 2 \ell)\ket{\chi_2} \nonumber\\
 &+ \left(-\frac{1}{2} (5 + k - 2 \ell) (1 + 
       2 \ell) + (3 + k + 2 \ell) \beta\right)\ket{\chi_3}\,,
\end{align}
\end{subequations}
up to an irrelevant $-(2\pi^2)^{-1}$ factor. Again, while lengthy, these equations simply define a $3\times3$ mixing matrix. We can now compute the commutator of the matrices representing $\Iq^{(3)}$ and $\Kq^{(3)}$, and find that  it vanishes \textit{if and only if $\alpha=1$ and $\beta=1$}. We thus recover the expected expressions for the local and non-local IMs $\Iq^{(3)}$ and $\Kq^{(3)}$ by imposing their mutual commutativity (in a natural ansatz). This was the conclusion announced in Subsection \ref{subsec:non_loc_quantum_IMs}.
\newpage 

\bibliographystyle{JHEP.bst}
\bibliography{refs}

\end{document}